\documentclass[longauth]{aa}

\usepackage{graphicx}
\usepackage{natbib}
\usepackage{scalerel}
\usepackage{makecell}
\usepackage{multirow}
\usepackage{txfonts}
\usepackage[table]{xcolor}
\usepackage[autopunct=true]{csquotes}
\usepackage{placeins}
\usepackage{txfonts}
\usepackage[pdfencoding=auto,psdextra]{hyperref}
\hypersetup{
    colorlinks=true,
    linkcolor=blue,
    filecolor=magenta,      
    urlcolor=blue,
    citecolor=blue
}
\FloatBarrier

\usepackage[utf8]{inputenc}
\usepackage{multirow}

\begin{document} 

\title{Pushing JWST to the extremes: search and scrutiny of bright galaxy candidates at z$\simeq$15-30}
   \author{M. Castellano\inst{1}, A. Fontana\inst{1}, E. Merlin\inst{1}, P. Santini\inst{1}, L. Napolitano\inst{1,2}, N. Menci\inst{1}, P. G. P\'erez-Gonz\'alez\inst{3},  A. Calabr\`{o}\inst{1}, D. Paris\inst{1}, L. Pentericci\inst{1}, J. A. Zavala \inst{4}, M. Dickinson\inst{5}, S. L. Finkelstein \inst{6}, T. Treu \inst{7}, R. O. Amorin \inst{8}, P. Arrabal Haro \inst{9}, P. Bergamini\inst{10,11}, L. Bisigello\inst{12}, M. Catone\inst{12}, E. Daddi\inst{13}, P. Dayal\inst{14}, A. Dekel\inst{15}, A. Ferrara\inst{16}, F. Fortuni\inst{1}, G. Gandolfi\inst{12,17}, M. Giavalisco\inst{4}, C. Grillo\inst{10,18}, S. T. Guida \inst{1,19}, N. P. Hathi \inst{20}, B. W. Holwerda \inst{21}, A. M. Koekemoer\inst{20}, V. Kokorev\inst{6}, Z. Li\inst{15}, M. Llerena\inst{1}, R. A. Lucas\inst{20}, S. Mascia\inst{1}, B. Metha\inst{22}, T. Morishita\inst{23}, T. Nanayakkara\inst{24}, F. Pacucci\inst{25}, G. Roberts-Borsani\inst{26}, G. Rodighiero\inst{11,16}, P. Rosati\inst{27}, V. Salazar\inst{3}, R. Schneider \inst{2,1}, R. S. Somerville\inst{28}, A. Taylor\inst{5}, M. Trenti\inst{21,29}, A. Trinca\inst{30,1,2}, X. Wang \inst{31,32,33}, P. J. Watson \inst{11}, L. Yang\inst{34}, L. Y. A. Yung \inst{19}}

   \institute{INAF - OAR, via Frascati 33, 00078 Monte Porzio Catone (Roma) - Italy 
   \and Dipartimento di Fisica, Sapienza, Universit\`a di Roma, Piazzale Aldo Moro 5, 00185 Roma, Italy 
      \and Centro de Astrobiología (CAB), CSIC–INTA, Cra. de Ajalvir km. 4, E28850, Torrejón de Ardoz, Madrid, Spain 
   \and University of Massachusetts, 710 N. Pleasant St, LGRB-520, Amherst, MA 01003, USA 
   \and NSF’s NOIRLab, Tucson, AZ 85719, USA 
    \and Department of Astronomy, The University of Texas at Austin, Austin, TX, USA 
   \and Department of Physics and Astronomy, University of California, Los Angeles, 430 Portola Plaza, Los Angeles, CA 90095, USA 
   \and Instituto de Astrof\'{i}sica de Andaluc\'{i}a (CSIC), Apartado 3004, 18080 Granada, Spain 
   \and Astrophysics Science Division, NASA Goddard Space Flight Center, 8800 Greenbelt Rd, Greenbelt, MD 20771, USA 
    \and Dipartimento di Fisica, Università degli Studi di Milano, via Celoria 16, I-20133 Milano, Italy 
    \and INAF – OAS, Osservatorio di Astrofisica e Scienza dello Spazio di Bologna, via Gobetti 93/3, I-40129 Bologna, Italy 
   \and INAF-Osservatorio Astronomico di Padova, Via dell'Osservatorio 5, 35122 Padova, Italy 
    \and CEA, IRFU, DAp, AIM, Universit\'e Paris-Saclay, Universit\'e Paris Cit\'e, Sorbonne Paris Cit\'e, CNRS, 91191 Gif-sur-Yvette, France 
    \and Kapteyn Astronomical Institute, University of Groningen, PO Box 800, 9700 AV Groningen, The Netherlands 
    \and Racah Institute of Physics, The Hebrew University of Jerusalem, Jerusalem 91904, Israel 
    \and Scuola Normale Superiore, Piazza dei Cavalieri 7, 56126 Pisa, Italy 20 
   \and Dipartimento di Fisica e Astronomia \textquote{G. Galilei}, Università di Padova, Via Marzolo 8, 35131 Padova, Italy 
      \and INAF – IASF Milano, via A. Corti 12, I-20133 Milano, Italy 
   \and Universit\`a di Napoli \textquote{Federico II}, C.U. Monte Sant’Angelo, Via Cinthia, 80126 Napoli, Italy 
   \and Space Telescope Science Institute, Baltimore, MD, USA 
   \and University of Louisville, Department of Physics and Astronomy, 102 Natural Science Building, 40292 KY Louisville, USA 
      \and School of Physics, The University of Melbourne, VIC 3010, Australia 
  \and  IPAC, California Institute of Technology, MC 314-6, 1200 East California Boulevard, Pasadena, CA 91125, USA 
     \and Centre for Astrophysics and Supercomputing, Swinburne University of Technology, PO Box 218, Hawthorn, VIC 3122, Australia 
   \and Center for Astrophysics, Harvard \& Smithsonian, 60 Garden St, Cambridge, MA 02138, USA 
   \and Department of Physics \& Astronomy, University College London, Gower St., London WC1E 6BT, UK 
      \and Dipartimento di Fisica e Scienze della Terra, Universit\`a degli Studi di Ferrara, Via Saragat 1, I-44122 Ferrara, Italy 
   \and Center for Computational Astrophysics, Flatiron Institute, 162 5th Avenue, New York, NY 10010, USA 
   \and ARC Centre of Excellence for All Sky Astrophysics in 3 Dimensions (ASTRO 3D), Australia 
   \and Como Lake Center for Astrophysics, DiSAT, Universit\`a degli Studi dell'Insubria,  via Valleggio 11, I-22100, Como, Italy 
   \and School of Astronomy and Space Science, University of Chinese Academy of Sciences (UCAS), Beijing 100049, China 
\and National Astronomical Observatories, Chinese Academy of Sciences, Beijing 100101, China 
\and Institute for Frontiers in Astronomy and Astrophysics, Beijing Normal University, Beijing 102206, China 
\and Laboratory for Multiwavelength Astrophysics, School of Physics and Astronomy, Rochester Institute of Technology, 84 Lomb Memorial Drive, Rochester, NY 14623, USA 
   }

   \date{...}

%
 
  \abstract
   {}
   {We investigate the galaxy UV Luminosity Function at $z\simeq15-30$ to constrain early galaxy formation scenarios aimed at explaining the mild evolution of the UV LF bright-end found by JWST at z$\approx$10-15.}
   {We designed customized Lyman-break color selection techniques to identify galaxy candidates in the redshift ranges $15 \leq z \leq 20$ and $20 \leq z \leq 28$. The selection was performed on the ASTRODEEP-JWST multi-band catalogs of the CEERS, Abell-2744, JADES, NGDEEP, and PRIMER survey fields, covering a total area of $\sim0.2$ sq. deg.}
   {We identify five candidates at $15 \leq z \leq 20$, while no objects are found based on the $z\gtrsim20$ color selection criteria. Despite exhibiting a $>$1.5 mag break, all the objects display multimodal redshift probability distributions across different SED-fitting codes and methodologies. The alternative solutions correspond to poorly understood populations of low-mass quiescent or dusty galaxies at z$\sim$3-7. This conclusion is supported by the analysis of five F200W-dropout objects that we find to be interlopers on the basis of NIRSpec PRISM spectra: four dusty star-forming galaxies at z$\sim$2.2-6.6, and a passive galaxy at z=4.91  with log$(M_{\rm star}/{\rm M}_{\odot}) \lesssim$ 9. We measured the UV luminosity function under different assumptions on the contamination level within our sample. We find that if even a fraction of the candidates is indeed at $z\gtrsim15$, the resulting UV LF points to a very mild evolution compared to estimates at $z<15$, implying a significant tension with existing theoretical models. In particular, confirming our bright ($M_{\text{UV}}<-21$) candidates would require substantial revisions to the theoretical framework. In turn, if all these candidates will be confirmed to be interlopers, we conclude that future surveys may need ten times wider areas to select $M_{\text{UV}}\lesssim-20$ galaxies at $z>15$. Observations in the F150W and F200W filters at depths comparable to those in the NIRCam LW bands are also required to mitigate contamination from rare red objects at z$\lesssim$8.
   }
   {}
\authorrunning{M. Castellano et al.}   
\titlerunning{Search and scrutiny of bright galaxy candidates at z$\simeq$15-30}   

\maketitle
\section{Introduction}\label{sec:intro}
Since the beginning of its operations, JWST has easily enabled the detection of galaxies beyond $z\sim10$, breaking the redshift barrier that was the consequence of the limited infrared (IR) sensitivity of HST, Spitzer and of ground--based telescopes. Several bright galaxies, more than expected on the basis of the observed evolution at $z=5-9$ or from theoretical models, were quickly detected by the very first Early Release Science observations \citep[e.g.,][]{Naidu2022,Castellano2022b,Finkelstein2022b}. This result has been later statistically corroborated by wider-area surveys \citep[e.g.,][]{Harikane2022b,Castellano2023a,McLeod2024,Donnan2024} and on the basis of spectroscopically confirmed samples \citep{Harikane2024,Napolitano2025}. In fact, spectroscopic confirmation and characterization have been extremely efficient at z$\sim$10-12 \citep[e.g.,][]{ArrabalHaro2023a, ArrabalHaro2023b,RobertsBorsani2024,Roberts-Borsani2025,Castellano2024,Napolitano2024b} and up to $z\sim14$ \citep{Carniani2024a,Carniani2024b,Naidu2025}. 

This situation is in stark contrast with the poor constraints available at $z>14.5$. Despite the fact that UV rest frame emission is, in principle, within NIRCam spectral coverage up to z$\sim$30, attempts to identify galaxies at $z>15$ in existing surveys have led to very few candidates \citep{Yan2023a,Yan2023b,Leung2023,Austin2023,Conselice2024,Robertson2024,Kokorev2024,Whitler2025,Gandolfi2025,PerezGonzalez2025}, none of which has yet been confirmed spectroscopically. The most striking example is object CEERS-93316, which appeared to be a strong $z\sim16$ candidate based on CEERS imaging data \citep{Donnan2023,Harikane2022b}, but which eventually proved to be a red galaxy at $z=4.9$ whose photometry appears consistent with a $z=16.2$ Lyman-break galaxy due to a very unfortunate combination of a red continuum with extremely strong rest-optical line emission \citep{ArrabalHaro2023b}.

From an observational point of view, several effects conspire to make selection at $z\gtrsim$15 more difficult. On the one hand, objects become fainter and are detected in a smaller number of photometric bands, making the detection of spectral breaks and the constraints on the UV continuum less significant. On the other hand, the contamination in photometric samples is expected to increase with redshift, as true sources become rarer relative to potential contaminants \citep{Vulcani2017}, and this trend may be worsened by new classes of poorly characterized low/intermediate-redshift objects entering the selection criteria \citep{Zavala2022,PerezGonzalez2023a,PerezGonzalez2024c,Glazebrook2023,Bisigello2023,Bisigello2025,Rodighiero2023,Gandolfi2025}. 

Breaking the z$\sim$15 barrier is fundamental for testing theoretical models of galaxy evolution and for approaching the epoch of formation of the first stars and first black holes. In fact, the different explanations that have been invoked to explain the mild evolution of the UV LF at z$\gtrsim$10 differ in their predictions at earliest times \citep{Kokorev2024,PerezGonzalez2025}. For instance, sustained high luminosity density beyond $z=15$ is favoured by changes in the initial mass function (IMF) or by an increased star-formation efficiency \citep[e.g.][]{Dekel2023,Trinca2024,hutter2025,Mauerhofer2025}, as well as by a rapid assembly of baryons \citep{Haslbauer2022,McGaugh2024}, while the prediction of an earlier phase of dusty-enshrouded star-formation \citep[e.g.][]{Ferrara2022,Ziparo2022} or alternative dark energy or dark matter scenarios \citep{Menci2024,Gandolfi2022} result in a sharp decline with redshift of the luminosity density.

In this paper we analyse the ASTRODEEP-JWST photometric sample \citep[M24 hereafter]{Merlin2024}, which provides consistent measurements on the major JWST deep surveys, to select bright galaxy candidates at z$\sim$15-30. We briefly present the dataset in Sect.~\ref{sec:obs}. In Sect.~\ref{sec:LBGDIAGRAMS} we describe our specific renditions of the Lyman-break technique and the results of our search for galaxy samples at $15 \leq z \leq 20$ and $20 \leq z \leq 28$. We investigate in detail alternative low-redshift solutions of our candidates, and the spectroscopic properties of five confirmed interlopers at z$\sim$2-7 in Sect.~\ref{sec:scrutiny}. The implications on the evolution of the UV LF and a comparison with theoretical models are presented in Sect.~\ref{sec:LF}, while Sect.~\ref{sec:future} explores the lessons learned from our analysis for designing future observations of galaxies at z$>$15. The results are summarised in Sect.~\ref{sec:summary}.

Throughout the paper we adopt AB magnitudes \citep{Oke1983}, a \citet{Chabrier2003} initial mass function (IMF) in the range 0.1-100 M$_{\odot}$, the \citet{Calzetti2000} attenuation law, and a flat $\Lambda$CDM concordance model (H$_0$ = 70.0~km~s$^{-1}$ Mpc$^{-1}$, $\Omega_M=0.30$).

\section{Observations and data analysis}\label{sec:obs}

We used JWST and HST photometric measurements from the ASTRODEEP-JWST catalogs presented in M24. We analysed the seven surveys comprising the public catalog release\footnote{\url{http://www.astrodeep.eu/astrodeep-jwst-catalogs/}}: CEERS \citep[ERS 1345, P.I. Finkelstein,][]{Finkelstein2025}; the JADES-GS (data release v2.0) and JADES-GN (v1.0) fields on the GOODS-South and GOODS-North footprints, respectively \citep[GTO 1180 and GTO 1210, P.I. Eisenstein,][]{Eisenstein2023} including FRESCO data \citep[GO 1895, P.I. Oesch,][]{Oesch2023}; the first-epoch imaging of the NGDEEP field \citep[Co-PIs Finkelstein, Papovich, Pirzkal,][]{Bagley2024}; the PRIMER (GO-1837, P.I. Dunlop) observations of the UDS and COSMOS fields in CANDELS \citep{Grogin2011,Koekemoer2011}; the A2744 field including JWST observations from GLASS-JWST \citep[ERS 1324, P.I. Treu,][]{Treu2022}, UNCOVER \citep[GO 2561, P.I. Labb\'e,][]{Bezanson2022}, DDT 2756 (P.I. Chen), and GO 3990 \citep[P.I. Morishita,][]{Morishita2024}. The considered observations cover a significant range in both area and depth, from relatively wide surveys as PRIMER-UDS ($\sim$250 sq. arcmin, 50\% completeness at mag$_{50} \sim$28.8 in the detection band), to deep pencil-beam pointings such as NGDEEP ($\sim$9.5 sq. arcmin, mag$_{50}\sim$30.8) and the lensed field A2744 ($\sim$46 sq. arcmin, mag$_{50}\sim$29.7).  We refer to M24 for details on the survey properties and photometric techniques. Briefly, sources in all fields were detected with \textsc{SExtractor} \citep{Bertin1996} on a weighted average of the NIRCam F356W and F444W images, which is also used to measure total fluxes in \citet{Kron1980} apertures using \textsc{A-PHOT} \citep{Merlin2019b}. Fluxes in the other bands were measured by scaling the aforementioned total flux according to the colors measured within optimal apertures on PSF-matched images \citep[see also][]{Merlin2022,Paris2023}. In the present paper we use total fluxes based on the optimal apertures as defined by M24, the signal-to-noise ratios are evaluated in 2 times the PSF full-width at half maximum (FWHM) for each band. The seven fields comprise a total of 531173 objects in an area of $\sim$615 sq. arcmin, making the ASTRODEEP-JWST the largest publicly released JWST catalogs available to date. The available imaging datasets are slightly different in the various fields. In particular, JWST NIRCam F090W is missing, or was not public at the time of the catalog, in the CEERS, NGDEEP and over most of the A2744 area, while the medium band F410M filter is not available in NGDEEP and in the GLASS-JWST observations of A2744. Most importantly, as discussed in M24, the HST coverage is even less uniform both in terms of depth and number of available filters. For the present work we built the weighted average stacks of all the ACS and WFC3 HST bands available in each field, respectively. The SNR of all ASTRODEEP-JWST sources was measured within an aperture with a diameter of 2 times the PSF FWHM on both the HST stacks, and used as described in the following section to constrain non-detection blueward of the Lyman-break.

\section{Lyman-break selection at z$\simeq$15-30}\label{sec:LBGDIAGRAMS}
We describe here the approach that we have used to select high-$z$ candidates, that is a modified version of Lyman--break color selection criteria \citep[][]{Giavalisco2002}. The selected objects are then inspected through a full photometric redshift analysis to investigate their reliability. 

\subsection{Color selection criteria}\label{sec:ColorSelec}

We defined color selection criteria for galaxies in two different redshift ranges, z$\sim$15-20 and z$\sim$20-30 on the basis of mock catalogues of objects at z$=$0-30 tailored to match the noise properties of our observations, as previously done for the z$\sim$9-15 range by \citet{Castellano2022b}. Namely, we use two different simulations. The first simulation is based on a catalogue comprising objects at 0$<$z$\leq$5 over an area of $\sim$0.12 sq. deg. generated with the Empirical Galaxy Generator (EGG) code \citep{Schreiber2017EGG}, which exploits empirical relations to reproduce the observed number counts and color distributions of galaxies at low and intermediate redshifts, including quiescent and dusty populations. The second simulation is based on the mock catalogs from the JAdes extraGalactic Ultradeep Artificial Realizations \citep[JAGUAR,][]{Williams2018}, including predicted NIRCam fluxes for objects at 0.2$<z<$15 and stellar mass log(M/M$_{\odot}$)$>$6 over an area of $\sim$0.34 sq. deg. JAGUAR provides a complementary test with respect to EGG also thanks to the inclusion of emission lines in the predicted SEDs.
We added sources at z$>$5 (z$>$15) to the EGG (JAGUAR) simulation following the evolving UV LF at z$\sim$5-10 \citep{Bouwens2021}, assuming no evolution beyond $z=10$ and artificially boosting the number counts at z$>$10 by a factor of 20 in order to provide sufficient statistics to design appropriate selection criteria. These high-redshift galaxies have been generated by randomly associating to each object a template from a library based on \citet[][BC03 hereafter]{Bruzual2003} models with metallicities 0.02 or 0.2 Z$_{\odot}$, 0$<$E(B-V)$<$0.2 and a constant star-formation history (SFH) to predict the relevant photometry.
The over-representation of high-redshift sources in the mock catalogs is taken into account by consistently scaling the relevant number counts when evaluating the selection criteria in terms of purity and completeness.  Finally, we assessed the potential contamination by late-type dwarf stars using synthetic JWST photometry for the models by \citet[][]{Marley2021} which include brown dwarfs and self-luminous extrasolar planets with $200 \leq T_{eff} \leq 2400$ and metallicity [M/H] from 0.5 to + 0.5. The brown dwarf models were normalized at 26.0$\leq$F444W$\leq$28.0 in 0.5 mag steps.
\begin{figure}
\centering
\includegraphics[trim={1.5cm 0.5cm 4.5cm 0.5cm},clip,width=\linewidth,keepaspectratio]{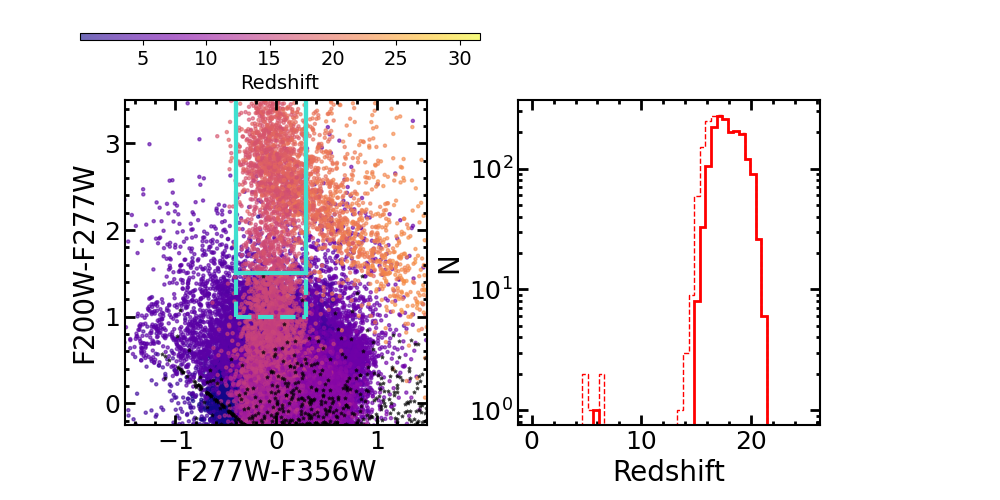}
\includegraphics[trim={1.0cm 0.5cm 4.5cm 0.5cm},clip,width=\linewidth,keepaspectratio]{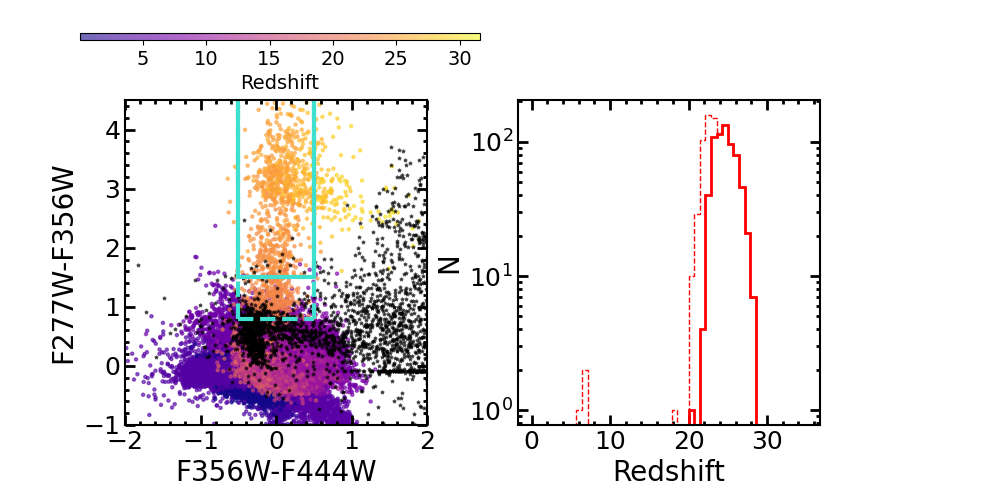}
\caption{Color selection diagrams (left panels) for the selection of galaxies at z$\sim$15-20 (top) and z$\sim$20-30 (bottom).  The cyan solid lines enclose the regions in which the reference sample to estimate the luminosity functions are selected. The cyan dashed lines enclose regions where the additional \textquote{extended samples} are selected. The relevant redshift distributions of the selected reference  (extended) samples are shown in the right panels as continuous (dashed) histograms. The points color-coded according to the relevant redshift show objects from a mock generated over an area of 0.12 sq. deg, with low-redshift populations generated through the EGG software \citep{Schreiber2017EGG}. Black stars show the position of brown dwarf models from \citet[][]{Marley2021}. All fluxes have been perturbed with realistic noise properties to reproduce the typical depth of the JADES-GS field. Similar diagrams have been analysed for all fields using both the EGG- and JAGUAR-based simulations described in Sect.~\ref{fig_LBGSIM}.} 
\label{fig_LBGSIM}
\end{figure} 
All the catalogues were perturbed by adding noise in order to reproduce the expected relation and scatter between magnitudes and errors in each band and in each of the analysed fields. 
 
After extensive testing, we first define a detection threshold corresponding to  SNR$>$10 in the detection band used by M24, i.e. F356W+F444W. We also define the  following selection criteria to identify objects at z$\sim$15-20 minimizing contamination from low-redshift sources:

\begin{equation}\label{col1}
\begin{aligned}
    &(F200W-F277W)>1.5\\
    &-0.4<(F277W-F356W)<0.4\\
    &(F356W-F444W)<0.5\\
\end{aligned}
\end{equation}

We require a signal-to-noise ratio SNR$<$2.0 in the F090W (where available), F115W and F150W bands, and in both the ACS and WFC3 stack, with at most one of these bands blueward of the break having SNR$>$1.5. In order to limit our sample to objects with continuous coverage redward of the Lyman break and to avoid spurious, single-band detections, we also require SNR$>$2 in each of the F277W, F356W, F444W bands. All the adopted signal-to-noise ratios are measured in 2$\times$FWHM apertures.

Similarly, we find that objects at z$\sim$20-28 are well identified by the following selection criteria, as shown in the bottom panel of Figure~\ref{fig_LBGSIM}:

\begin{equation}\label{col2}
\begin{aligned}
    &(F277W-F356W)>1.5\\
    &-0.5<(F356W-F444W)<0.5
\end{aligned}
\end{equation}

As above, we require SNR$>$2 in the bands redward of the Lyman break (F356W, F444W), SNR$<$2.0 blueward of the break (stacked ACS and WFC3 images, F090W, F115W, F150W and F200W bands), with at most one band blueward of the break having SNR$>$1.5.

When analysing the observed dataset (Sect.~\ref{sec:SAMPLES}), we will also exclude objects classified as spurious by M24, or after visual inspection, such as hot pixels and stellar spikes.

We find that the proposed diagrams efficiently select high-redshift targets up to z$\sim$28, where the F356W-F444W color becomes $\gtrsim$0.5 due to the Lyman-break entering the F356W band. Some contamination from low-redshift galaxies is evident from the redshift distribution of the selected objects (right panels in Fig.~\ref{fig_LBGSIM}). We find a contamination rate of $<$0.05 objects/arcmin$^{2}$ in the EGG-based simulations, $<$0.01 objects/arcmin$^{2}$ in the JAGUAR-based simulation for the z$\sim$15-20 selection, and $<$0.01 arcmin$^{-2}$ in both simulations for the z$\gtrsim$20 selection. We do not find contamination by late-type dwarf stars in any of the proposed selection criteria, consistently with their expected colours and the detectable emission at $\lambda_{obs}\sim$1$\mu$m \citep{Holwerda2018,Holwerda2024}.

Our baseline selection criteria are meant to isolate a \textquote{reference sample} that will be adopted to estimate the UV LF. In addition, we find that by lowering the color thresholds we can select additional very high-redshift targets, although with a significantly higher contamination rate ($\gtrsim$0.1/arcmin$^{-2}$). An additional \textquote{extended sample} of targets with $(F200W-F277W)>1.0$ (z$\sim$15-20) and $(F277W-F356W)>0.8$ (z$\gtrsim$20) has been selected for potential follow-up and to characterize the properties of interloper populations in Sect.~\ref{sec:INTERLOPERS}.

\begin{figure*}
\centering
\includegraphics[trim={1.2cm 0.8cm 1.5cm 1.5cm},clip,width=0.49\linewidth,keepaspectratio]{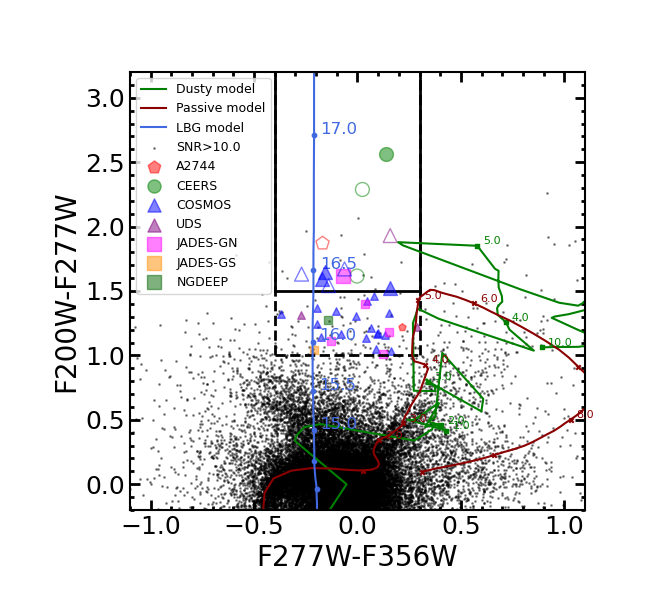}
\includegraphics[trim={1.2cm 0.8cm 1.5cm 1.5cm},clip,width=0.49\linewidth,keepaspectratio]{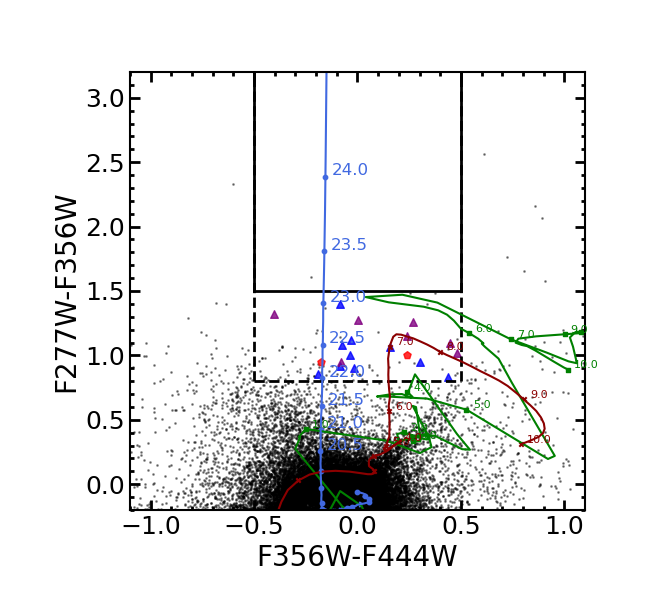}
    \caption{Observed color selection diagrams for LBGs at z$\sim$15-20 (left) and z$\sim$20-30 (right). The black continuous lines enclose the region where \textquote{reference} samples are selected.  Large, filled markers show the position of the objects selected from the various fields, while the seven selected interlopers are shown as open symbols: CEERS-93316 \citep{ArrabalHaro2023b}, the transient A2744\_27713, and the five objects observed with NIRSpec PRISM by the CAPERS survey (Sect.~\ref{subsec:CAPERS_interloper}). Small markers show the position of candidates in the \textquote{extended} samples selected within the color region enclosed by dashed lines. Objects detected at SNR$>$10 at any redshift in the JADES-GS field are shown as black points to highlight the parameter space where the bulk of ASTRODEEP-JWST objects are found. The colored tracks mark the expected colors of stellar plus nebular BC03 templates at the different redshifts indicated by the relevant labels: high-redshift star-forming galaxies at z$\geq$10 with Age=20 Myr, Z=0.02 Z$_{\odot}$, E(B-V)=0 (blue); passively evolving galaxies at 0$\leq z\leq $10 with Z=0.2 Z$_{\odot}$ formed with an instantaneous burst at z=15 (dark red); dusty objects at 0$\leq z\leq $10 with Age=100 Myr, Z=0.2 Z$_{\odot}$, E(B-V)=0.8 (dark green).} 
\label{fig_LBGF200W}
\end{figure*} 

\subsection{Selected sources over $\sim$0.2 deg$^2$}\label{sec:SAMPLES}

Our baseline F200W dropout criteria for z$\sim$15-20 yields a total of 12 objects. Among them, and consistently with previous works, we re-select as a z$\sim$16 candidate the strong line-emitter object CEERS-93316 (ID=84213 in M24) with $z_{spec}=$4.9 \citep{ArrabalHaro2023b}. We also find that the only candidate selected in the A2744 field (ID=27713 in M24) is a transient source. In fact, object A2744\_27713 is not detectable in the first epoch observations (June 2022) of the GLASS-JWST NIRCam parallel, but it is clearly detected in all LW bands from the second epoch dataset observed in November 2022. This object, whose redshift remains undetermined, masquerades as a F200W dropout because it falls in the chip gap of NIRCam SW second-epoch observations. 

After excluding these two interlopers, we are left with an initial sample of 10 F200W dropouts. 
Of these, five candidates (1 in UDS, 1 in CEERS and 3 in COSMOS) have been recently observed with NIRSpec PRISM by the CAPERS survey (GO-6368, P.I. M. Dickinson), finding that they are interlopers with a spectroscopic redshift in the range $z_{spec}\sim 2-7$, as discussed in more detail in Sect.~\ref{subsec:CAPERS_interloper}.

The final \textquote{reference} sample of F200W dropout candidates that we shall discuss in this paper is therefore made of 5 sources: 3 candidates in PRIMER-COSMOS, 1 in CEERS, 1 in JADES-GN, and no sources selected in both JADES-GS and NGDEEP. As described above, neither A2744 nor PRIMER-UDS, which is the widest/shallowest among the considered fields, contribute to the final sample as the two selected sources are confirmed interlopers. Interestingly, candidates in  COSMOS are close on the sky plane, such that if their high-redshift nature will be confirmed they would be at $\sim$1-5 physical Mpc distance from each other, possibly implying that they are part of distant overdensities. In addition, we select 26 F200W-dropout objects meeting the less conservative color thresholds described in Sec.~\ref{sec:LBGDIAGRAMS} (\textquote{extended sample}).

We do not find any object meeting our reference F277W-dropout criteria, while 19 sources are included in the relevant \textquote{extended sample}.

We show in Fig.~\ref{fig_LBGF200W} the position of all our candidates in the observed color-color diagrams. The IDs and main properties of objects in our reference F200W dropout sample are presented in Table~\ref{tab:candidates}, while their SED and NIRCam thumbnails are shown in Fig.~\ref{fig_SEDs_ALL}. The SEDs and main properties of the objects in the extended samples are presented in the Appendix~\ref{sec:appendix-extended}.

We performed additional checks on the reliability of the five objects in the reference sample of F200W-dropout sources. First of all, we measured the SNR within an aperture of 2 times the PSF FWHM on a stack of the NIRCam F090W, F115W and F150W bands available for each of them, finding that they are all non-detected at SNR$<$2. We then computed the SNR on the NIRCam F090W, F115W and F150W bands in two other apertures measured by M24, namely an aperture with a radius of 0.1 arcsec (R01), and the one with a diameter of 3 times the PSF-FWHM. All objects are non-detected in these bands at SNR$<$2 in all cases.

We then searched for MIRI imaging observations at the position of our candidates. The three COSMOS sources are covered by observations by the PRIMER and COSMOS-3D (GO-5893, PI K. Kakiichi) programs, while CEERS\_17384 has been observed by the MIRI EGS Galaxy and AGN survey (MEGA; GO-3794, PI A. Kirkpatrick). We reduced the MIRI imaging datasets in all fields using the Rainbow JWST pipeline as described in \citet{PerezGonzalez2024a}, and measured the SNR at the position of the candidates on the final mosaics in apertures with a radius of 0.3 (F770W), 0.4 (F1000W) and 0.6 arcsec (F1500W, F1800W, F2100W). We find non-detections in all analysed images for COSMOS\_84213 (F770W$>$26.4, F1800W$>$23.4, at 2$\sigma$), COSMOS\_118438 (F770W$>$26.2, F1800W$>$23.5), COSMOS\_107923 (F770W$>$26.3, F1000W$>$24.8). CEERS\_17384 is undetected in the F1000W ($>$25.3), F1500W ($>$24.5), and F2100W ($>$23) bands, but shows a $\sim$3$\sigma$ detection in the F770W one (26.6 $\pm$ 0.3 AB), which we consider as tentative, because of potential contamination by two bright sources in its vicinity.

Finally, we used the DAWN JWST Archive (DJA)\footnote{\url{https://dawn-cph.github.io/dja/index.html} to check whether candidates have imaging observations in NIRCam bands that were not included in M24. We find that candidate JADES-GN\_9538 is covered by NIRCam medium-band observations in the F182M, F210M and F335M bands. It has a marginal ($\sim$2$\sigma$) detection in the F182M band, it is not detected in the F210M one, and a 6$\sigma$ detection in F335M, resulting in a flat F335M-F356W color. We will discuss in Sect.~\ref{sec:PofZ} the implication of MIRI and NIRCam medium-bands  observations on the redshift probability distribution of our candidates.}

We estimated the M$_{UV}$ of the five candidates by converting to the rest-frame the observed F277W magnitude assuming a redshift z=18 where our redshift selection function peaks, and their UV slope $\beta$ by fitting the F277W, F356W and F444W bands. We find two candidates brighter than  M$_{UV}\sim$-21, with the brightest object in the sample COSMOS\_107923 having M$_{UV}\sim$-22.7. Three of the objects have a red $\beta>$-2, the remaining ones being consistent with a flat or moderately blue UV slope ($\beta \lesssim -2$).  We have measured their half-light radius by fitting the light distribution in the F277W band with \textsc{GALIGHT} \citep{Ding2020,Birrer2021} assuming a \citet{Sersic1968} profile with free index $n$ and fixing the redshift at the best-fit solution at $z>10$. We find half-light radii consistent with those of observed in galaxies at z$\sim$10-15 \citep{Westcott2024,Ono2025}. Our candidates have a typical half-light radius of $\simeq$0.2-0.3 kpc.

We compared our samples with $z\gtrsim$15 samples selected by other groups in the fields analysed here. Our F200W-dropout candidate in CEERS is not included in the sample of very red sources by \citet{Gandolfi2025}, while 4 of their 5 objects potentially at z$>$15 have indeed colours compatible with our selection window, but have not been included here because of SNR$<$10 in the detection band, and in one case (their A-22691) a marginal (SNR$\sim$2.2) detection in the ACS stacked image. The remaining source (U-53105) is not detected in the M24 catalog. Object NGD-z15a/NGDEEP-1369 presented as a z$\sim$15.6 candidate by both \citet{Austin2023} and \citet{Leung2023} is matched to object ID=1301 in M24. It has colours consistent with our inclusive selection window but is not part of our \textquote{extended sample} because of a detection at SNR$\sim$2.5 in the stack of the ACS bands. Five of the sources selected by \citet{PerezGonzalez2025} in the MIDIS+NGDEEP observations, including the z$\sim$19.6 candidate MDS025593 by \citet{PerezGonzalez2023}, are not covered by the first-epoch NGDEEP imaging used by M24, the remaining ones being non-detected except their MIDIS-z17-7 (ID=8412 in M24) which is at SNR$\sim$4, i.e. well below our SNR=10 threshold. The $z\gtrsim$15 sources selected by \citet{Hainline2024} in the JADES fields with a counterpart in M24 do not fall within our colour selection window, and are detected at SNR$<$10 except their \textsc{JADES-GS-53.12692-27.79102} (ID=51718 in M24). Finally, the z$\sim$15 candidate JADES-GN+189.16733+62.31026 by \citet{Whitler2025} (ID=56718 in M24) is not selected due to its $(F200W-F277W)\sim 0.9$, i.e. slightly below the threshold to enter our \textquote{extended} sample presented in Sect.~\ref{sec:appendix-extended}.

\begin{figure*}
\centering
\includegraphics[trim={0.5cm 0.5cm 0.5cm 0.1cm},clip,width=0.3\linewidth,keepaspectratio]{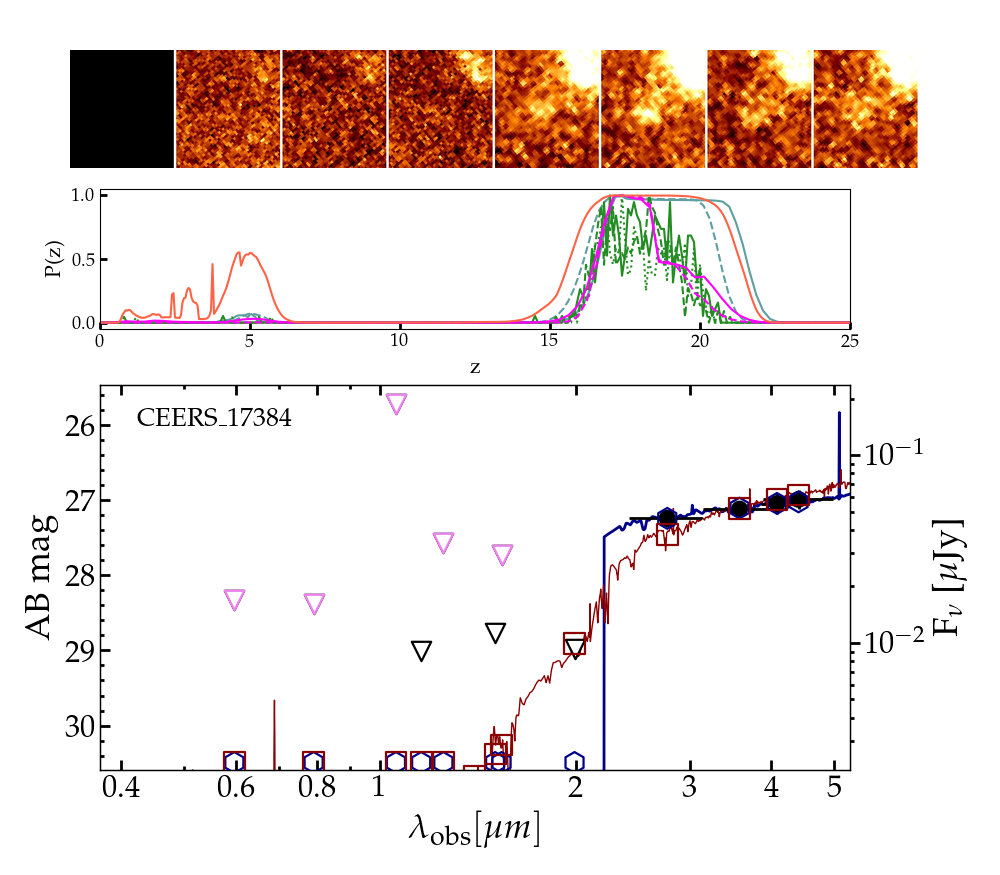}
\includegraphics[trim={0.5cm 0.5cm 0.5cm 0.1cm},clip,width=0.3\linewidth,keepaspectratio]{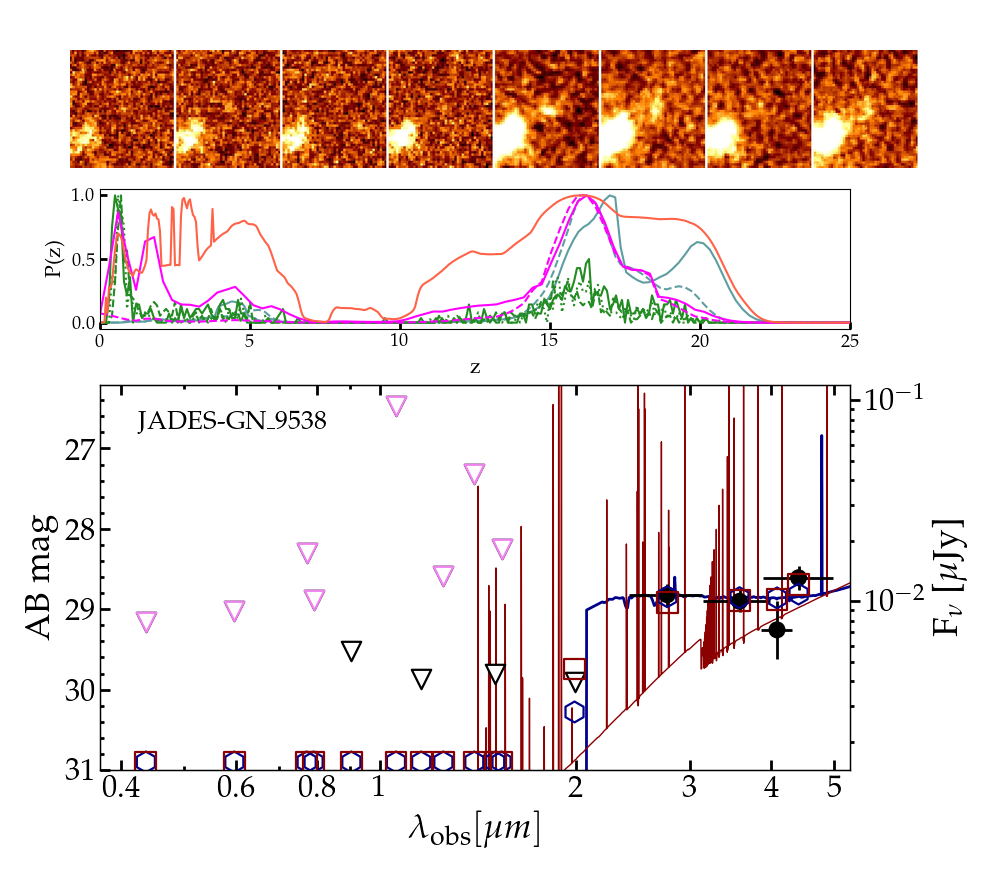}
\includegraphics[trim={0.5cm 0.5cm 0.5cm 0.1cm},clip,width=0.3\linewidth,keepaspectratio]{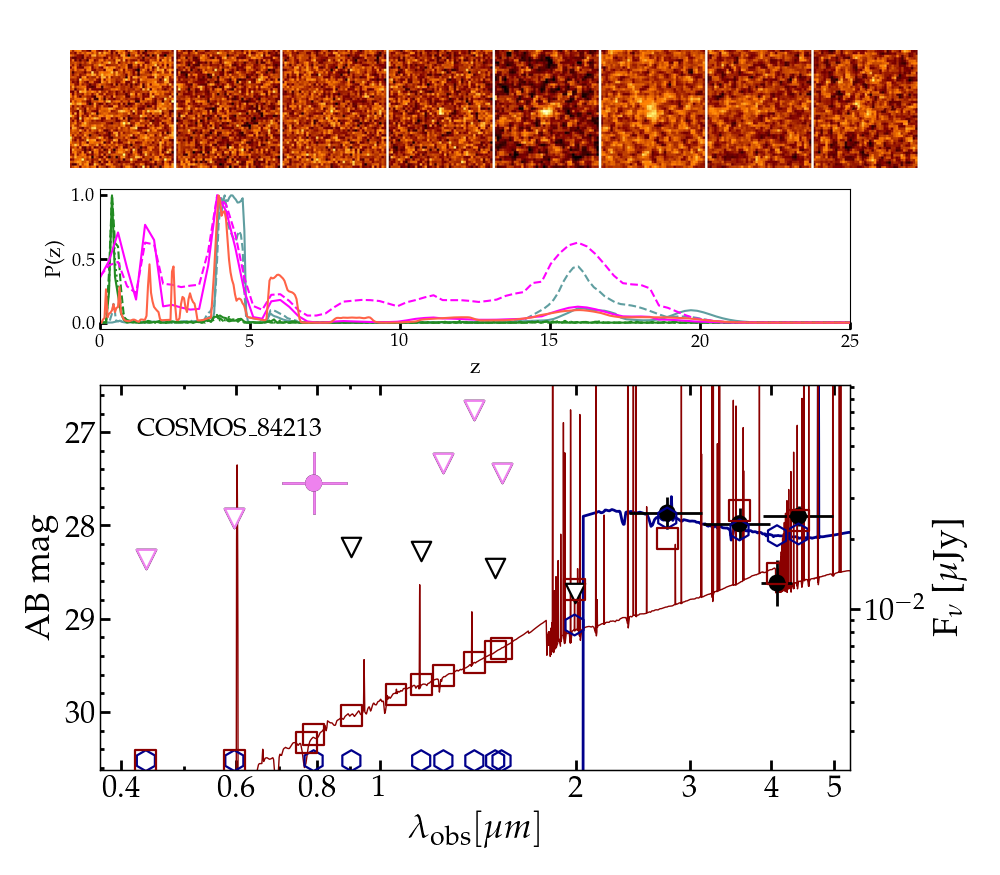}
\includegraphics[trim={0.5cm 0.5cm 0.5cm 0.1cm},clip,width=0.3\linewidth,keepaspectratio]{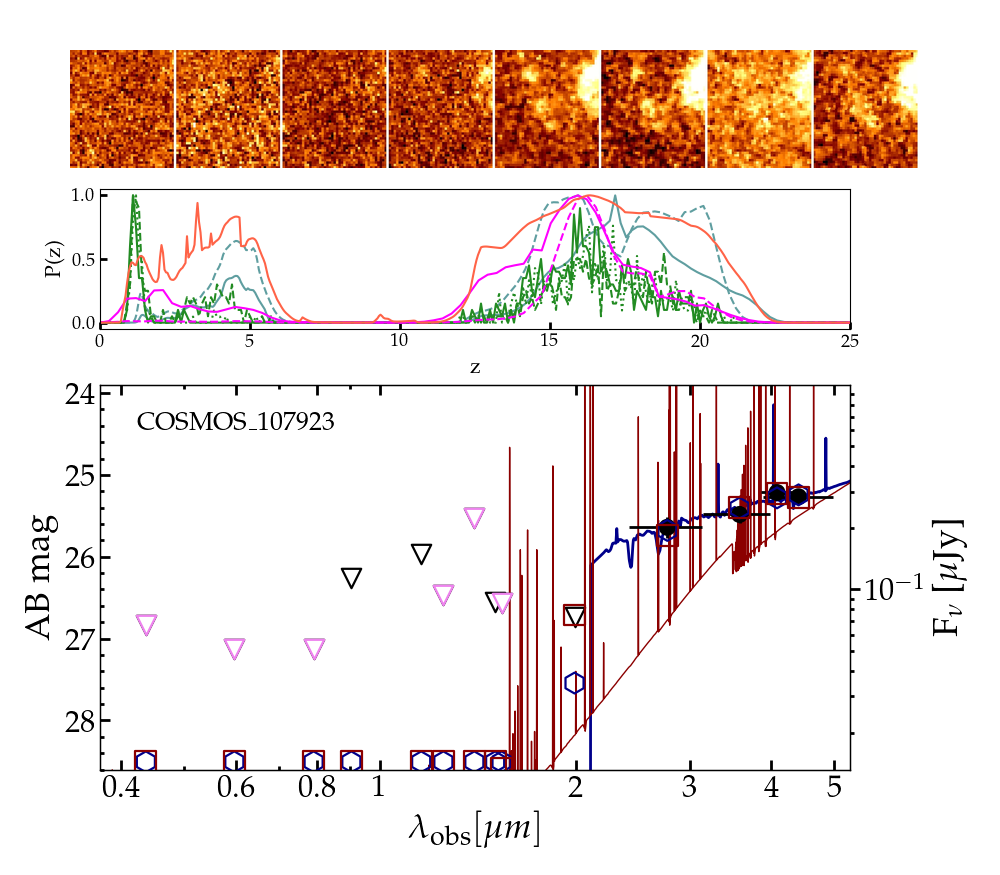}
\includegraphics[trim={0.5cm 0.5cm 0.5cm 0.1cm},clip,width=0.3\linewidth,keepaspectratio]{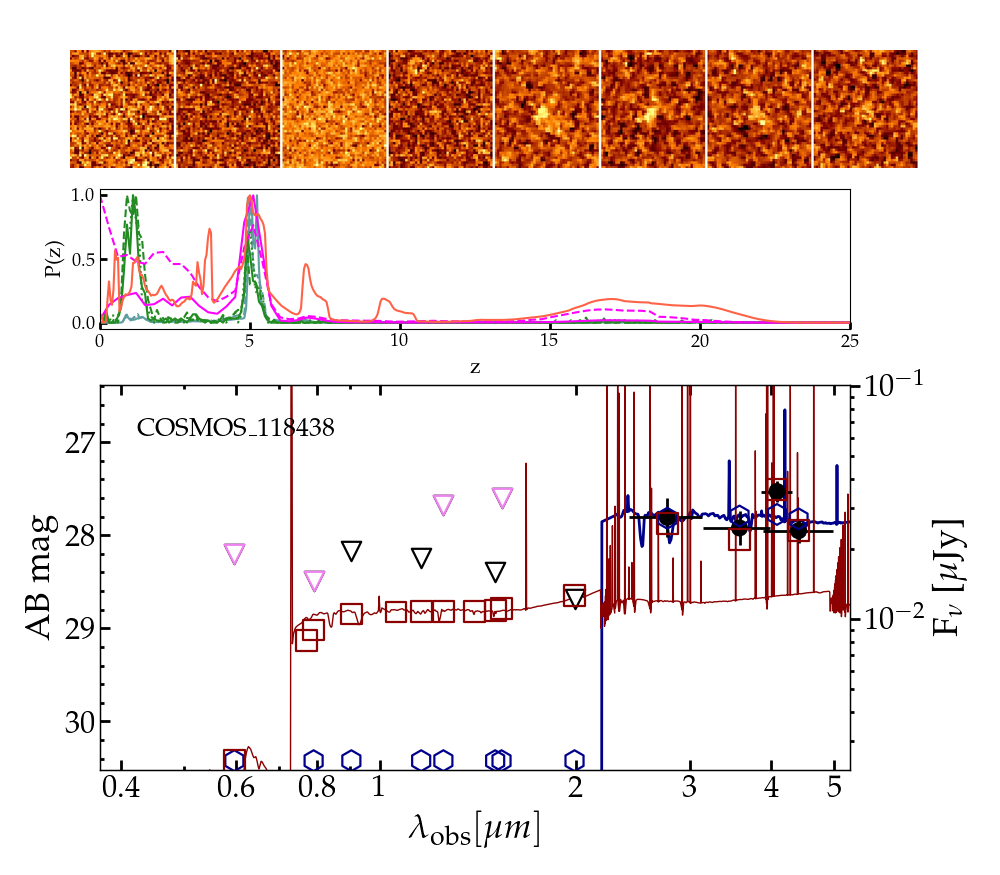}
\caption{Spectral energy distributions, $P(z)$ and NIRCam thumbnails of the five F200W dropout candidates. For each object the best-fit templates at high- and low-redshift from the \textsc{zphot} run are shown in blue and red, respectively. The relevant predicted magnitudes are indicated by blue empty squares and red empty hexagons, respectively. The photometric measurements are from M24, with black (magenta) circles and error-bars indicating JWST (HST) bands. The 2$\sigma$ upper limits are shown as triangles. The $P(z)$ from \textsc{zphot}  are shown as orange lines, the ones from \textsc{EAzY} adopting standard (standard plus Larson) templates are shown as continuous (dashed) light blue lines, the $P(z)$ from \textsc{BAGPIPES} are shown in green with continuous, dashed and dotted lines respectively assuming a delayed, double power-law and exponential SFH, and the $P(z)$ from \textsc{CIGALE} using a star-forming (star-forming+AGN) fit are shown as continuous (dashed) magenta lines. All curves are normalized to have $P(z)$=1 at the peak. The 1.2 $\times$ 1.2 arcsec thumbnails, from left to right, respectively show the objects in the F090W (where available), F115W, F150W, F200W, F277W, F356W, F410M and F444W bands used for the ASTRODEEP-JWST measurements. } 
\label{fig_SEDs_ALL}
\end{figure*} 

\begin{table*}[ht]
\caption{F200W dropout candidates in the ASTRODEEP-JWST fields$^a$}\label{tab:candidates}
\centering

\begin{tabular}{ccccccccccc}
ID &          R.A. &        Dec &  F356W  &  M$_{1500}$ &  R$_{e}$  & $\beta$ & z$_{high}$ & z$_{low}$ & $\chi^2_{high}$ &$\chi^2_{low}$\\

 & deg. & deg. & AB & & kpc & & &  & &\\
\hline

CEERS\_17384 &  214.853243 &  52.77368 &  27.11 $\pm$   0.08 & -21.06 &0.317	$\pm$ 0.05 & -1.50 $\pm$ 0.33&17.2	&4.7 & 0.36 & 0.90\\  
 JADES-GN\_9538 &  189.191052 &  62.17421 &  28.90 $\pm$  0.13 & -19.49 & 0.227	$\pm$ 0.043& -1.63 $\pm$ 0.40&16.1 & 2.8 & 0.62  & 0.64\\
 COSMOS\_84213 &  150.167211 &  2.368995 & 28.14 $\pm$ 0.18 & -20.35& 0.199	$\pm$ 0.053& -2.03 $\pm$ 0.40&15.6& 4.0 & 2.31& 1.73\\
 COSMOS\_107923 &  150.107337 &  2.428780  &  25.48 $\pm$   0.10  & -22.68& 0.219	$\pm$ 0.024& -1.22 $\pm$ 0.26&16.3& 3.3 & 0.59 & 0.65\\
 COSMOS\_118438 &  150.196541 &  2.464192 &  27.86 $\pm$   0.19& -20.61  & 0.256	$\pm$ 0.077&  -2.27 $\pm$ 0.47&17.1 & 5.0 & 1.37 & 0.37\\
\hline
\end{tabular}
\small \\a) ID, coordinates and F356W magnitudes from M24.  The M$_{UV}$ and half-light radius ($R_h$) have been measured from the observed F277W band. The  UV slope $\beta$ is obtained by fitting the F277W, F356W and F444W bands. The last four columns show the best-fit solutions obtained with \textsc{zphot} at $z>10$ ($z_{high}$) and $z<10$ ($z_{low}$), and the relevant $\chi^2$ values.
\end{table*}

\section{A closer scrutiny of the selected candidates}\label{sec:scrutiny}
In this section we analyse in more detail the selected candidates to assess their reliability and evaluate the possibility that their peculiar colours are instead indicative of rare lower redshift interloper populations. Although we focus here on the 5 candidates, we remark that very similar arguments can be made for the 5 sources that have been confirmed to be interlopers, as discussed in Sect.~\ref{subsec:CAPERS_interloper}.

\subsection{The photometric redshift probability distribution}\label{sec:PofZ}
Following a well-established practice, we have derived the photometric redshift of our candidates analysing their multi-wavelength photometry with a set of standard SED
fitting tools. To alleviate the impact of specific flavours of the adopted techniques, and to broaden the range of spectral libraries explored, we have exploited four different codes: \textsc{zphot} \citep{Fontana2000}, \textsc{EAzY} \citep{Brammer2008},
\textit{BAGPIPES} \citep{Carnall2018,Carnall2019b} and \textsc{CIGALE} \citep{Boquien2019}. Rather than focusing on the best-fitting photometric redshift, we have used all these
codes to compute the redshift probability distribution ($P(z)$). This distribution encapsulates the full information on the different potential solutions for an object with given photometry.  

\begin{figure*}[ht]
\centering
\includegraphics[trim={1.2cm 0.5cm 2.0cm 1.0cm},clip,width=0.45\linewidth,keepaspectratio]{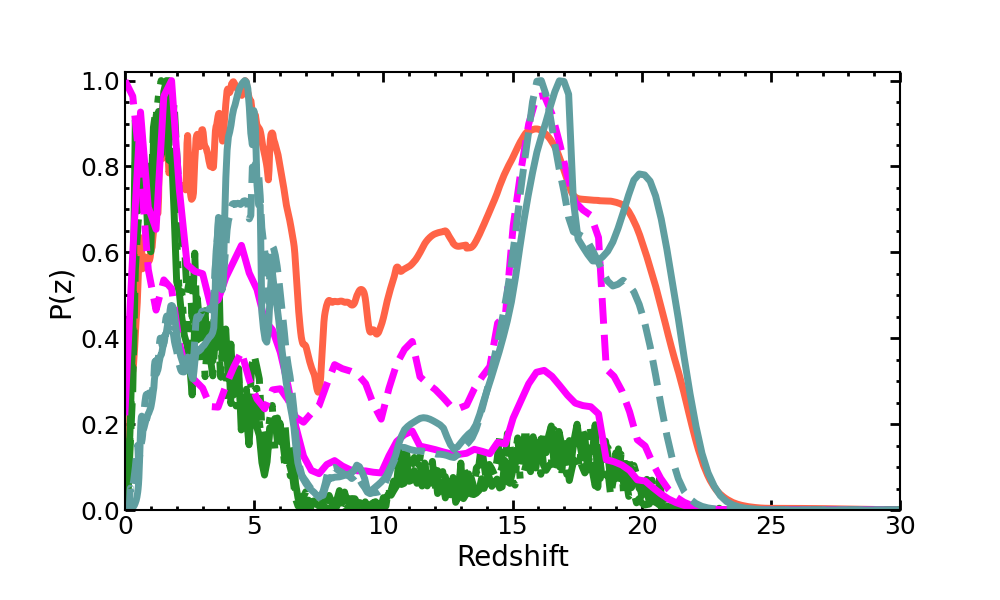}
\includegraphics[trim={1.2cm 0.5cm 2.0cm 1.0cm},clip,width=0.45\linewidth,keepaspectratio]{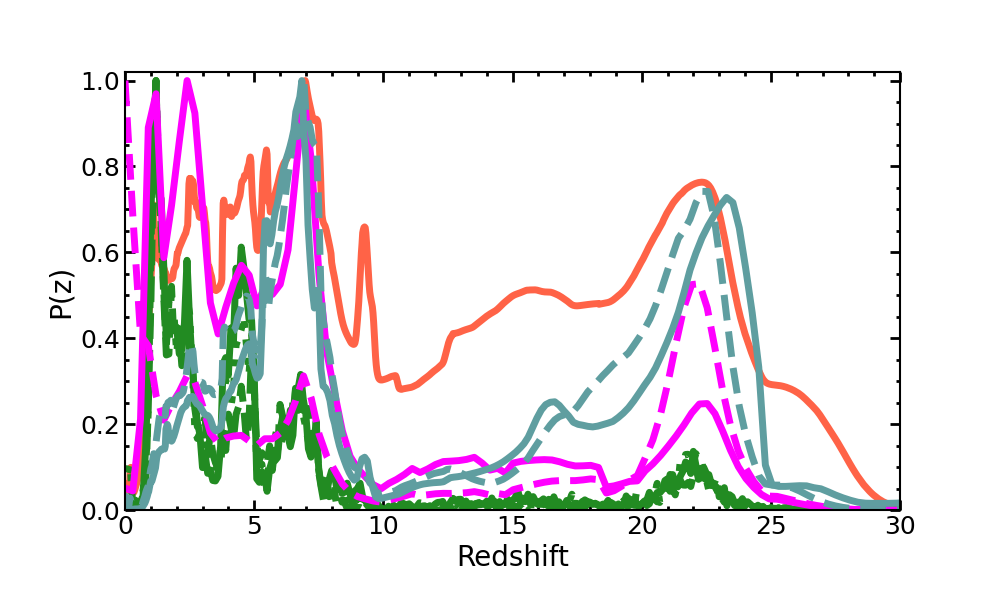}
\caption{The average redshift probability distribution functions $P(z)$ for objects in the extended samples of F200W dropouts (left) and F277W dropouts (right) computed with \textsc{zphot}, \textsc{EAzY}, \textsc{BAGPIPES}, and \textsc{CIGALE} (same color conventions as in Fig.~\ref{fig_SEDs_ALL}). The curves are normalized to have $P(z)$=1 at the peak.} 
\label{fig_Pz_inclusive}
\end{figure*}

Briefly, we have run \textsc{zphot} adopting a set of synthetic models drawn from the BC03 library with a range of metallicities from
$0.02~Z_\odot$ to $2.5~Z_\odot$,
$0\leq E(B-V)\le 1.1$, and a \textquote{delayed} ($\phi\propto
t^2 e^{-t/\tau}$) star--formation
history \citep[see][ and M24]{Santini2023}. Nebular emission is
self--consistently included following \citet{Schaerer2009} \citep[see also][]{Castellano2014}, based on the template 
luminosity at ionizing frequencies. We have analysed our candidates with \textsc{EAzY} in two different ways as described in M24, i.e. 1) using only its standard set of
semi-empirical templates, and 2) including also the recent set of templates from \citet{Larson2022} which are designed explicitly to reproduce the colors of high redshift galaxies. 
The \textsc{Bagpipes} runs exploit BPASS v. 2.2.1 stellar models with an upper-mass cutoff of the IMF of 300~M$_{\odot}$ \citep{Stanway2018}, and nebular emission computed self-consistently with \textsc{CLOUDY} \citep{Ferland2013} as described by \citet{Carnall2018}. Following \citet{Gandolfi2025}, we increased the number of live points (i.e., the walkers used by \textsc{Bagpipes} in the Markov Chain Monte Carlo sampling) from the default 400 to 2000 to enhance sensitivity to strong line emitter solutions, and allowed the ionization parameter to reach
log\,$U = -1$.  We assume three different star-formation histories: delayed, double power law, and exponential SFH. 
Finally, we run \textsc{CIGALE} in two configurations. In the first case, we assume a SFH with a delayed component of age between 100 Myr and the age of the universe at each redshift, plus a constant burst of 10 Myr duration. The fraction of stellar mass formed in the recent burst is allowed to vary between 0 and 50\% of the total assembled mass. We use BC03 templates including nebular emission, metallicity 0.02, 0.2, 1 Z$_{\odot}$, and V-band extinction 0$\leq A_V \leq$5. The second \textsc{CIGALE} configuration exploits star-formation plus AGN templates where the stellar component is parametrized as described above and the AGN emission is based on the \textsc{dale2014} module \citep{Dale2014}. The AGN fraction (f$_{AGN}$), defined as the ratio of AGN luminosity to the total AGN and dust luminosities, is set as a free parameter. 
To summarise, from the combination of codes and assumptions adopted we have obtained eight different $P(z)$ for all our candidates. The $P(z)$ of the five objects in our reference sample of F200W dropout candidates are shown in the upper panels of Fig.~\ref{fig_SEDs_ALL}. We show in Fig.~\ref{fig_Pz_inclusive} the average $P(z)$ for the extended samples of F200W- and F277W-dropout candidates.

Although there are specific differences among the various objects and the different codes, a number of general conclusions can be drawn from this analysis.

First, it is clear that all objects exhibit a $z>15$ solution, consistent with the color selection criteria adopted. The inferred redshifts tend to be reasonably similar among the codes, as the main spectral feature determining the high-redshift solution is the Lyman-break (coupled with the shape of the star--forming continuum immediately redward) that is essentially
common in all recipes. However, \textit{all} our candidates also show a lower-redshift
solution, in \textit{all} the runs analysed here. The low--redshift
solutions are typically peaked at $z=3-7$, suggesting that the strong
observed break in the F200W band can also be ascribed to a break around
the Balmer break/4000$\AA$ rest-frame region, as we shall describe better
in the following. In several cases z$\lesssim$2 solutions are also
viable, and generally preferred by \textsc{BAGPIPES}. Detailed inspection shows that this is generally due to a
peculiar combination of strong emission lines that may conspire to
reproduce the observed colours. We remark that, given the resulting $P(z)$, none of our candidates
would pass a selection criterion such as $\Delta\chi^2 >$4 between different redshift solutions that has been often
adopted to build luminosity functions at $z>9$ \citep[e.g.,][]{Finkelstein2023b,Harikane2022b}.  Unfortunately, our candidates are simply too faint, and their photometry is built on a too small number of photometric bands with solid detections, to be unambiguously selected in the same manner.
It is also clear that, while they overall provide a consistent picture of \textquote{double peaked} solutions, there
are significant differences between the adopted codes, both in terms of
the breadth of the low-redshift solution and of the relative
weight between the low- and high--redshift peaks. These differences
arise from the different libraries adopted and probably from slight
differences in the fitting procedure. This picture is confirmed by the average $P(z)$ of the extended samples of F200W- and F277W-dropout candidates (Fig.~\ref{fig_Pz_inclusive}).

We note that $P(z)$ also depends crucially on the details
of the adopted photometry. Clearly, but somewhat counterintuitively, the additional information provided by the HST
photometry decreases the constraining power of
$P(z)$. Because of the faintness of our candidates, in fact, the HST
upper limits are easily satisfied by a wide class of solutions at low
redshifts, eventually adding little contribution to the global $\chi^2$ while
increasing the number of degrees of freedom $n$ and hence increasing
the probability at low redshifts. These results are based on the ASTRODEEP-JWST catalogs, which provide homogeneous photometric measurements for all sources. We find that the MIRI constraints discussed in Sect.~\ref{sec:SAMPLES} do not significantly change the scenario described above. The $P(z)$ maintain the same double-peaked nature with both high- and low-redshift solutions when including the MIRI upper limits, and, in the case of CEERS\_17384, the tentative F770W detection. In fact, the MIRI non-detections are poorly constraining for all our sources, with the exception of COSMOS\_107923 for which the limit in the F770W increases the probability of passive solutions with respect to dusty star-forming ones in the low-redshift peak. In fact, the MIRI imaging should be 1-2 mag deeper in the F770W and F1000W bands, and $>$3-4 mag deeper at $\lambda_{obs}>$15$\mu$m to yield meaningful constraints for the other  candidates.  Finally, when including additional NIRCam medium-band photometry for JADES-GN\_9538 we find an overall similar $P(z)$ but with a slightly increased probability at z$\sim$13-15 due to the marginal detection in the F182M band.

We take from these results three main lessons. The first is that all our
objects are, in principle, credible candidates at $15<z<20$, but none of
them are solid enough to be unambiguously assigned these extreme
redshifts.  In addition, the differences among the various $P(z)$
suggest that a detailed and sophisticated analysis built on their shape
should be taken with caution, as they may depend on subtle
details in the photometric measurements and on the spectral libraries
adopted.
Finally, as we will discuss below, the most important factor preventing a robust use of the $P(z)$ in the selection process is our limited knowledge of the populations of faint, red interlopers.

\subsection{The nature of potential interlopers}\label{sec:INTERLOPERS}
The photometric redshift distributions described above help us
unveil the physical properties of the galaxies that may contaminate
our selection criteria. Adopting for simplicity the results of the
\textsc{zphot} code, we have inspected the physical properties
corresponding to the models that populate the low redshift peaks in
the $P(z)$.  The best-fit SEDs at $z<10$ of the F200W dropout candidates are shown in Fig.~\ref{fig_SEDs_ALL}. 

In the case of the candidate in CEERS, the low-redshift best fit solution is a $z\simeq 5$ \textquote{quiescent} galaxy with low specific star--formation rate and large stellar age. The remaining sources, instead, are best reproduced by star--forming models with a very high specific star--formation rate and large dust attenuation,  whose red continuum and strong emission lines yield colors compatible with our selection criteria. 

\begin{figure}
\centering
 \includegraphics[trim={3cm 0.5cm 1.0cm 3.0cm},clip,width=\linewidth,keepaspectratio]{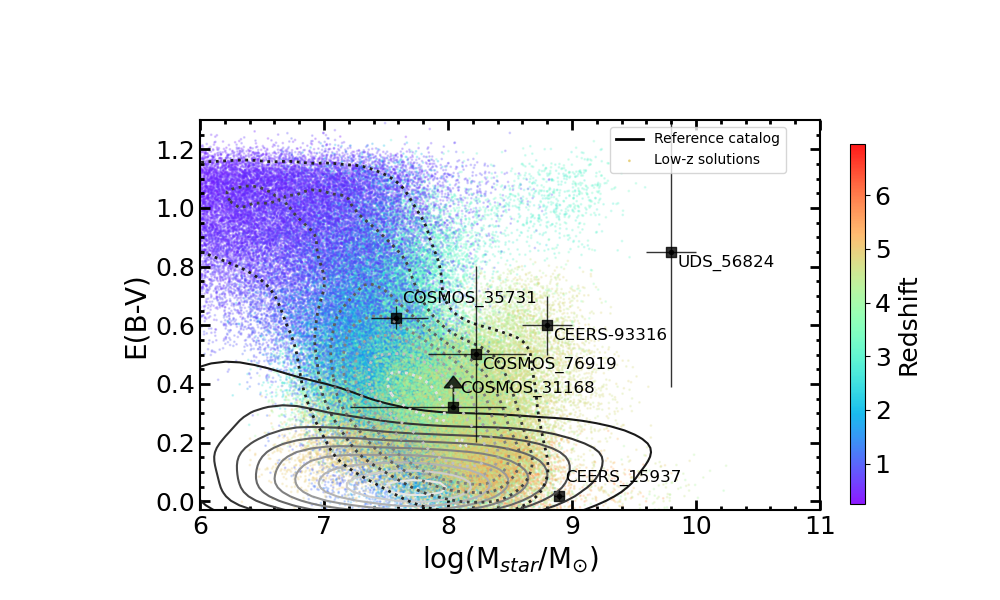}
  \includegraphics[trim={3cm 0.5cm 1.0cm 3.0cm},clip,width=\linewidth,keepaspectratio]{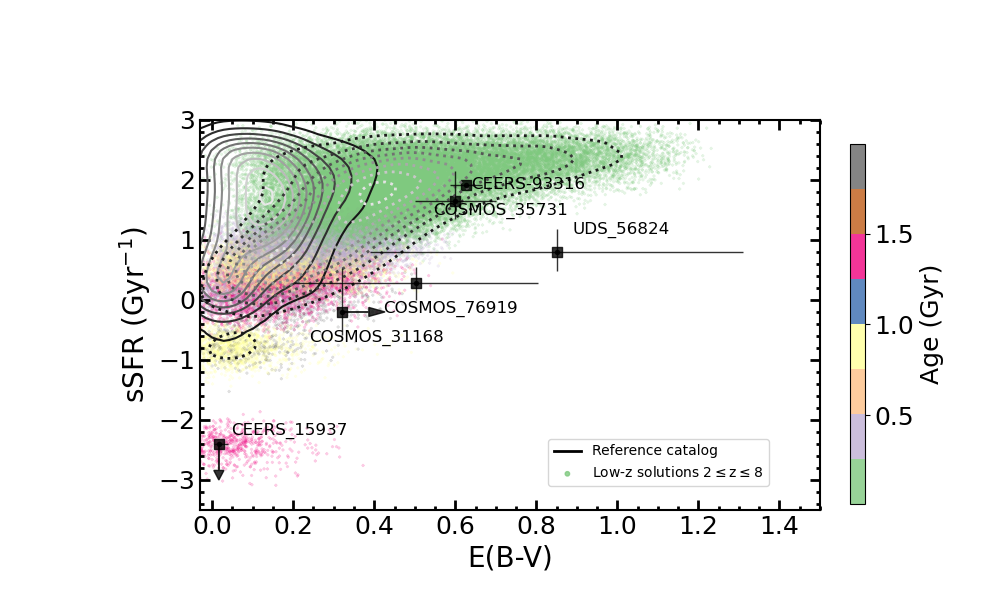}
\caption{\textbf{Top}: the position in the $E(B-V)$ vs. $M_{star}$ plane of galaxy templates (points colour-coded according to the redshift) that provide an acceptable fit with probability $P(z)>0.5$ to the F200W dropout candidates. The regions occupied by the 90\% to 10\%, at 10\% steps, of the aforementioned templates are enclosed by dotted curves. The continuous curves enclose the regions occupied by the 90\% to 10\%, at 10\% steps, of the objects in the same redshift range from the JADES-GS field.  The black square and error-bars mark the positions of the confirmed interlopers.  \textbf{Bottom}: same as top panel for low-redshift solutions at $2\leq z \leq 8$ in the $sSFR$ vs. $E(B-V)$ plane, colour-coded according to the stellar age.} 
\label{fig_lowz_phys}
\end{figure}

These two kinds of solutions are representative of the general properties of galaxy templates that typically yield low-redshift solutions for our objects, as shown in Fig.~\ref{fig_lowz_phys}. We compared the solutions that provide an acceptable fit with probability $P(z)>0.5$ for the F200W-dropout candidates to the locus of objects in the same redshift range from the JADES-GS field.
The statistically acceptable models populate regions that have a small overlap with the ones occupied by the bulk of sources in the same redshift range. Consistently with the SEDs shown in Fig.~\ref{fig_SEDs_ALL}, the $2<z<8$ templates cover a 
region in the $sSFR$ versus $E(B-V)$ plane that connects quiescent, low
dust models (lower left corner) with highly star-forming, dusty (upper right corner) ones. Instead, the general distribution of galaxies in this plane shows that most
of the objects tend to populate the region of intermediate $sSFR$ and
$E(B-V)$.  The $E(B-V)$ vs $M_{star}$ plane shows that acceptable solutions include z$\sim$0-4 templates with a higher dust-extinction than \textquote{typical} sources in the same redshift range, and, in particular, are consistent with potential contamination of the F200W dropout selection by very low-mass, dusty galaxies, as previously noted by \citet{Bisigello2023} and \citet{Gandolfi2025}.

All these templates can reproduce the sharp break observed around $\lambda_{obs} \sim 2~\mu m$ in our
objects. Because of the faintness of our candidates, the amplitude of
the break (which is much larger in $z>15$ galaxies) cannot be properly
measured with the existing photometry, leaving room for the ambiguity
between the two redshift solutions.

We remark, however, that these potential low-redshift solutions
correspond to objects that would be extremely interesting to investigate.
These models have stellar masses $M_*$ in the range $10^7-10^9 M_\odot$, sometimes even as low as $10^6 M_\odot$. As long as quiescent galaxies are concerned, 
only  objects with stellar mass above $10^{10}M_\odot$ have been confirmed  at $z>4$ 
\citep{Carnall2024,Glazebrook2024,Weibel2024,PerezGonzalez2024b}. The available estimates of the
stellar mass function of quiescent galaxies do not extend to these low masses, especially at these redshifts \citep{Santini2021}. These sources may also host the low-mass SMBH that are fundamental to constrain early AGN-galaxy co-evolution \citep{Pacucci2023}.

Concerning dusty solutions, on the other hand, there is a general consensus that faint, low-mass galaxies
at $z>2$ are essentially dust-free according to the strong correlations between dust-extinction and stellar mass \citep{McLure2018b,Bouwens2020}. If (even some of) these candidates are instead low-mass, dusty galaxies as suggested by one class of solutions, they would correspond to a phase in galaxy evolution that has not been
widely investigated so far. Recently, \cite{Bisigello2025} has
spectroscopically confirmed a low mass, dusty galaxy at
$z\simeq 5$, suggesting that other similar objects (1-15\% of the sources of similar mass and redshift) might be hidden in the photometric sample. Although rare, galaxies with rest-frame colors that approach our color selection criteria do exist in nature. As a result, we cannot exclude the possibility that some - or even all - our candidates are indeed intermediate redshift interlopers.

We cannot exclude that contamination from AGN is also present, although it seems less likely than from dusty or quiescent galaxies. In fact, when including AGN emission, \textsc{CIGALE} tends to increase rather than decrease the probability of $z>15$ solutions, in some cases indicating them as best-fit. Finally, we do not expect significant contamination from Little Red Dots (LRD) at z$\sim$3-8, except possibly for the reddest objects with strong emission lines whose predicted colours are similar to those of dusty, high-sSFR galaxies  \citep{Killi2024,PerezGonzalez2024a}.

\begin{figure}
\centering
\includegraphics[trim={3.5cm 2.5cm 3.5cm 2.cm},clip,width=\linewidth,keepaspectratio]{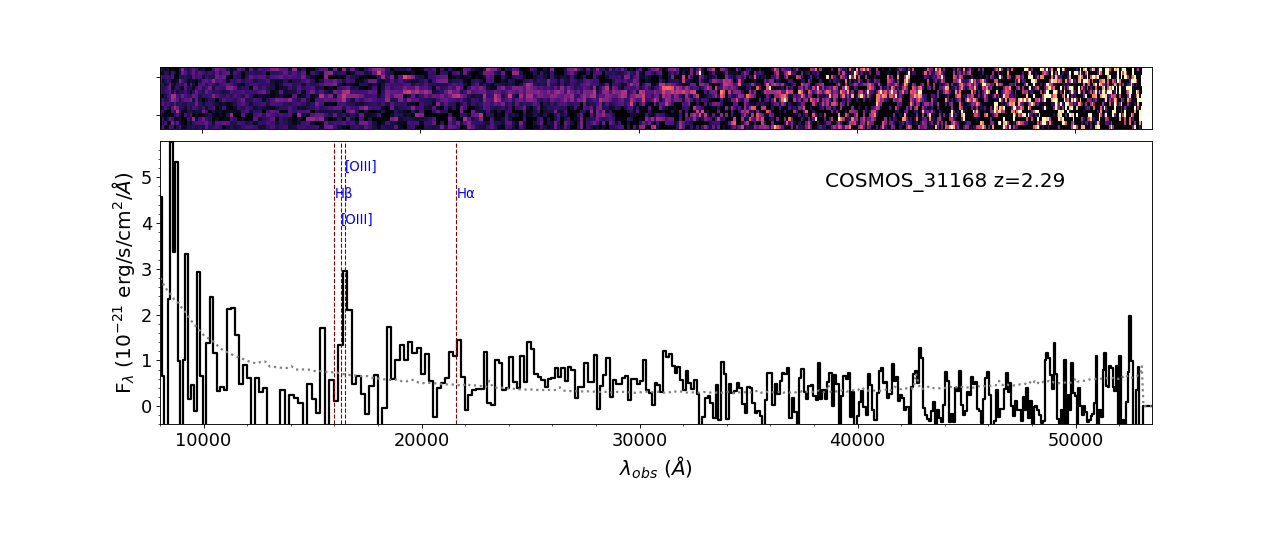}
\includegraphics[trim={3.0cm 2.5cm 3.5cm 2.cm},clip,width=\linewidth,keepaspectratio]{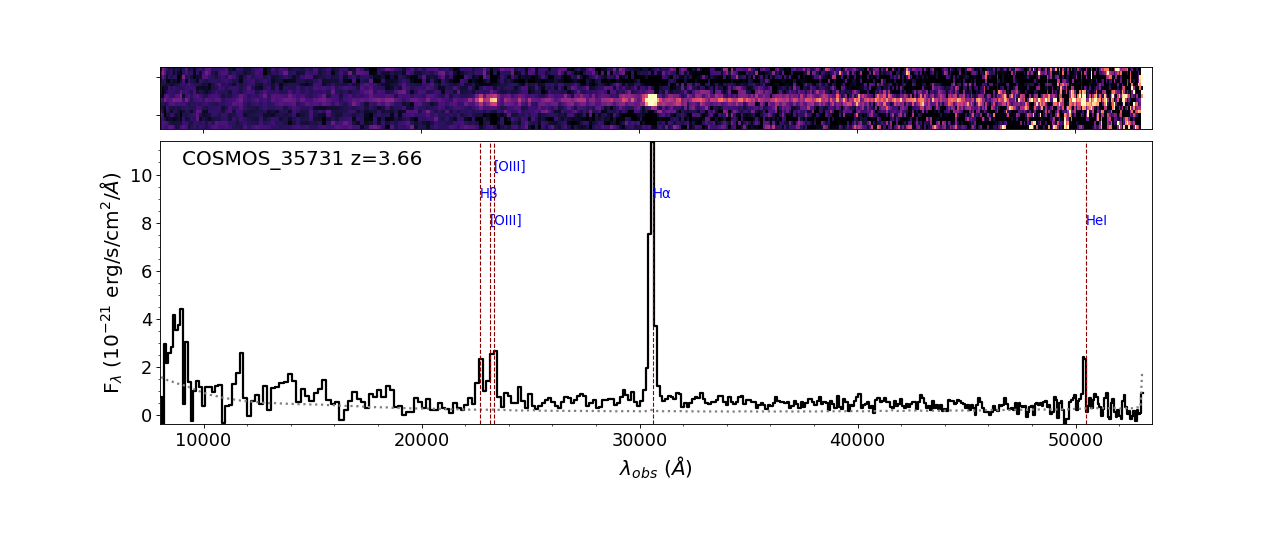}
\includegraphics[trim={3.5cm 2.5cm 3.5cm 2.cm},clip,width=\linewidth,keepaspectratio]{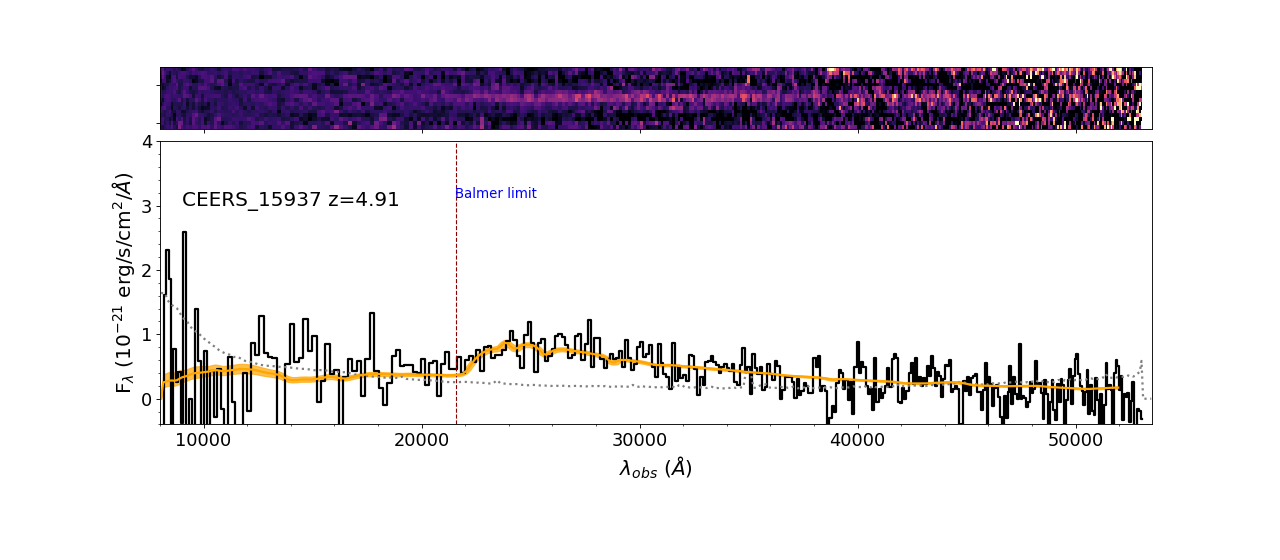}
\includegraphics[trim={3.5cm 2.5cm 3.5cm 2.cm},clip,width=\linewidth,keepaspectratio]{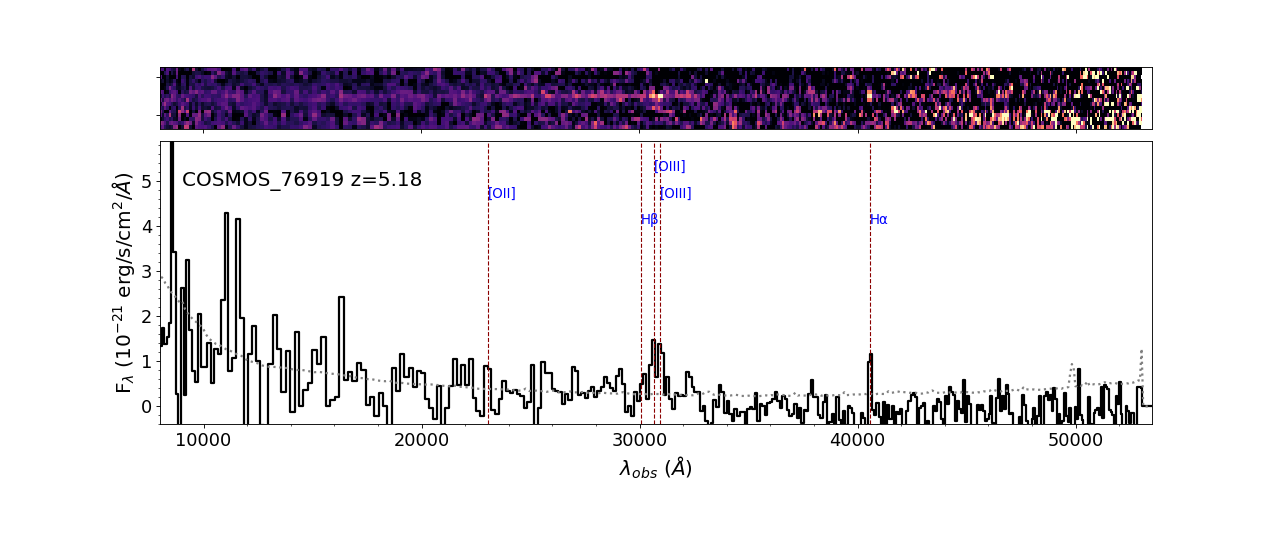}
\includegraphics[trim={3.0cm 2.5cm 3.5cm 2.cm},clip,width=\linewidth,keepaspectratio]{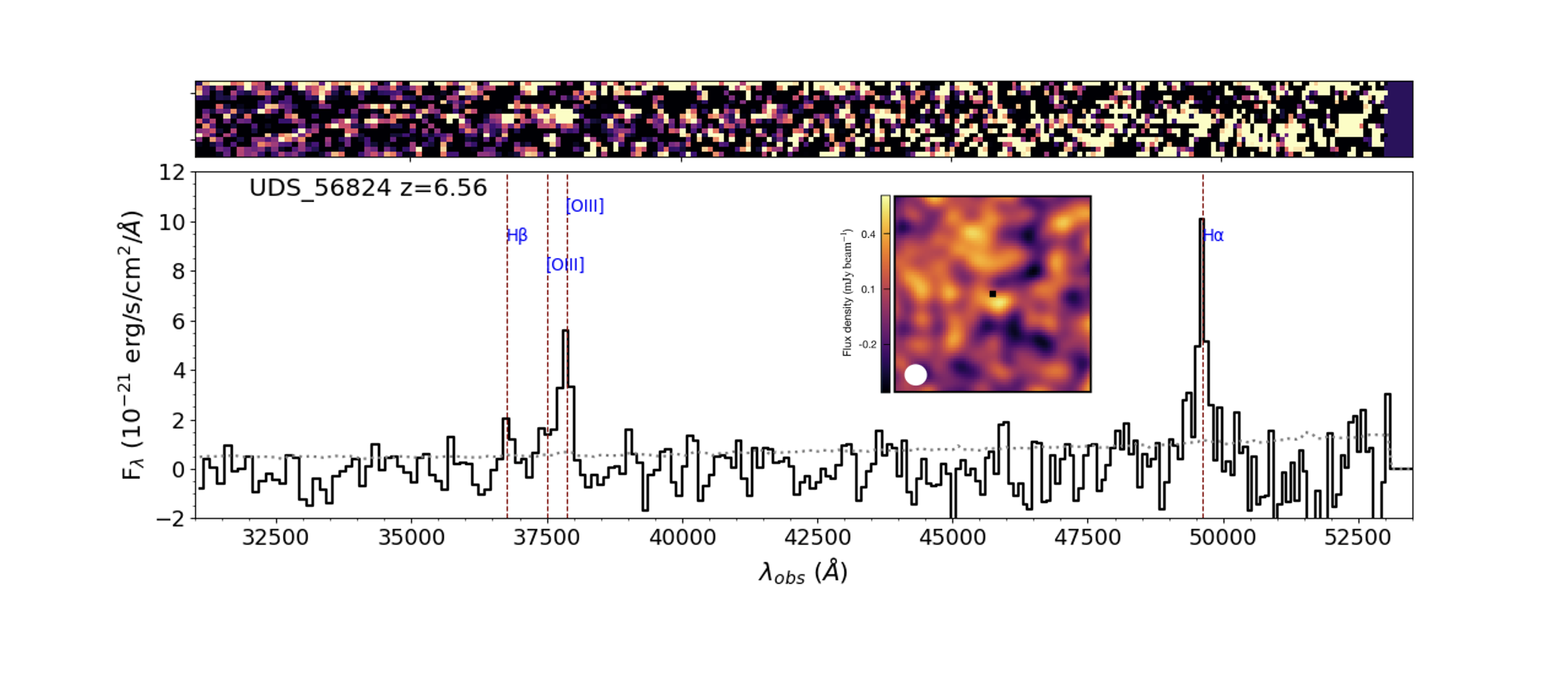}
\caption{The observed 2D (top) and 1D (bottom) NIRSpec PRISM spectra of the five interlopers observed by the CAPERS survey. In the bottom panels the gray dashed line shows the noise RMS, and red dashed lines highlight the wavelength of the detected features. The best-fit template obtained by fitting the CEERS\_15937 spectrum with BAGPIPES is shown in yellow in the relevant panel. The inset in the UDS\_56824 panel shows the 3 $\times$ 3 arcsec ALMA Band 7 map centered at the position of the object (black square), with the relevant beam size shown on the lower-left corner.}
\label{fig_CAPERS_spec}
\end{figure}

\subsection{The confirmed interlopers}\label{subsec:CAPERS_interloper}

Five of  the ten objects originally selected as F200W-dropouts have been recently observed with the NIRSpec PRISM in the framework of the CAPERS program, and found to be interlopers at z$\sim$2-7 (empty symbols in Fig.~\ref{fig_LBGF200W}). We describe in the following their physical properties as inferred from the combination of NIRSpec spectroscopy (Fig.~\ref{fig_CAPERS_spec}) and NIRCam photometry (Sect.~\ref{sec:appendix-interlopers}). These observations confirm that contaminants of our selection criteria are found in the populations presented in Sect.~\ref{sec:INTERLOPERS}, namely dusty star—forming or quiescent galaxies with low stellar mass.

The spectroscopic observations have been carried out by adopting a NRSIRS2 readout pattern, standard 3-shutter ``slits'', and a 3-point nodding. The data were reduced with the STScI Calibration Pipeline\footnote{\url{https://jwst-pipeline.readthedocs.io/en/latest/index.html}} version 1.13.4 as described in detail in \citet{ArrabalHaro2023a} \citep[see also][]{Castellano2024,Napolitano2025}. We determine the spectroscopic redshift from the weighted average centroid of the detected emission lines, and measured line fluxes with a Gaussian fit after linearly extrapolating the continuum emission at the line position \citep[see][for details]{Napolitano2024b,Napolitano2025}.

\subsubsection{COSMOS\_31168 at $z_{spec}$=2.29} The spectrum of COSMOS\_31168 (R.A.= 150.1809331 deg., Dec=2.2607514 deg.) shows one clear emission line at 1.64$\mu$m, a lower significance emission feature at 2.15$\mu$m and a faint continuum. The most plausible redshift solution is $z_{spec}$=2.294, interpreting the aforementioned emission lines as the unresolved [O~III]$\lambda_{4959,5007}$ doublet (with flux 9.7$\pm$1.9 10$^{-19} erg/s/cm^2$) and H$\alpha$ (=3.1$\pm$0.5 10$^{-19} erg/s/cm^2$). The H$\beta$ line is not detected, with a 2$\sigma$ flux upper limit of 0.75 10$^{-19} erg/s/cm^2$. The corresponding limit on the Balmer ratio results in $E(B-V) >$=0.32. We performed an SED-fitting of the object with \textsc{zphot} by fixing the redshift at the spectroscopic value and adopting the aforementioned constraint on $E(B-V)$, finding that the object is fitted as a low-mass (log$(M_{\rm star}/{\rm M}_{\odot})$ = 8$^{+0.4}_{-0.8}$), low SFR (=0.07 $^{+1.9}_{-0.06}$ M$_{\odot}$~yr$^{-1}$) dusty object.

\subsubsection{COSMOS\_35731 at $z_{spec}$=3.66} The CAPERS spectrum of this source (R.A.= 150.133574 deg., Dec=2.2710158 deg.) has prominent H$\alpha$ (28.6$\pm$0.2 10$^{-19} erg/s/cm^2$), [O~III]$\lambda_{4959,5007}$ (8.7$\pm$0.2 10$^{-19} erg/s/cm^2$) and H$\beta$ (4.8$\pm$0.2 10$^{-19} erg/s/cm^2$) emission, yielding a robust spectroscopic redshift $z_{spec}$=3.657. The Balmer decrement implies a high extinction ($E(B-V)=$0.6$\pm$0.04), with a corrected SFR measured from the H$\alpha$ luminosity of 1.9 $\pm$ 0.1 M$_{\odot}$~yr$^{-1}$. The SED-fitting performed at fixed redshift and constraining $E(B-V)$ within the 2$\sigma$ range inferred from spectroscopy yields a stellar mass of only log$(M_{\rm star}/{\rm M}_{\odot})$ = 7.6$^{+0.3}_{-0.2}$.

\subsubsection{CEERS\_15937 at $z_{spec}$=4.91} The spectrum of CEERS\_15937 (R.A.= 214.9442826 deg., Dec= 52.8358477 deg.) shows continuum emission with a clear break at $\sim$ 2.15$\mu$m. The shape of the continuum and the faint emission detectable down to $\sim$ 1$\mu$m  shows that the object is not at very high-redshift. Due to the lack of evident emission features, we fitted the spectrum with \textsc{BAGPIPES} assuming a double power-law SFH, and leaving the redshift as a free parameter. We find that the object is fitted by a low-mass (log$(M_{\rm star}/{\rm M}_{\odot})$ = 8.9 $\pm$ 0.05)  galaxy at $z_{spec}$=4.91$^{+0.03}_{-0.05}$, with low extinction ($E(B-V)<$0.02) and negligible SFR, corresponding to an upper limit of sSFR$<$-11.4 Gyr$^{-1}$. 

\subsubsection{COSMOS\_76919 at $z_{spec}$=5.18} This object (R.A.= 150.1845568 deg., Dec= 2.3535079 deg.) is constrained to be at $z_{spec}$=5.182 from the detection of the H$\alpha$ (2.2$\pm$0.5 10$^{-19} erg/s/cm^2$), [O~III]$\lambda_{4959,5007}$ (4.8$\pm$0.3 10$^{-19} erg/s/cm^2$), and H$\beta$ (0.4 $\pm$ 0.2 10$^{-19} erg/s/cm^2$) lines. Similarly to COSMOS\_31168 and COSMOS\_35731, the object turns out to be a dusty ($E(B-V)=$0.5$\pm$0.3), low-mass (log$(M_{\rm star}/{\rm M}_{\odot})$ = 8.2 $\pm$ 0.4), and low SFR (=0.3 $\pm$ 0.1 M$_{\odot}$~yr$^{-1}$) object on the basis of the Balmer decrement, H$\alpha$ luminosity and constrined SED-fitting performed with \textsc{zphot}.

\subsubsection{UDS\_56824 at $z_{spec}$=6.56} The spectrum of object UDS\_56824 (R.A.=34.454893 deg., Dec=-5.215586 deg) covers only the region at $\lambda_{obs}>$3$ \mu$m, with a total integration time of 5690 secs. The redshift can be accurately measured to be $z_{spec}$=6.56$\pm$0.01 from the H$\alpha$, H$\beta$, [O~III]$\lambda_{4959,5007}$ lines, which are clearly detected with fluxes F$_{H\alpha}$=16.2 $\pm$ 2.1 10$^{-19} erg/s/cm^2$,  F$_{H\beta}$=2.1 $\pm$ 0.9 10$^{-19} erg/s/cm^2$, F$_{[OIII]4959}$=2.1 $\pm$ 1.1 10$^{-19} erg/s/cm^2$, and F$_{[OIII]5007}$=1.26 $\pm$ 0.18 10$^{-19} erg/s/cm^2$. The Balmer ratio implies a high dust attenuation with $E(B-V)$=0.85$\pm$0.46. Its [O~III]$\lambda_{5007}$/H$\beta$=6.0$\pm$2.7 puts the object in the star-forming region of the mass-excitation diagram \citep{Juneau2014}, although at the border with the AGN locus, such that a contribution from a dust-obscured active nucleus cannot be excluded. We checked the ALMA archive finding that the position of UDS\_56824 has been observed by project \#2015.1.01074.S (PI H. Inami) that we analyse as follows. We start from the calibrated measurement set and use the \textsc{CASA} \textit{tclean} function to create a continuum map using the four 2 GHz-wide spectral windows, with an effective central frequency of $\sim$343 GHz (i.e., 870 $\mu$m). We apply natural weighting to the visibilities and test the effect of adding a \textit{uvtaper} value to artificially increase the beam size. A $\sim$3$\sigma$ detection is found at the NIRCam position of UDS\_56824, with an integrated flux density of $S_{870_{\rm \mu m}}=0.54 \pm 0.19\,$mJy (see Fig.~\ref{fig_CAPERS_spec}). 

The measured ALMA flux density was then combined with NIRCam photometry to perform an energy-balance SED fitting using \textsc{BAGPIPES}. During the fitting, the redshift was fixed to the spectroscopic value of z=6.56, while the dust extinction, A$_V$, was allowed to vary freely between 1.8 and 5, as constrained by the Balmer decrement. The best-fit stellar mass and star formation rate are log$(M_{\rm star}/{\rm M}_{\odot})$ = 9.8 $\pm$ 0.18 and SFR= 40 $^{+20}_{-10}$ M$_{\odot}$~yr$^{-1}$, respectively, implying that UDS\_56824 lies above recent estimates of the main sequence at $6 \leq z \leq 7$  \citep{Rinaldi2024,Cole2025}.  Additionally, we estimated the dust mass using standard relationships (e.g., \citealt{2019ApJ...887...55C}) and the following assumptions: $\kappa_{450\rm \mu m}=0.13\,\rm m^2\,kg^{-1}$, T$_d$=25\,K, and $\beta_{dust}$=1.8. This yields a dust mass of $\sim$5.5$\times 10^8\,$M$_{\odot}$, close to 10\% of the stellar mass - a remarkably high value given the redshift of the galaxy. Assuming a higher dust temperature of T$_d=50$\,K would lower the dust-to-stellar mass ratio to log$(M_d/M_{star}) \sim$ -2.0, bringing it into better agreement with other high-redshift dusty galaxies \citep[e.g.,][]{Ferrara2025b, Algera2024} and supporting a mild evolution toward higher dust temperatures at high redshifts \citep[e.g.,][]{Mitsuhashi2024}. Nevertheless, even in this scenario, this galaxy would stand out due to its high dust attenuation compared to UV-selected galaxies.

\subsection{Critical assessment}\label{subsec:assessment}
These findings are consistent with our analysis in Sect.~\ref{sec:INTERLOPERS} indicating dusty star-forming galaxies and low-mass passive objects up to z$\sim$7-8 as a potential source of contamination of the F200W-dropout sample. However, it is apparent that a proper characterization of these interloper populations is extremely difficult and requires spectroscopic follow-up observations. In fact, CEERS\_15937 is the only case for which the $P(z)$ yielded an alternative solution in good agreement with the spectroscopic value (Fig.~\ref{fig_CAPERS_SED}). Instead, the redshifts of the other four objects do not match the primary, low-redshift peaks in our $P(z)$, although the most likely alternative solutions are similarly dusty, star-forming objects whose broad-band fluxes are boosted by different combinations of emission lines.

Similarly to CEERS-93316 \citep{ArrabalHaro2023b}, the SEDs of COSMOS\_31168, COSMOS\_35731, COSMOS\_76919 and UDS\_56824 resemble that of z$>$15 LBGs due to the combination of a red, attenuated continuum and line emission. Instead, CEERS\_15937 enters the F200W-droput selection due to the well-known ambiguity between Lyman and Balmer break detection through broad-band photometry. 
Their position in the $E(B-V)$ vs. $M_{star}$, and in the $sSFR$ vs. $E(B-V)$ planes are in line with the expectations based on the $P(z)$ (Fig.~\ref{fig_lowz_phys}). The four dusty star-forming interlopers are significant outliers in the $A_{1600}-M_{\rm star}$ relation \citep{McLure2018b}, similarly to CEERS-93316 and CEERS-14821 \citep{Bisigello2025}.  Object UDS\_56824 appears to be an extreme case in terms of redshift and stellar mass compared to the locus of alternative solutions of the $z>$15 candidates. The other known interlopers at z$\sim$2-5 are similar to UDS\_56824 in terms of extinction and sSFR, but with a lower stellar mass log$(M_{\rm star}/{\rm M}_{\odot})\lesssim$9.  In fact, UDS\_56824 shows that even at z$\sim$6.5, some galaxies may be dominated by dust-obscured star formation (note that the uncorrected H$\alpha$-based SFR is $\sim$ 2-4 M$_{\odot}$~yr$^{-1}$). Finally, this ALMA-detected galaxy is fainter than all SCUBA-2-selected sources in UDS \citep{Geach2017} and all previous ALMA detections in the field reported in \citet{Dudzeviciute2020}, suggesting that a significant population of dusty galaxies at z$>$6 may have been missed by previous submillimeter surveys.

Object CEERS\_15937 at z=4.91 is remarkable, being among the most distant confirmed passive objects. Most importantly, it is the least massive known to date, the only two comparable cases being GS-z5-Q1 and COS-z5-Q1 with a $>$3 times larger stellar mass \citep[log$(M_{\rm star}/{\rm M}_{\odot})\sim$9.5-9.6,][]{Baker2025}. Similarly to  GS-z5-Q1 and COS-z5-Q1, CEERS\_15937 is a member of a known $z\simeq$5 overdensity \citep{Naidu2022c,ArrabalHaro2023b} together with other known quiescent \citep{deGraaff2025} and dusty \citep{Zavala2022,Bisigello2025} galaxies. The number density and formation pathway of low-mass quiescent galaxies at high-redshift are presently unknown \citep[e.g.,][]{Merlin2025}. An in-depth investigation of CEERS\_15937 and similar objects may shed light on the role of enviroment at early times and on their connection with the recently discovered class of \textquote{mini-quenched} galaxies at high-redshift \citep{Strait2023,Looser2024}.

We exploit the analysis described above to perform a detailed evaluation of the objects in our reference sample of F200W-dropouts. We first note that two objects in our sample consistently show a probability for the high-redshift solution which is significantly lower than any of the low/intermediate redshift peaks: COSMOS\_84213 and COSMOS\_118438. In the case of COSMOS\_84213, this is likely due to a marginal detection (SNR$\sim$2) in one HST band each, albeit it is non-detected in both the ACS and WFC3 stacks that we used for our selection. In addition, this source show a drop in the F410M band similarly to COSMOS\_31168, COSMOS\_76919, and UDS\_56824, thus making dusty star-forming solutions more likely. On the contrary, COSMOS\_118438 has a slightly higher flux in F410M than in F356W and F444W, which also leads SED-fitting codes to prefer strong-line emitting templates at z$\lesssim$6. In fact, we tested that the probability of the $z>15$ solution increases after removing the F410M from the fit for all the aforementioned sources. The remaining three sources in our sample have high-redshift solutions as significant as the low-redshift ones in most of the SED-fitting runs. In particular, CEERS\_17384 and COSMOS\_107923 show a high consistency among all codes and recipes. However, there are compelling reasons to consider their reliability with caution. The CEERS object appear somewhat similar to CEERS\_15937, as it has a clear peak corresponding to a passive solution at $z_{phot}\sim 5$, leading us to suspect that it is a member of the same $z\simeq$5 overdensity \citep{Naidu2022c,ArrabalHaro2023b}. Finally, COSMOS\_107923, albeit being the brightest in the sample falls in a region of relatively shallow image depth and therefore has a corresponding limited sampling of the spectral break. This source also has $\beta \sim$-1.2 which is redder than the UV slopes measured in spectroscopically confirmed objects at $z>10$ \citep{RobertsBorsani2024,RobertsBorsani2024b}, with the only notable exception of the X-ray emitting AGN GHZ9 at z=10.145 \citep[$\beta$=-1.1$\pm$0.12,][]{Napolitano2024b}.

\begin{figure*}[ht]
\centering
\includegraphics[trim={4cm 0.5cm 3cm 9.0cm},clip,width=\linewidth,keepaspectratio]{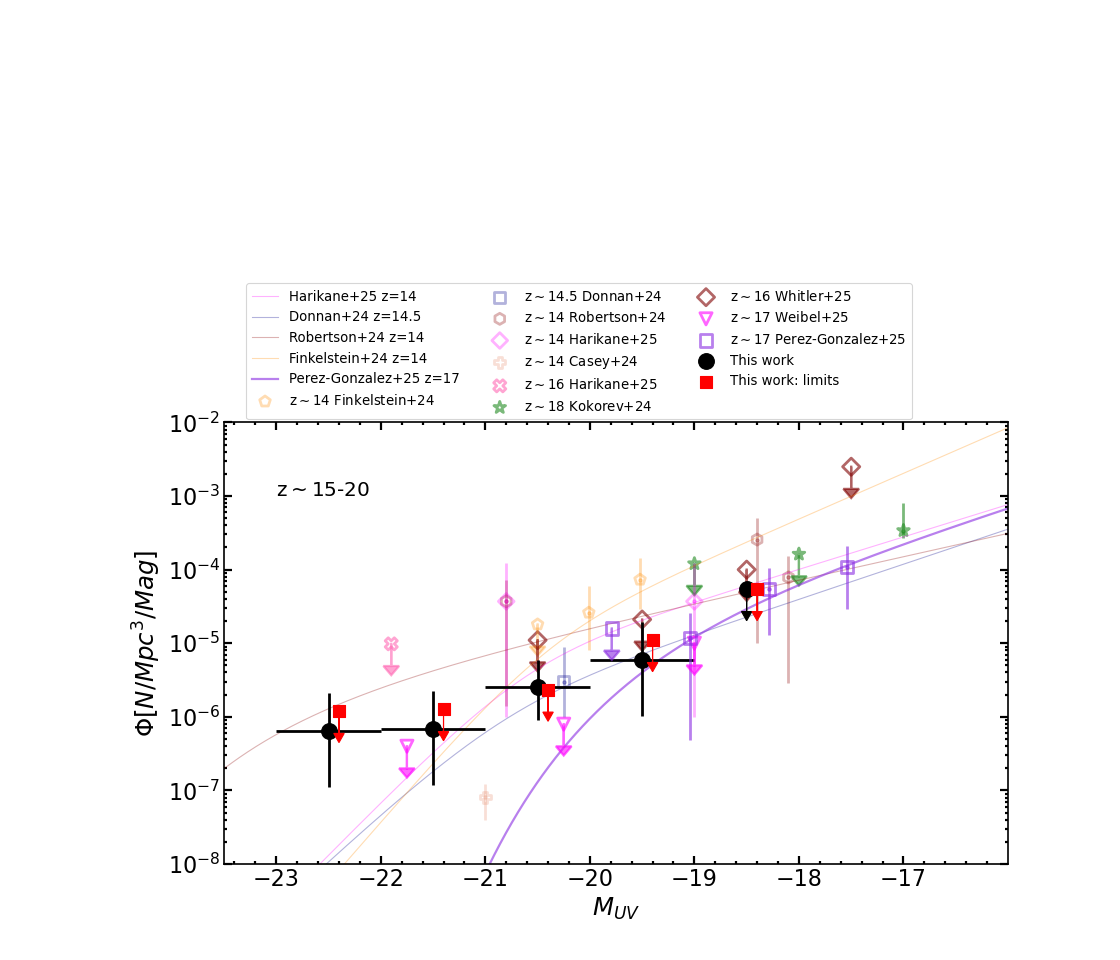}
\caption{The UV LFs at $15 \leq z \leq 20$ based on the F200W-dropout selection of the ASTRODEEP-JWST catalogues, compared to results in the literature by \citet{Harikane2025,Donnan2024,Robertson2024,Casey2024,Finkelstein2024,Whitler2025,Kokorev2024,Weibel2025} (see label for details). The UV LF at $15 \leq z \leq 20$ is shown for the 2 scenarios discussed in Sect.~\ref{sec:LF}: assuming that all candidates are at $z>15$ (\textit{Case 1}, black circles and error-bars), or that they are all interlopers (\textit{Case 2}, red squares and 1-$\sigma$ upper limits).} 
\label{fig_LF_ALL_cons_F200}
\end{figure*} 

\begin{figure}[ht]
\centering
\includegraphics[trim={1.5cm 0.5cm 1.5cm 4.5cm},clip,width=\linewidth,keepaspectratio]{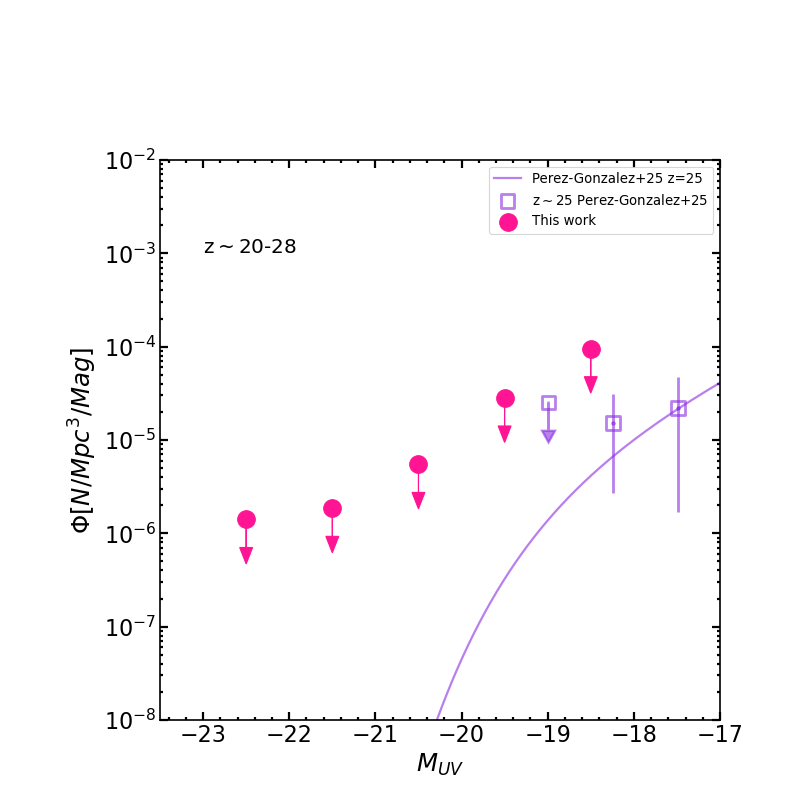}
\caption{The 1-$\sigma$ upper limits on the UV LFs at $20 \leq z \leq 28$ based on the non-detection of F277W-dropout candidates in the ASTRODEEP-JWST catalogues, compared to binned and Schechter estimates at z=25 by \citet{PerezGonzalez2025}.} 
\label{fig_LF_ALL_cons_F277}
\end{figure} 

\section{Constraints on the UV Luminosity Function beyond z=15}\label{sec:LF}

We explore in this section the implications that our findings may have on the evolution of the Luminosity Function (LF) at $z>15$. In order to take into account the caveats discussed above on the potential contamination from rare classes of interlopers, we consider two opposite scenarios regarding the reliability of our F200W-dropout sample. In \textit{Case 1}, we assume that all our five candidates are at $15<z<20$, as indicated by the selection function of Fig.~\ref{fig_LBGSIM}. As we have described above, all our candidates are selected following a standard, self-consistent approach which has proved to be effective at lower redshifts. Despite the concerns described above, we stress that all our candidates  are at least consistent with the expected properties of galaxies in this redshift range, and albeit contamination is expected both on the basis of our simulations and selection results (Sect.~\ref{sec:LBGDIAGRAMS}), there are no compelling reasons to reject them \textit{a priori}. In the opposite \textit{Case 2} we assume that all candidates are interlopers, corresponding to a non-detection of $z>15$ sources in every field, and compute the corresponding upper limits on the LF.

\subsection{The computation of the LF}\label{sec:LFcalculation}

The LF has been computed following a standard approach described in detail in \citet{Castellano2023a}, which takes into account incompleteness and selection effects through imaging simulations that have been performed separately for each of the fields. Briefly, we inserted in blank regions of the observed images $2.5 \times 10^5$ mock Lyman-break galaxies at $15<z<30$ and with a uniform distribution at $-23.0 < M_{\rm UV}< -18.0$ mag. The observed magnitudes are obtained by randomly associating a model from a library based on BC03 models with metallicity $Z =0.02\,Z_{\odot}$, $0 < E(B-V) < 0.2$ mag and a constant star-formation history. We assume that objects follow a circular \citet{Sersic1968} light profile with index $n=1$. Considering the lack of estimates of the size distribution at these redshifts we have assumed a fixed size of 0.2 kpc, which is consistent with the typical R$_h$ of our candidates. In order to avoid overcrowding, simulations are performed by inserting 500 objects each time. Detection, photometry, and color selection on the simulated galaxies are performed in the same way as for the real catalogs. The simulated populations are used to estimate the completeness of our colour selections in each of the considered magnitude bins, hence the effective volume accessible in each field.  
In the case of A2744, which is affected by lensing, we adopt the approach in Eq. 1 of \citet{Castellano2023a}, namely the effective volumes in each bin are obtained by taking into account the area at different magnification levels computed on the basis of the model by \citet{Bergamini2022}, and the relevant completeness for the selection of objects with the considered UV rest-frame magnitudes. While we remark that no candidates are found in this field, we limit our LF analysis to regions with $\mu<5$, to avoid the small strongly lensed regions where systematic uncertainties may be significant, and source multiplicity would need to be taken into account in the simulation process.

We underline that our procedure for estimating the binned LF does not attempt to include the effect of contamination by leveraging the information contained in $P(z)$ to weight the number of observed objects \citep[see e.g.][]{Donnan2024}. However, for the reasons described above, we believe that the knowledge of the population of potential interlopers in LBG selections at these extreme redshifts is too uncertain at the moment, and we prefer to bracket the various options with the two opposite scenarios described above.

The results are reported in Tables~\ref{tab_F200WLF} (z=15-20) and~\ref{tab_F277WLF} (z=20-30), and shown in Fig.~\ref{fig_LF_ALL_cons_F200} and~\ref{fig_LF_ALL_cons_F277}, respectively.

The comparison to available measurements of the LF at $z\simeq 14$ clearly shows that if all our 5 F200W-dropout candidates are at $z\simeq 15-20$ (\textit{Case 1}), the LF continues a trend of very slow evolution, similar to recent findings at $z=10-15$ \citep{Harikane2025,Donnan2024,Robertson2024,Casey2024,Finkelstein2024}. Our estimates would be in agreement with the measurements at z=16-18 by \citet{Harikane2024}, \citet{Whitler2025}, \citet{Kokorev2024} \citet{PerezGonzalez2025}, and \citet{Weibel2025} in the same luminosity range. Most importantly, a similar scenario remains valid if we consider partial, but not complete, contamination of our sample and/or successful confirmation of some candidates in our extended sample. In fact, every point in the LF is originated by a small number of observed galaxies falling in that luminosity bin (typically 1 or 2), such that even if a small fraction of candidates is indeed at $z>15$, the corresponding density in the considered bin would be comparable to the available estimates at $z\sim$14. In addition, two of our candidates are brighter than the highest-redshift secure galaxy JADES-GS-z14-0 \citep[M$_{UV}$=-20.81, ][]{Carniani2024b}: if confirmed, they would imply a number density at M$_{UV}<$-21 higher than current measurements at $z\sim$14.

Needless to say, if we assume that none of our candidates are genuinely at $z>15$ (\textit{Case 2}), the upper limits simply indicate that the actual LF is located at lower densities and luminosities, to an extent that we cannot establish with the existing data. The reader should not get confused by the fact that the upper limits of this scenario are not significantly different from the positive detections.  The 1$\sigma$ upper limit in case of non-detection, computed  on the basis of small number Poisson statistics \citep{Gehrels1986}, corresponds to $\simeq 1.8$, a value very close to the observed densities in \textquote{Case 1}.

The estimate of the galaxy number density at $20<z<30$, for which no reference F277W-dropout candidates have been selected, yields a similar conclusion. The derived upper limits are consistent with the z$\sim$25 estimates by \citet{PerezGonzalez2025}, and somewhat above the existing measurements at $z\simeq 14$ and at $15<z<20$, due to the reduced effective volume sampled by our surveys. This result demonstrates that significantly wider/deeper areas are in principle necessary to sample this redshift range - a point that will be further expanded below.

\begin{figure*}
\centering
\includegraphics[trim={3cm 0.5cm 2.0cm 4.0cm},clip,width=0.45\linewidth,keepaspectratio]{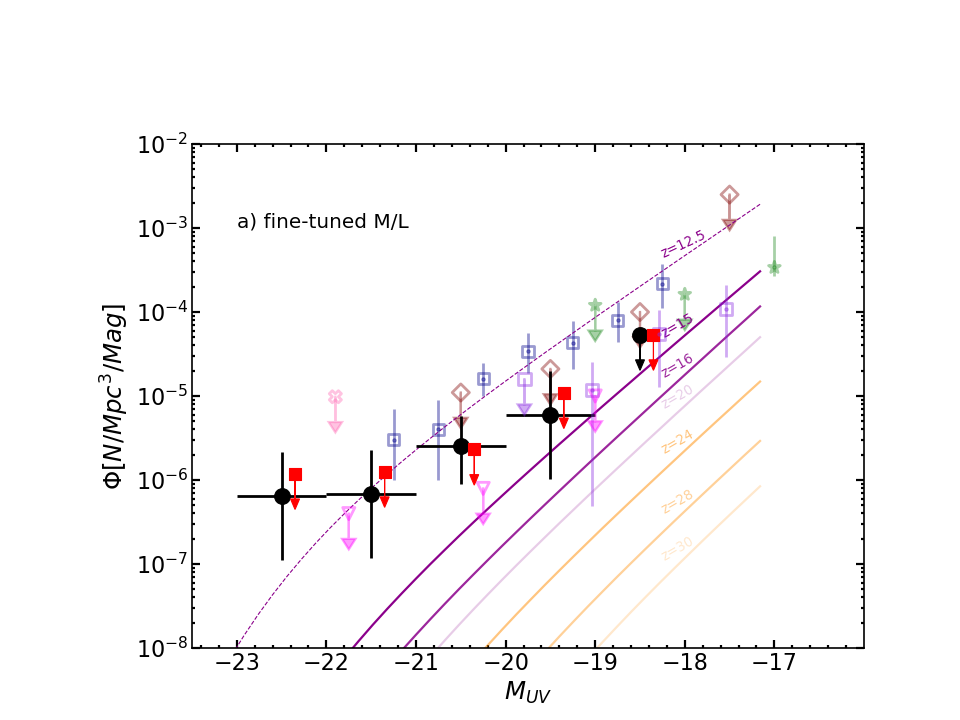}
\includegraphics[trim={3cm 0.5cm 2.0cm 4.0cm},clip,width=0.45\linewidth,keepaspectratio]{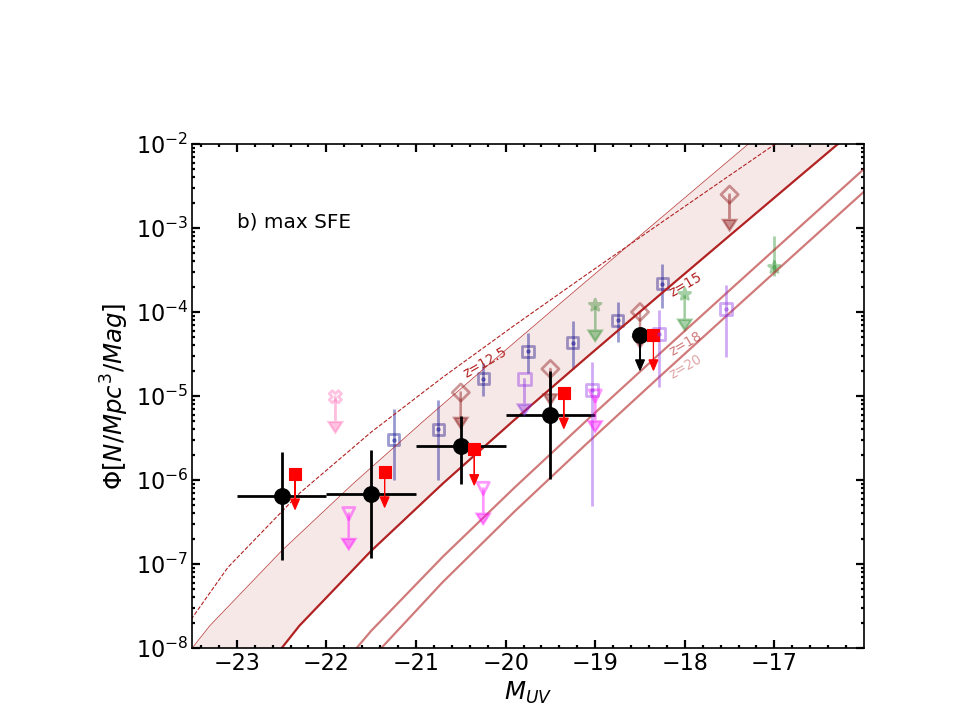}
\caption{Comparison between the UV LF at $15 \leq z \leq 20$ (symbols as in Fig.~\ref{fig_LF_ALL_cons_F200}) and two empirical models at different redshifts as indicated by the relevant labels: a) model based on the \citet{Sheth1999} HMF and the $L_{UV}/M_H$ at $z=5$ by \citet{Mason2015} brightened by 1 magnitude to match the z$\sim$12.5 UV LF; b) a model maximising the abundance of high-redshift galaxies (see Sect.~\ref{sec:LFteo}). In both panels are included for reference the binned LFs measured by \citet{Donnan2024} (z=12.5), \citet{Whitler2025} (z=16), \citet{Kokorev2024} (z=18), \citet{PerezGonzalez2025} (z=17 and z=25), with symbols as in Fig.~\ref{fig_LF_ALL_cons_F200}. } 
\label{fig_LF_theor1}
\end{figure*} 

\begin{figure*}
\centering
\includegraphics[trim={3cm 0.5cm 2.0cm 4.0cm},clip,width=0.45\linewidth,keepaspectratio]{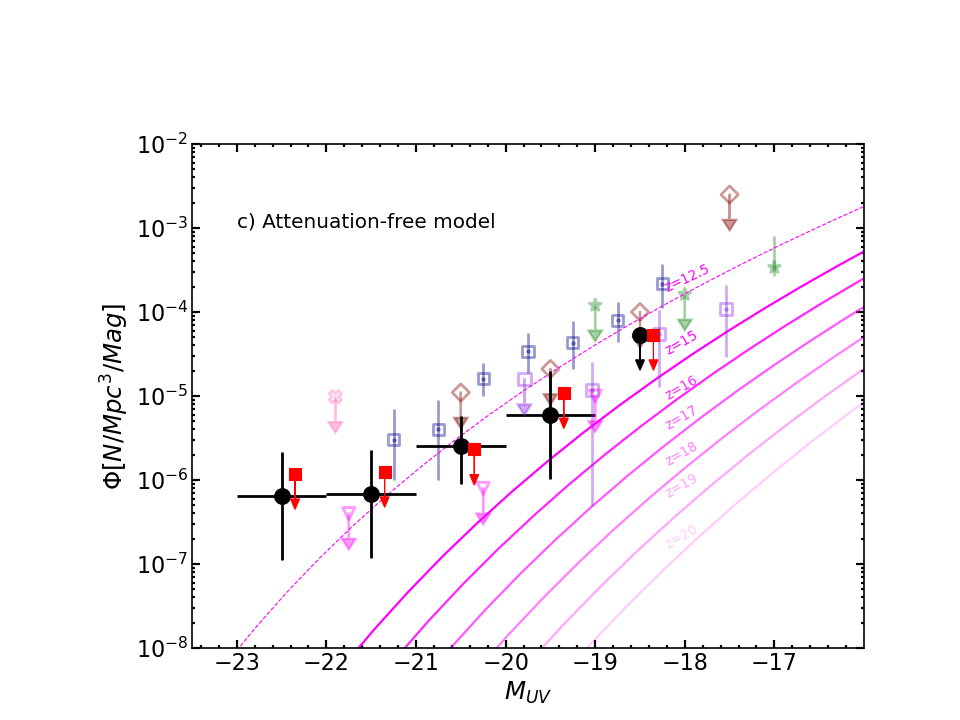}
\includegraphics[trim={3cm 0.5cm 2.0cm 4.0cm},clip,width=0.45\linewidth,keepaspectratio]{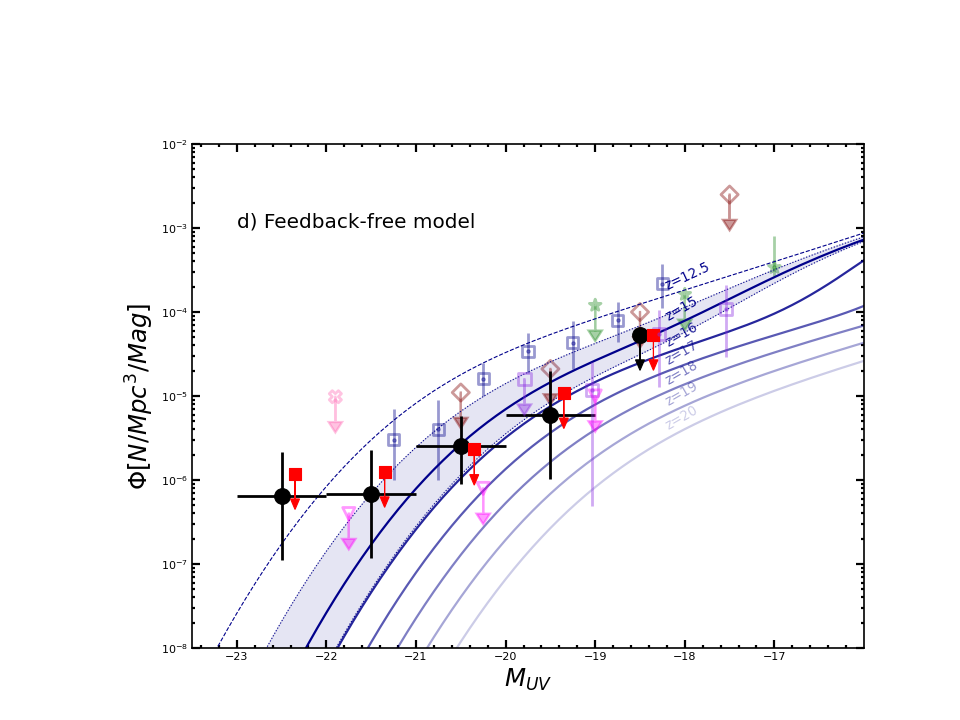}
\includegraphics[trim={3cm 0.5cm 2.0cm 4.0cm},clip,width=0.45\linewidth,keepaspectratio]{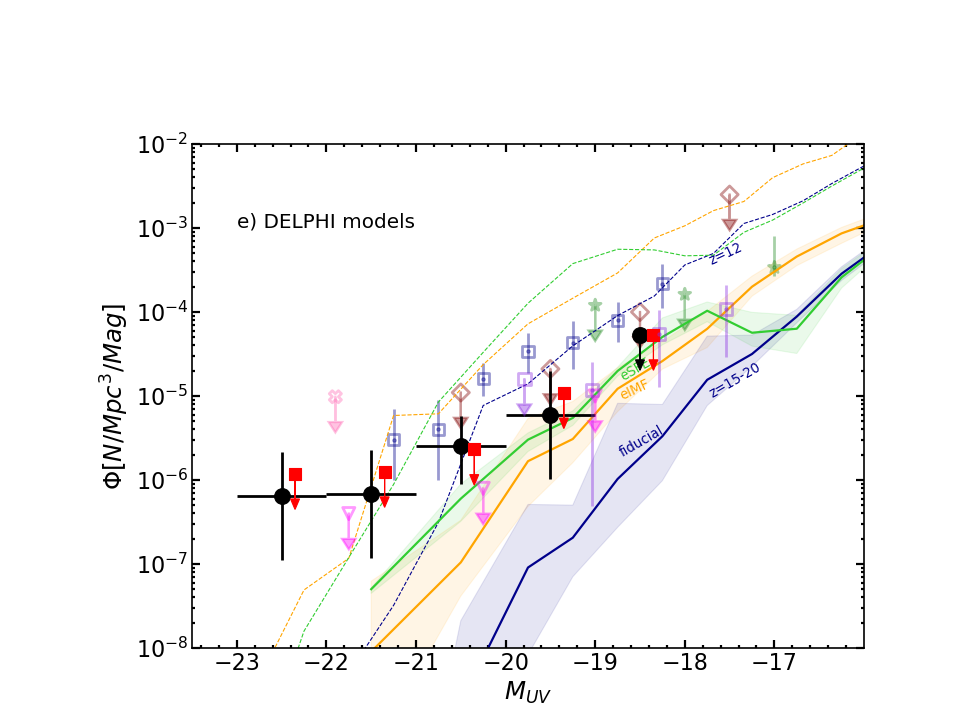}
\includegraphics[trim={3cm 0.5cm 2.0cm 4.0cm},clip,width=0.45\linewidth,keepaspectratio]{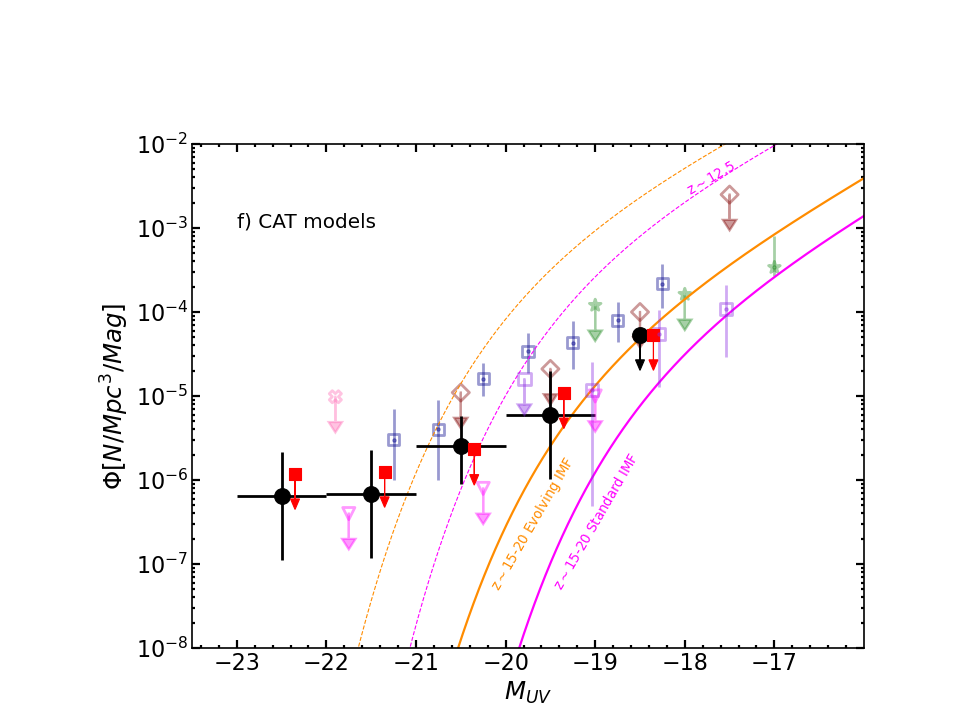}
\includegraphics[trim={3cm 0.5cm 2.0cm 4.0cm},clip,width=0.45\linewidth,keepaspectratio]{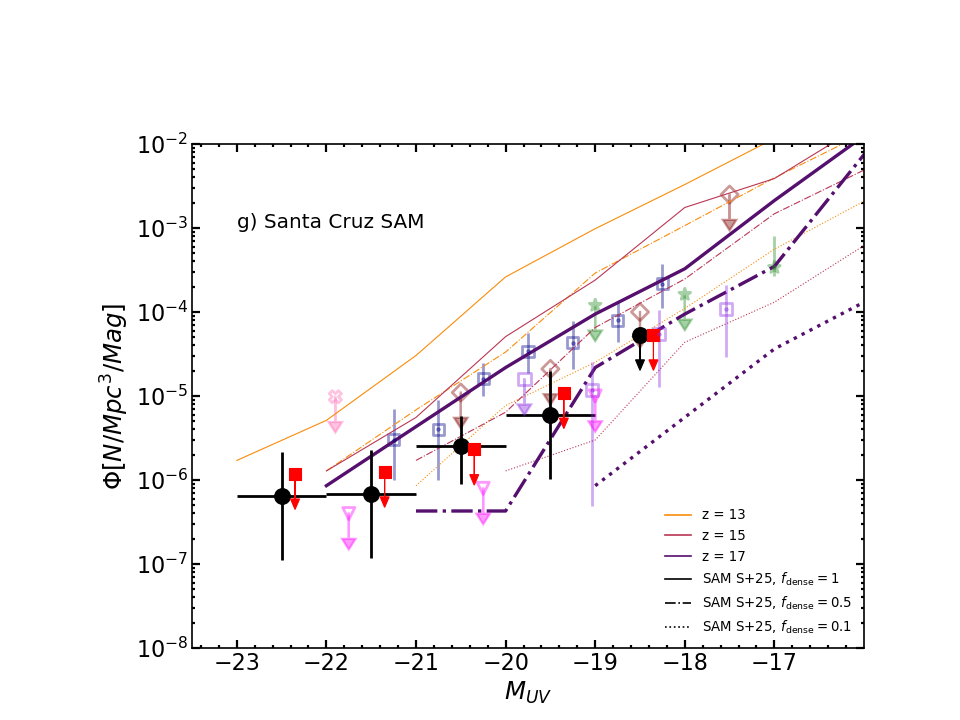}
\caption{Same as Fig.~\ref{fig_LF_theor1} for self-consistent theoretical models: c) Attenuation-free model \citep{Ferrara2022,Ferrara2025,Ziparo2022}; d) the feedback-free model  with $\epsilon$=0.3 (shaded region encloses predictions for $ 0.2 \leq \epsilon \leq 0.5 $ at z=15) \citep{Dekel2023,Li2024}; e) the DELPHI fiducial (blue), eSFE (green) and eIMF (orange) models by \citet{Mauerhofer2025}; f) the CAT models \citep{Trinca2024} with standard (magenta) and evolving IMF (orange); g) the Santa Cruz SAM \citep{Somerville2025}, with dense gas fraction $f_{dense}$ = 0.1, 0.5, 1.} 
\label{fig_LF_theor2}
\end{figure*}

\begin{table}[ht]
\caption{Binned Luminosity Function at z=15-20$^a$}\label{tab_F200WLF}
\begin{tabular}{cccc}
$M_{\rm UV}$ & $N_{obj}$ &   $\phi$ (\textit{Case 1})  &  $\phi$ (\textit{Case 2}) \\
 & &$10^{-5}$ Mpc$^{-3}$ mag$^{-1}$ & \\
\hline
-22.5 & 1&0.06 $^{+0.14}_{-0.05}$ & $<$0.12 \\
-21.5 & 1&0.07 $^{+0.16}_{-0.05}$  & $<$0.13\\
-20.5 & 2&0.25 $^{+0.3}_{-0.16}$ & $<$0.23\\
-19.5 & 1&0.6 $^{+1.4}_{-0.5}$ & $<$1.1\\
-18.5 & 0&$<$5.4 &$<$5.4\\
\hline
\end{tabular}
\small \\a) The number $N_{obj}$ of sources in the reference F200W-dropout sample in each rest-frame $M_{\rm UV}$ bin, and the resulting number densities assuming they are all at $z>15$ (\textit{Case 1}) or that they are all interlopers (\textit{Case 2}). 
\end{table}

\begin{table}[ht]
\caption{Binned Luminosity Function at z=20-30}\label{tab_F277WLF}
\begin{tabular}{cc}
$M_{\rm UV}$ &     $\phi$  \\
 & $10^{-5}$ Mpc$^{-3}$ mag$^{-1}$  \\
\hline
-22.5 & $<$ 0.14 \\
-21.5 & $<$ 0.19 \\
-20.5 & $<$ 0.55 \\
-19.5 & $<$ 2.8 \\
-18.5 & $<$ 9.5 \\
\hline
\end{tabular}
\end{table}

\subsection{Comparison to theoretical predictions}\label{sec:LFteo}
We explore the implications of our results by showing in Fig.~\ref{fig_LF_theor1} and  Fig.~\ref{fig_LF_theor2} a comparison with a variety of theoretical models aimed at explaining the mild evolution of the UV LF beyond z$\simeq$10.

\subsubsection{Empirical models }
To put our results in context, we first present a simple comparison with an empirically-adjusted, theoretically-motivated LF. The model is obtained starting from a
standard calculation of the Cold Dark Matter Halo Mass Function (CDM-HMF). We adopt the \citet{Sheth1999} form, assuming a CDM linear power spectrum. Compared to other expressions proposed so far for the halo mass function \citep[e.g.,][]{Yung2024b}, this form provides the most extended high-mass tail and thus constitutes the most conservative form for our goals. This HMF is converted directly into a UV LF assuming the $L_{UV}-M_H$  conversion curve at $z=5$ by \citet{Mason2015}, brightened by exactly 1 magnitude to broadly match the observed $z\sim12.5$ UV LF by \citet{Donnan2024}. We remark that we make no effort to physically motivate this brightening, which can be ascribed to a number of effects. We use it simply as a reference point to illustrate the evolution of the UV LF at higher redshifts under simple assumptions. 

We let then evolve up to $z=30$  the LF at $z>12$ under the assumption of a non-evolving $L_{UV}/M_H$, which is shown in the same panel of Fig.~\ref{fig_LF_theor1}. In practice, the entire evolution of the LF is driven by the corresponding evolution of the Press \& Schechter HMF. As can be seen, the resulting evolution is extremely accelerated beyond $z=12$, with a drop of up to two orders of magnitude at  $M_{UV}\sim-19$ from $z=12$ to $z=16$, and four orders of magnitudes up to $z=30$, or, equivalently, by a drop in luminosity at constant density of $\sim$2 magnitudes from $z=12$ to $z=16$ and more than 4~mags from  $z=12$ to $z=30$. By construction, this reflects the evolution of the critical mass for collapse in the standard $\Lambda$--CDM model, that indeed evolves dramatically at these redshifts \citep[e.g.,][]{Menci2024}. While the assumption of a constant $L_{UV}/M_H$ beyond $z\simeq 12$ is certainly coarse and inadequate, it is a useful exercise to demonstrate how hard it is to imagine physical mechanisms that may effectively compensate for this fast evolution and maintain the UV LF significantly higher.  

As an opposite case, we build an empirical model explicitly aimed at maximising the abundance of high-redshift galaxies under the extreme assumption that all baryons accreted onto a DM halo are instantaneously converted into stars (\textquote{max SFE} model). Specifically, we assume that the star formation rate equals the baryonic mass growth $f_b \dot m_h$, where $f_b=\Omega_b/\Omega_m$ is the baryon mass fraction and $\dot m_h$ is the dark-matter mass growth rate. The latter is computed after the fitting formula (based on N-body simulations) given in \citet{Correa2015}, which depends on the halo mass $m_h$ and on the redshift. The UV luminosity associated with the different halo masses is then computed by assuming a star-formation efficiency $\epsilon$=1 and a $L_{UV}/M_{star}$ ratio of a dust-free template with metallicity Z=0.02 Z$_{\odot}$, age=10 Myr, and a \citet{Chabrier2003} IMF. Notably, the resulting UV LFs at $z\geq$15 fall below our measurements (Fig.~\ref{fig_LF_theor1}, right panel). 
To further include effects that maximize the UV luminosity associated with a given dark matter halo, we also allowed for a stochastic fluctuation of the star formation rate, adopting the simplified description proposed by \citet{Kravtsov2024}. In this approach the star formation rate is multiplied by $10^{\Delta}$, where $\Delta$ is a correlated random number drawn from a Gaussian distribution with zero mean. 
Since our aim is not to provide a best-fit of the LFs but rather to derive a maximal UV luminosity associated with dark matter halos, we adopt a simplified treatment, where -  instead of extracting $\Delta$ from a proper distribution - we assume for it a fixed value $\Delta=0.5$. This is larger than the typical range for the rms value $\sigma_{\Delta}=0.08-0.4$ resulting from the analysis of stochasticity of the star formation rate in the high-resolution zoom-in simulations 
by \citet{Kravtsov2024}, and definitely larger than the value $\sigma_{\Delta}=0.15$ they assume in their best-fitting models for the UV luminosity function. Only the additional effect of extreme stochasticity, highlighted by a shaded region in Fig.~\ref{fig_LF_theor1}, allows this simple empirical model to match the number density inferred from our reference F200W-dropout sample \citep[consistently with the results by][]{Pallottini2023}. A similar result may be obtained by allowing for top-heavy, or flat IMFs, which can increase $L_{UV}$ by up to a factor of $\sim$10 compared to a Chabrier IMF. 

These simple tests suggest that any smoothly evolving theoretical extrapolation of the $z\sim 12$ UV LF would predict an evolution at $z>15$ stronger than our estimates, and that additional physical mechanisms must be at play to match such a high abundance of bright galaxies.

\subsubsection{Analytic and semi-analytic models }
A number of self-consistent physical models have been explored to understand the high-abundance of bright galaxies observed at z$\gtrsim$9 by JWST. In the \textquote{Attenuation-free model} \citep[AFM,][]{Ferrara2022,Ziparo2022,Fiore2023,Ferrara2025} radiation-driven outflows expel or lift the previously formed dust, thus boosting the UV luminosity to an extent that matches the observed LFs at z$\sim$10-14. In the \textquote{Feedback-free starbursts} scenario \citep[FFB,][]{Dekel2023,Li2024} the excess of bright galaxies is explained as the result of high densities and low metallicities yielding a extremly high star-formation efficiency at cosmic dawn\footnote{The predictions used in the present work incorporate the improved estimate of the UV luminosity based on the median SFH, as described in \citet{Dekel2025}.}. The DELPHI semi-analytic model (SAM) based on cold gas fractions and star formation efficiencies sampled from the \textsc{sphinx} simulations \citep{Mauerhofer2025} explored two different mechanisms boosting the abundance of galaxies at $z\gtrsim$9: a stellar initial mass function (IMF) that becomes increasingly top-heavy with decreasing metallicity and increasing redshift (eIMF model), and star formation efficiencies that increase with increasing redshift (eSFE model). Similarly, the CAT SAM invokes a gradual transition in the IMF, modulated by metallicity and redshift to match the UV LFs at very high-redshift \citep{Trinca2024}. Finally, the recently updated Santa Cruz semi-analytic model \citep[][]{Yung2024,Somerville2025} was run on dark matter halo merger trees extracted from the GUREFT simulations \citep{Yung2024b}, and incorporates a star formation efficiency that increases with increasing gas surface density, motivated by results from molecular cloud-scale simulations with radiative transfer. As overall galaxy surface densities are naturally predicted to be higher at early times in the $\Lambda$CDM picture, these models predict higher star formation efficiencies and therefore more UV luminous galaxies at early times. The free parameter $f_{\rm dense}$ represents the fraction of the ISM that is in dense, star forming clouds. 

Consistently with the simple empirical predictions described above, all these theoretical scenarios point towards a strong evolution at $z>15$ (Fig.~\ref{fig_LF_theor2}). Both the AFM model, and the FFB one with star-formation efficiency $\epsilon=0.3$, provide a good match to the z$\sim$12 UV LF but while the AFM predicts an abundance of galaxies at z$>$15 lower than our Case 1 LF, the FFB model with $0.2 \leq \epsilon \leq 0.5$ is consistent with our estimates at M$_{UV}>$-22. The CAT SAM based on a standard IMF \citep{Trinca2024}, the DELPHI \textquote{fiducial} model, and the Santa Cruz SAM with $f_{dense}$=0.1, fall below our estimates.

Interestingly, $\gtrsim$2$\sigma$ tensions are found at the brightest magnitudes, i.e. at M$_{UV}<$-22 in the case of the FFB model, and at M$_{UV}<$-20 for the others. The tensions are apparently alleviated when assuming a change in physical properties. The DELPHI eIMF and eSFE models, and the CAT model assuming an evolving IMF, are partially consistent with our estimates at $M_{UV}\sim$-18.5--19.5. An evolution of the dense gas fraction from $\sim$0.1 at z=13 to $f_{dense}>$0.5 could match our estimates at all luminosities according to the Santa Cruz SAM, while the FFB model with  $\epsilon = 1$ would match the observed Case 1 abundance in the brightest bin. 

These comparisons help us to put the results presented in this paper in context, leading to our main conclusions:

$\bullet$ If even a fraction of the candidates presented here is indeed at $z\gtrsim15$, the tension with existing theoretical models, would be significant. In particular, the confirmation of bright (M$_{UV}<$-21) candidates would require deep revisions of our theoretical framework. A high abundance at the bright-end would imply a SFE close to 100\%, or a substantial contribution from AGN or other very luminous sources, such as black holes \citep{pacucci2022} or primordial black holes \citep{Matteri2025}. In addition, a successful confirmation of any of the candidates in the extended F277W-dropout sample (Sect~\ref{sec:appendix-extended}) would imply a dramatic discrepancy with all theoretical models. 

$\bullet$ If instead all these candidates are interlopers, we are forced to conclude that future surveys will need to cover much wider areas to secure the selection of bright galaxies significantly beyond $z=15$ and to test predictions of current theoretical models. We will further investigate this point in the next Section.

\section{Designing a survey to break the z=15 barrier}\label{sec:future}

We built on the analysis of the ASTRODEEP-JWST fields to constrain the requirements for future observations designed to individuate robust samples of galaxies at $z>15$. 

We first discuss how to improve the robustness of candidate selection. We use the properties of potential, rare low-redshift contaminants discussed in Sect.~\ref{sec:INTERLOPERS} to constrain the relative depth between different JWST bands capable of discriminating between interlopers and genuine high-redshift galaxies. We note that in the surveys analysed here the bands immediately blueward of the Lyman break, i.e. F150W and F200W, are typically shallower by $\sim$0.5 mag and up to $\sim$1 mag than the LW bands F356W and F444W sampling the UV continuum up to z$\sim$30. We have thus used the expected fluxes of all templates at $0<z<8$ that provide a good fit to our candidates to build mock JWST catalogs varying the relative depth between the various filters. We consider templates of the low-redshift solutions of all our samples, i.e. the reference F200W dropout objects, and the extended samples of F200W- and F277W-dropouts, and perturb the predicted fluxes with random Gaussian noise. We compared a reference scenario in which the NIRCam SW bands are 0.5 mags shallower than the LW ones, similarly to the observed fields, to scenarios in which the SW bands are as deep or deeper than the LW ones. For simplicity, we normalize the templates to have mag=29 in the F356W band and fix a depth 29AB at SNR=10 in the same band in all our simulations. 
\begin{figure}
\centering
\includegraphics[trim={3cm 0.5cm 2.0cm 4.0cm},clip,width=\linewidth,keepaspectratio]{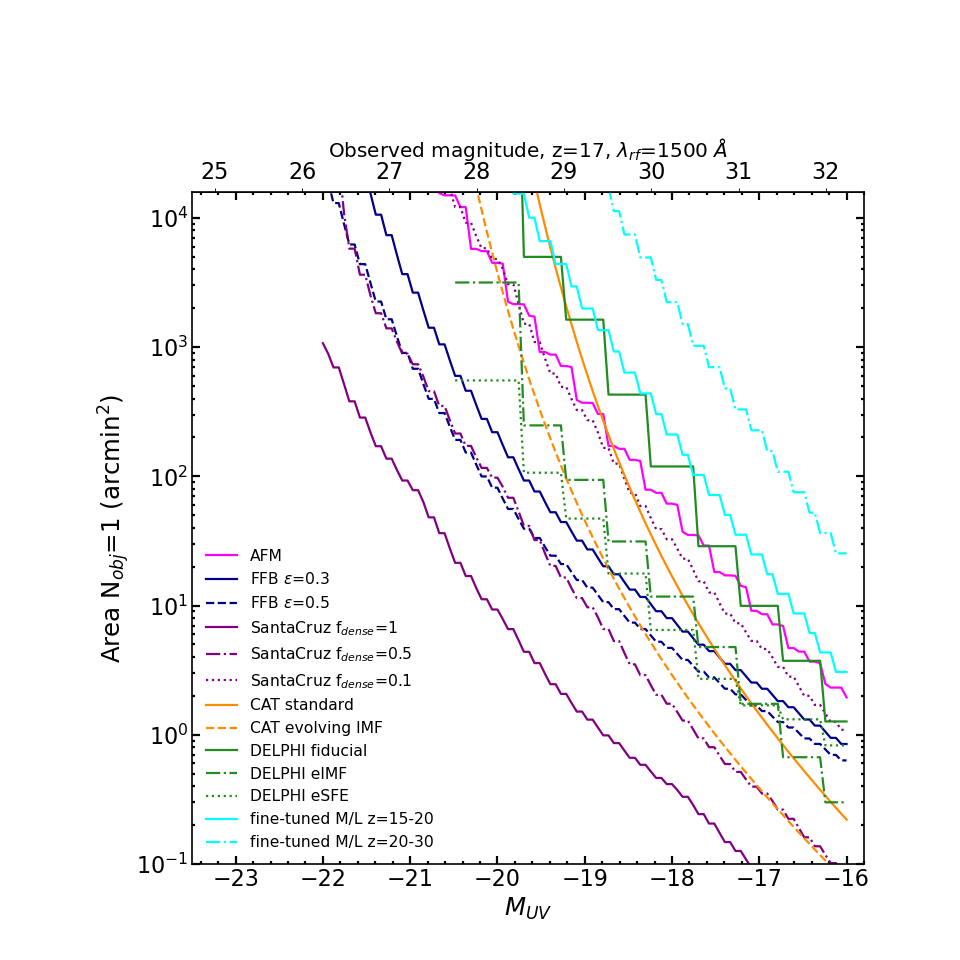}
\caption{The area at which at least one object brighter than M$_{UV}$ at $15 \leq z \leq 20$ is expected according to the following theoretical predictions (see label for details): Attenuation-free model \citep{Ferrara2022,Ferrara2025,Ziparo2022}; feedback-free model with $\epsilon$=0.3 and with $\epsilon$=0.5 \citep{Dekel2023,Li2024}; DELPHI fiducial, eSFE and eIMF models by \citet{Mauerhofer2025};  CAT models \citep{Trinca2024} with standard and evolving IMF; Santa Cruz SAM \citep{Somerville2025}, with dense gas fraction $f_{dense}$ = 0.1, 0.5, 1; empirical \textquote{fine-tuned M/L} model (see Sect.~\ref{sec:LFteo}). The top axis shows the corresponding observed continuum magnitude (at rest-frame wavelength $\lambda_{rf}=1500~\AA$) assuming z=17. The dash-dotted cyan line shows the N$_{obj}$=1 area versus M$_{UV}$ for the empirical \textquote{fine-tuned M/L} model at  $20 \leq z \leq 30$: the relevant observed magnitude at z=25 is fainter by 0.52 mag than the scale shown on top.} 
\label{fig_LF_areas}
\end{figure} 

As expected, we find that the bands immediately redward of the expected break are crucial for discriminating between low- and high-redshift solutions. Increasing the depth of the F090W and/or F115W bands is basically ineffective in reducing the contamination rate, while having both F150W and the F200W as deep as the LW bands reduces by a factor of 3 the fraction of templates contaminating the colour selections. The best advantage is obtained when the two bands are 0.5 mag deeper than the LW ones at fixed SNR. In such a case, the contamination fraction is reduced by a factor of $\sim$10. Most importantly, such a deeper imaging at 1.5-2 $\mu m$ would make it possible to adopt the more inclusive color threshold $(F200W-F277W)>1.0$ (z$\sim$15-20, see Fig.~\ref{fig_LBGSIM}) with a contamination rate of $<$0.5\% of the considered templates, with a significant gain in the accessible color space. Our analysis is in good agreement with the results by \citet{Adams2025}, showing that doubling the exposure time in the bands sampling below the break is the most efficient way to reduce contamination in a standard high-redshift survey.

We then addressed the question of the survey area required to detect a given number of sources as predicted by the various theoretical models discussed in the previous section. The results are shown in Fig.~\ref{fig_LF_areas}. Finding at least \textit{one} object at the very bright end (M$_{UV}<$-20) of the UV LF at $15\leq z \leq 20$ requires reaching a continuum magnitude around 28 (at SNR=10) over an area of $\gtrsim 3000$ sq.arcmin in most of the scenarios. Consistently with Figure~\ref{fig_LF_theor2}, only models with extreme star-formation efficiency predict the detection of one object on areas comparable to our current data set ($\simeq 600$ sq.arcmin). 
Conversely, if one aims at the faint side of the LF (M$_{UV}<$-17) at $15\leq z \leq 20$, an area of $\simeq 5$ sq. arcmin at a depth around $m=31$ is required to detect \textit{one} object. 
Following our tests, in both cases observations at least $0.5$ mags deeper are required in the F150W and F200W bands to perform an unambiguous selection. 

With these numbers in mind, it is intriguing to estimate the JWST observing time needed to reach these combinations of depth and size.
For the bright side (M$_{UV}\leq$-20), according to the JWST exposure time calculator, NIRCam can observe at a depth of $\simeq 28.5$ at SNR=10 both the F150W, F200W bands with a total of $\sim$2.5 hours of net exposure time per pointing. The simultaneous observation in two channels allows to observe F277W, F356W, and F444W at $\simeq 28$ (SNR=10). An area of 12000 sq. arcmin which, according to most of the predictions, enables the detection of at least 2-3 sources with M$_{UV}\leq$-20, would require an investment of no less than $\sim$3000 hours plus overheads. Similarly, a deep pencil-beam NIRCam pointing in the F150W, F200W, F277W and F356W bands to unambiguously detect at SNR=10 at least two ultra-faint objects of continuum mag $\simeq 31$ AB requires more than 1300 hours of net exposure.
While this simple exercise is only meant to provide an order-of-magnitude estimate of the time needed for robust photometric selections, it clearly highlights that a thorough characterization of the available candidates and of the potential contaminants shall be considered a prerequisite for any future effort in this direction.

\section{Summary and conclusions}\label{sec:summary}
The high abundance of galaxies beyond z$\approx$10 can potentially be explained by several competing scenarios. Extending the constraints on the UV LF to the poorly explored range at z$>$15 allows discriminating among different theoretical models, and individuating bright galaxies which are crucial to expand our knowledge on the first phases of galaxy formation. To this aim, we have analysed the ASTRODEEP-JWST photometric sample by \citet{Merlin2024}, which provides consistent measurements on the major JWST deep surveys, to select bright galaxy candidates at z$\sim$15-30. On the basis of mock observations mimicking the properties of our dataset, we have designed specific renditions of the Lyman-break selection technique that efficiently identify galaxy candidates in the redshift ranges $15 \leq z \leq 20$ and $20 \leq z \leq 28$. 

We isolated five candidates at $15 \leq z \leq 20$, while no objects are found at $z\gtrsim20$. A closer inspection of the selected candidates shows that despite exhibiting a $>$1.5 mag break, the selected objects consistently display multimodal redshift probability distributions $P(z)$ across different SED-fitting codes and methodologies. The alternative solutions cover regions in the $sSFR$ versus $E(B-V)$ and $E(B-V)$ versus $M_{star}$ planes different from the general populations at the same redshifts. Most importantly, they correspond to populations of low-mass ($\sim 10^7-10^9 M_\odot$) quiescent or dusty galaxies that have not been thoroughly investigated so far. This result is corroborated by the spectral properties of five confirmed interlopers of our F200W-dropout selection. Four objects are found to be dusty star-forming galaxies, including three low-mass ($\log M_*/M_\odot \simeq $7.6-8.8) objects at z$\sim$2-5, and a highly attenuated ($E(B-V)=0.8$) starburst galaxy at $z=6.56$ with mass $\log M_*/M_\odot = 9.8$. The fifth confirmed interloper, CEERS\_15937 at z=4.91, is found to be the first spectroscopically confirmed passive galaxy with log$(M_{\rm star}/{\rm M}_{\odot}) \lesssim$ 9 at high-redshift. These results imply that while our candidates are, in principle, credible objects at $15<z<20$, none of
them would pass stringent selection criteria based on $\Delta\chi^2$ between different redshift solutions. In addition, considering that the low-redshift templates populate a basically unexplored parameter space whose galaxy density is unconstrained, using their $P(z)$ as a weight to estimate the high-redshift UV LF appears to be an approach rife with uncertainties.

We adopted a more pragmatic approach of estimating the UV LF at z$\approx$15-20 by assuming different contamination levels. The UV LF based on the, admittedly extreme, assumption of negligible contamination indicates a very mild evolution compared to estimates at immediately lower redshifts, at odds with all theoretical predictions. In particular, the confirmation of bright (M$_{UV}<$-21) candidates would require deep revisions of our theoretical framework, and might be even in contrast with any plausible model under a standard $\Lambda$--CDM cosmological scenario. The tension with theoretical models implies that even if only a small fraction of the candidates is confirmed to be at $z>15$, a further evolution of physical properties, such as IMF or star-formation efficiency would be required to explain the observed number densities. If instead all the analysed candidates are interlopers, we are forced to conclude that future surveys will need to cover much wider areas to secure the selection of bright galaxies significantly beyond $z=15$. 

According to a variety of theoretical models, finding at least one object at the very bright end (M$_{UV}<$-20) of the UV LF requires surveying very large areas, ranging from $\sim$500 sq. arcmin. to $\gtrsim$1000 sq. arcmin at $15\leq z \leq 20$, and to more than 2000 sq. arcmin at $z\geq$20. However, our analysis shows that a large area coverage is not the only required ingredient, because the depth achieved by current surveys is not optimal to avoid contamination. A simple test based on the properties of the low-redshift solutions of our candidates indicates that NIRCam imaging in the F150W and F200W bands should be at least as deep as the observations in NIRCam LW bands, and possibly 0.5 mag deeper, to decrease the contamination rate significantly. This is a demanding requirement for both ultra-deep pencil beam observations targeting the LF faint-end and for large area surveys sampling the bright-end. We argue that a more pragmatic approach should aim, first of all, at a thorough spectroscopic characterization of the candidates available on current surveys and of potential contaminating populations to gather key information to plan future surveys aimed at breaking the current redshift records.

\begin{acknowledgements}
We  thank  the  referee  for  the  constructive comments that helped us improve the manuscript. We thank J. Dunlop and R. Ellis for the useful discussions.
We acknowledge financial support from NASA through grant JWST-ERS-1324. Support was also provided by the PRIN 2022 MUR project 2022CB3PJ3 – First Light And Galaxy aSsembly (FLAGS) funded by the European Union – Next Generation EU, by INAF Mini-grant ``Reionization and Fundamental Cosmology with High-Redshift Galaxies", and by INAF GO Grant "Revealing the nature of bright galaxies at cosmic dawn with deep JWST spectroscopy." P.G.P.-G. and L.C. acknowledge support from grant PID2022-139567NB-I00 funded by Spanish Ministerio de Ciencia e Innovaci\'on MCIN/AEI/10.13039/501100011033, FEDER {\it Una manera de hacer Europa}. L.N. acknowledges support from grant ``Progetti per Avvio alla Ricerca - Tipo 1, Unveiling Cosmic Dawn: Galaxy Evolution with CAPERS" (AR1241906F947685). RA acknowledges financial support from grants PID2023-147386NB-I00 "XTREM" and PID2022-136598NB-C32 “Estallidos8”, funded by MCIU/AEI/10.13039/501100011033 and ERDF/EU, and the State Agency for Research of the Spanish MCIU through ‘Center of Excellence Severo Ochoa’ award to the IAA-CSIC (SEV-2017-0709) and CEX2021-001131-S. F.P. acknowledges support from a Clay Fellowship administered by the Smithsonian Astrophysical Observatory. The Flatiron Institute is supported by the Simons Foundation. 
This work is based on observations made with the NASA/ESA/CSA James Webb Space Telescope. The data were obtained from the Mikulski Archive for Space Telescopes at the Space Telescope Science Institute, which is operated by the Association of Universities for Research in Astronomy, Inc., under NASA contract NAS 5-03127 for JWST. These observations are associated with programs \#1324, 1345, 1180, 1210, 1837, 2561, 2756, 3794, 3990, 5893, and 6368. This paper makes use of the following ALMA data: ADS/JAO.ALMA\#2015.1.01074.S. ALMA is a partnership of ESO (representing its member states), NSF (USA) and NINS (Japan), together with NRC (Canada), NSTC and ASIAA (Taiwan), and KASI (Republic of Korea), in cooperation with the Republic of Chile. The Joint ALMA Observatory is operated by ESO, AUI/NRAO and NAOJ. Some of the data products presented herein were retrieved from the Dawn JWST Archive (DJA). DJA is an initiative of the Cosmic Dawn Center (DAWN), which is funded by the Danish National Research Foundation under grant DNRF140.
\end{acknowledgements}

\appendix

\section{The extended samples of z$\gtrsim$15 dropout candidates}\label{sec:appendix-extended}
\begin{figure*}[ht]
\centering
\includegraphics[trim={0.5cm 0.5cm 0.5cm 0.1cm},clip,width=0.3\linewidth,keepaspectratio]{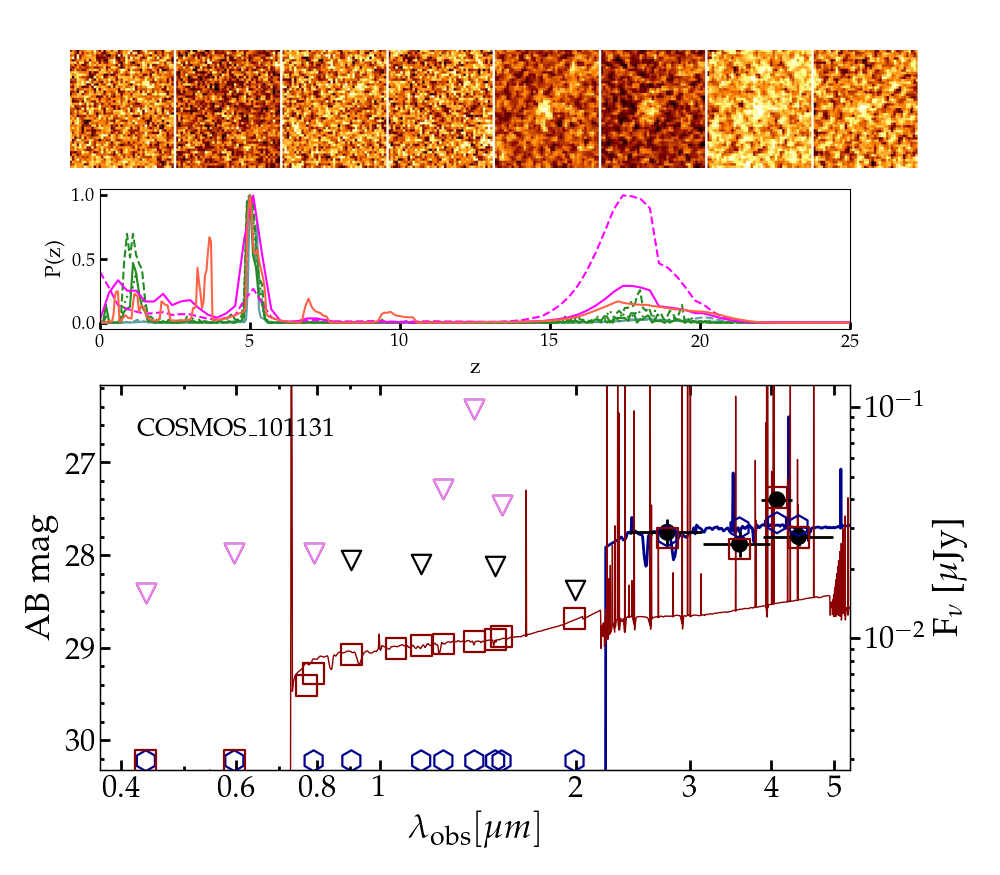}
\includegraphics[trim={0.5cm 0.5cm 0.5cm 0.1cm},clip,width=0.3\linewidth,keepaspectratio]{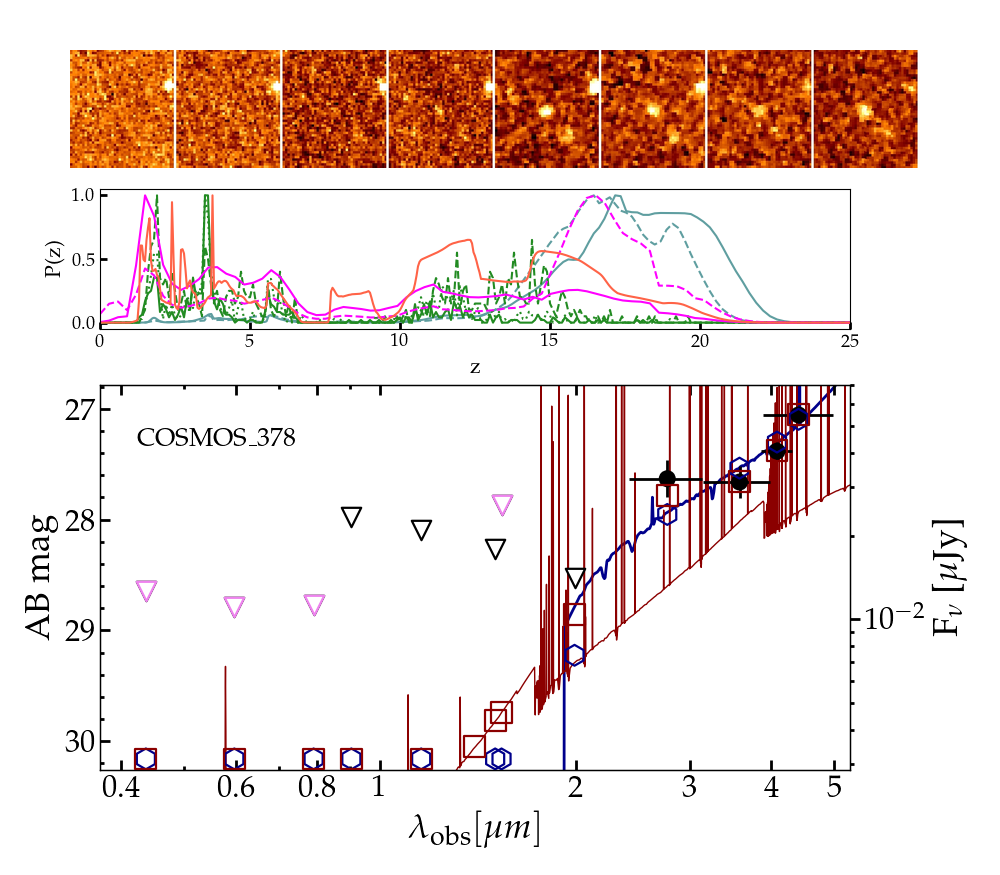}
\includegraphics[trim={0.5cm 0.5cm 0.5cm 0.1cm},clip,width=0.3\linewidth,keepaspectratio]{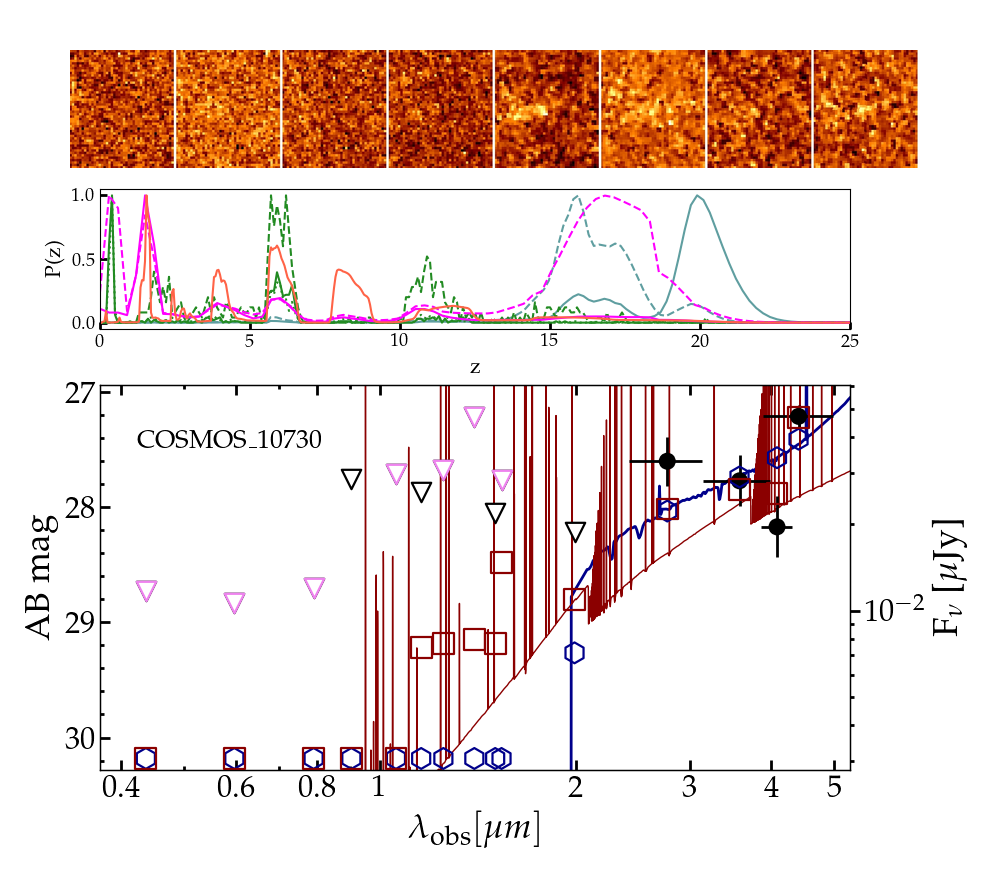}
\includegraphics[trim={0.5cm 0.5cm 0.5cm 0.1cm},clip,width=0.3\linewidth,keepaspectratio]{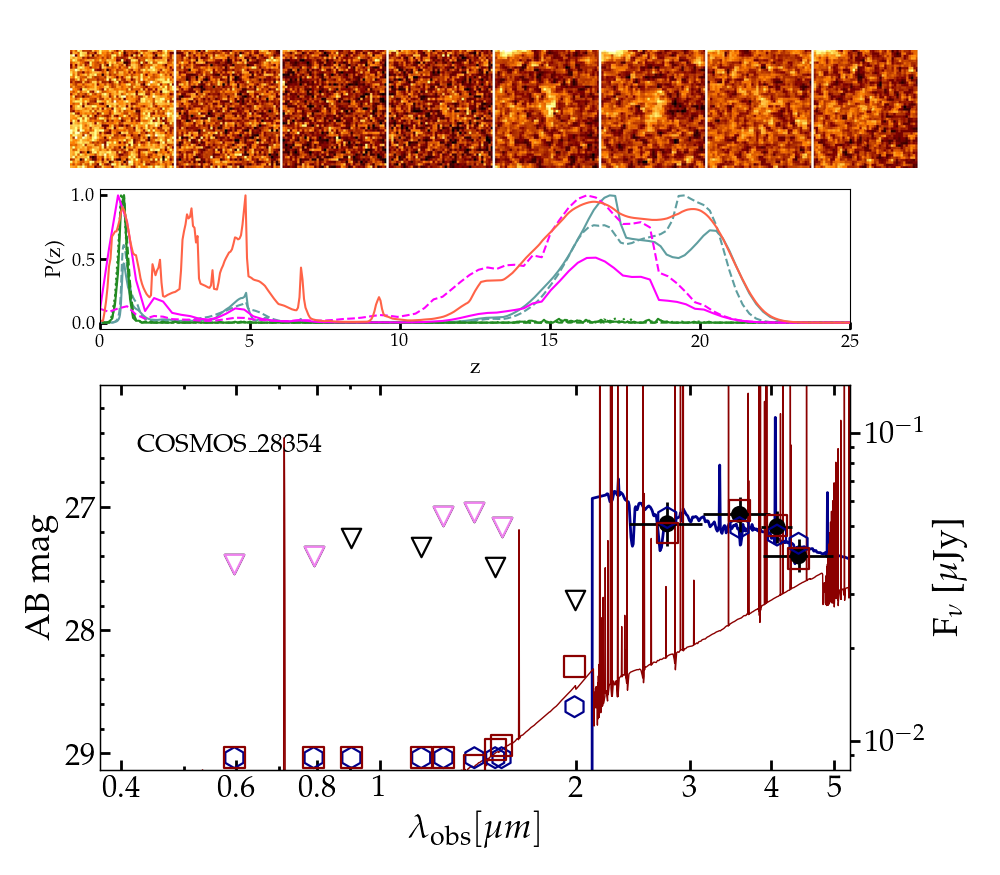}
\includegraphics[trim={0.5cm 0.5cm 0.5cm 0.1cm},clip,width=0.3\linewidth,keepaspectratio]{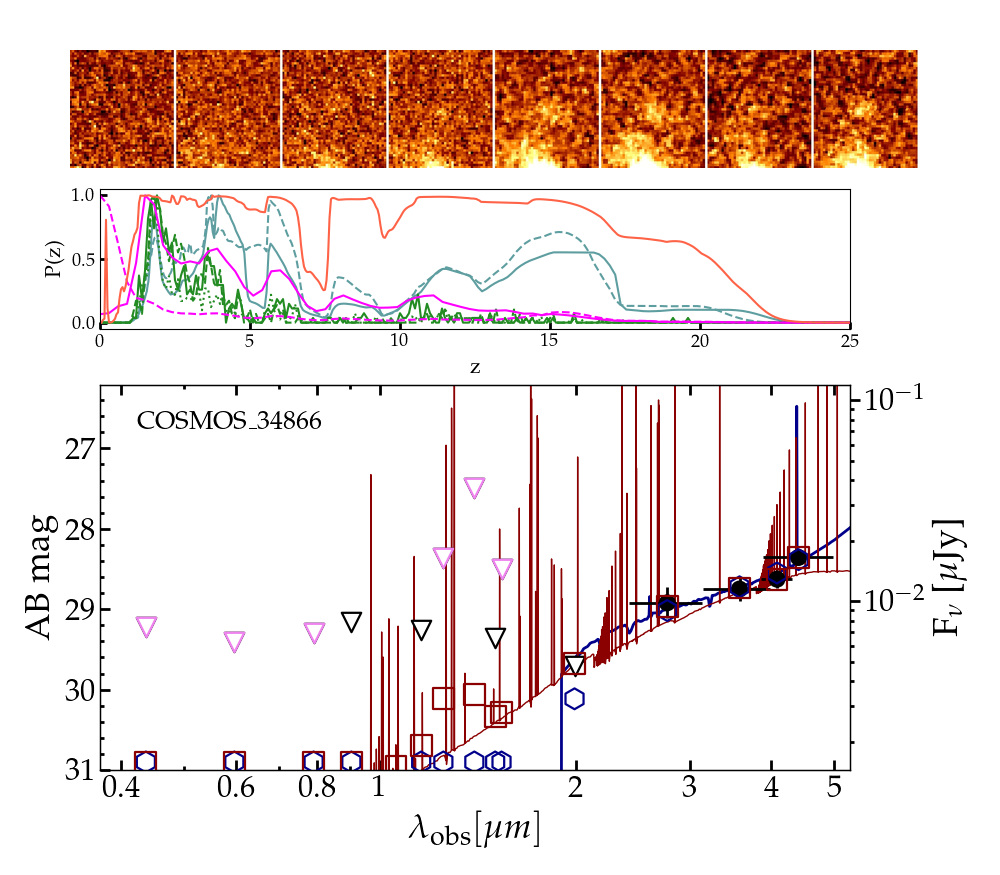}
\includegraphics[trim={0.5cm 0.5cm 0.5cm 0.1cm},clip,width=0.3\linewidth,keepaspectratio]{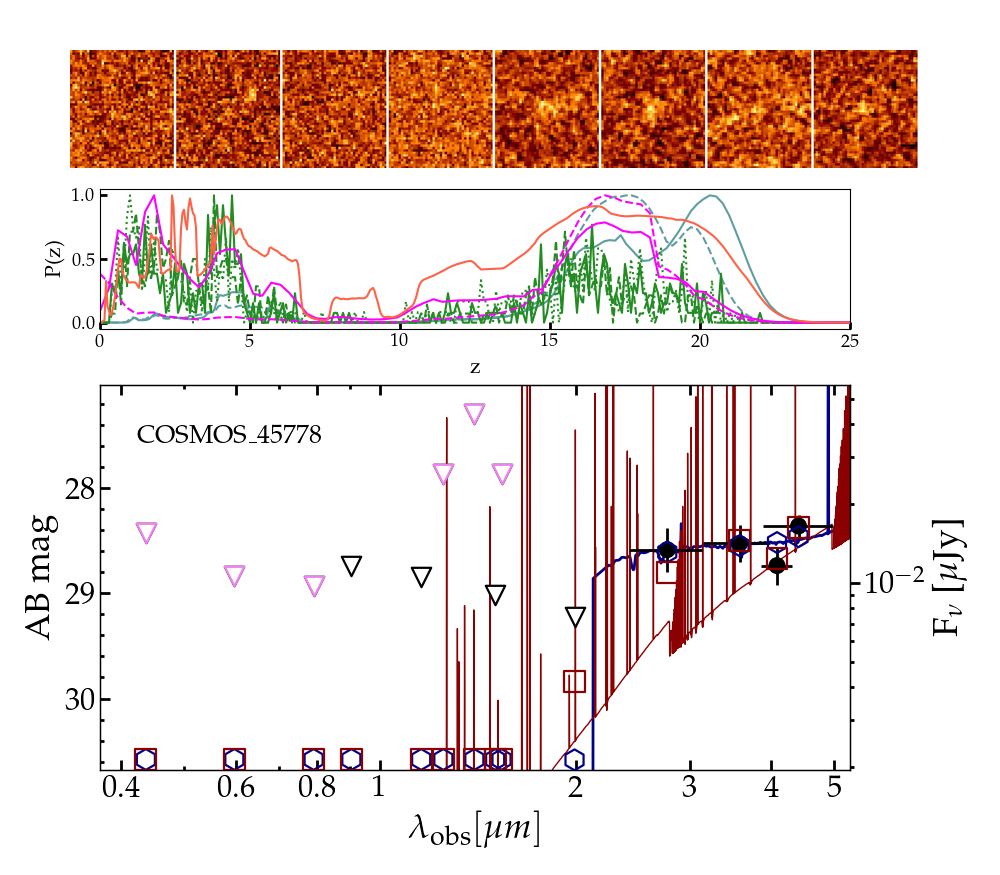}
\includegraphics[trim={0.5cm 0.5cm 0.5cm 0.1cm},clip,width=0.3\linewidth,keepaspectratio]{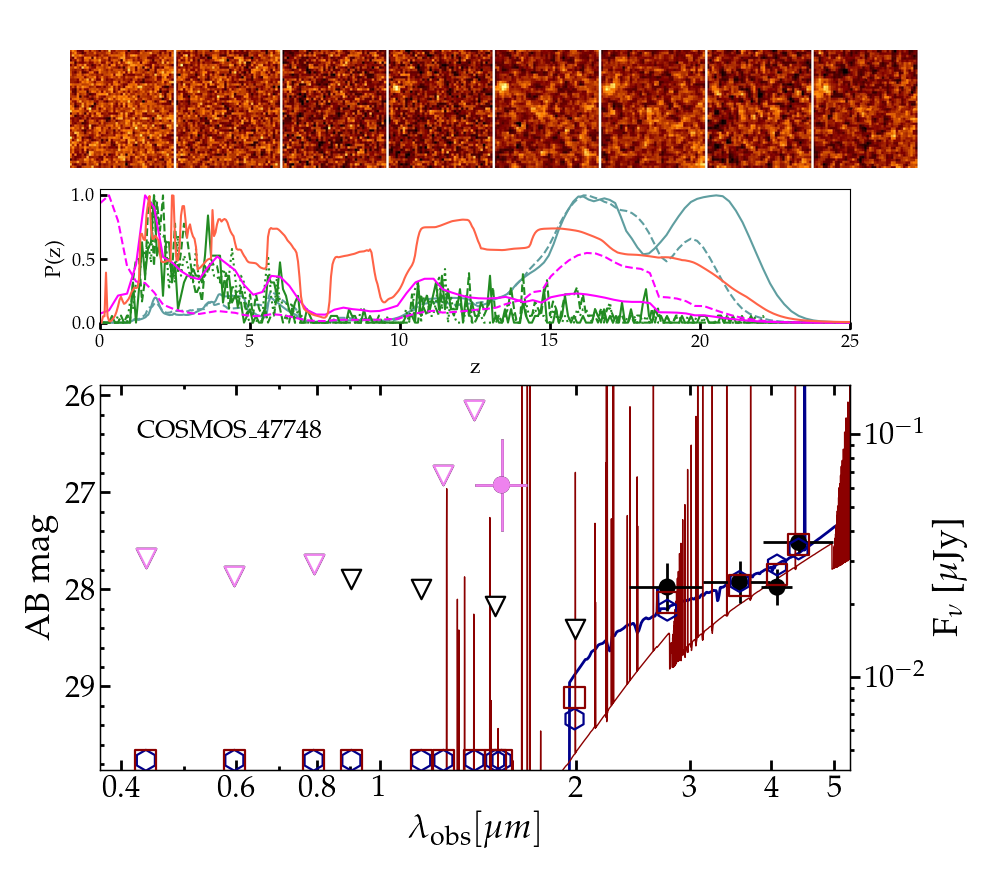}
\includegraphics[trim={0.5cm 0.5cm 0.5cm 0.1cm},clip,width=0.3\linewidth,keepaspectratio]{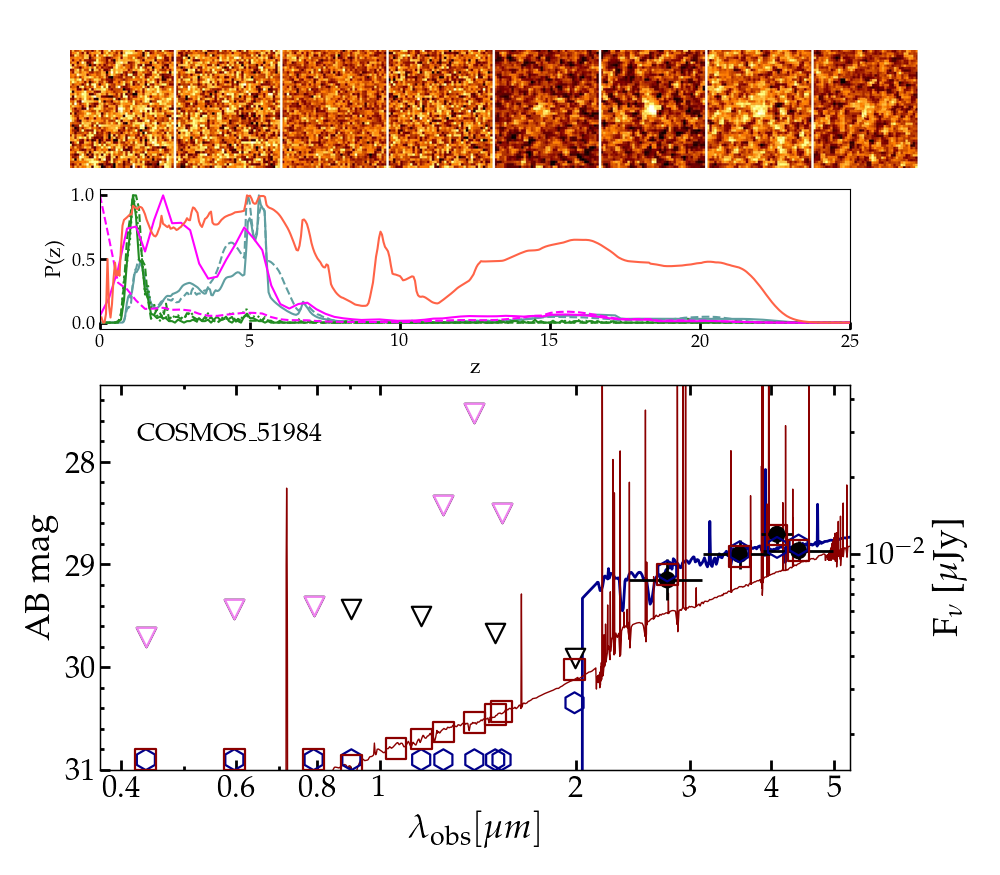}
\includegraphics[trim={0.5cm 0.5cm 0.5cm 0.1cm},clip,width=0.3\linewidth,keepaspectratio]{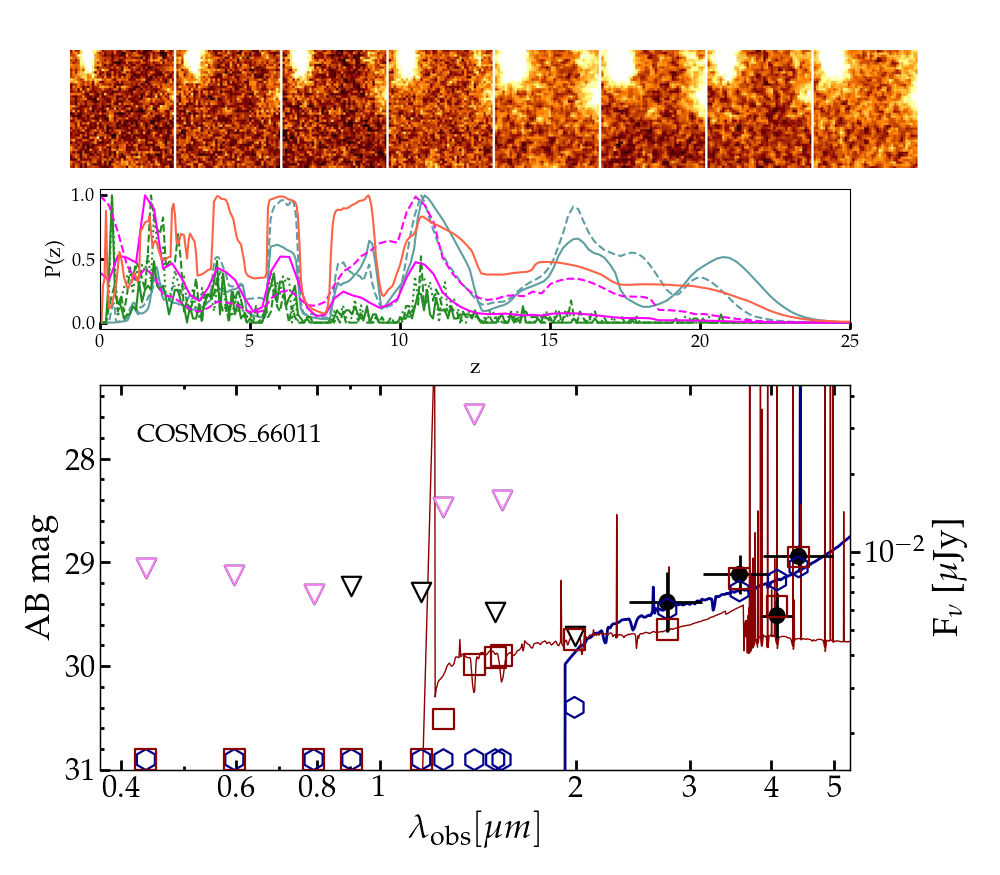}
\includegraphics[trim={0.5cm 0.5cm 0.5cm 0.1cm},clip,width=0.3\linewidth,keepaspectratio]{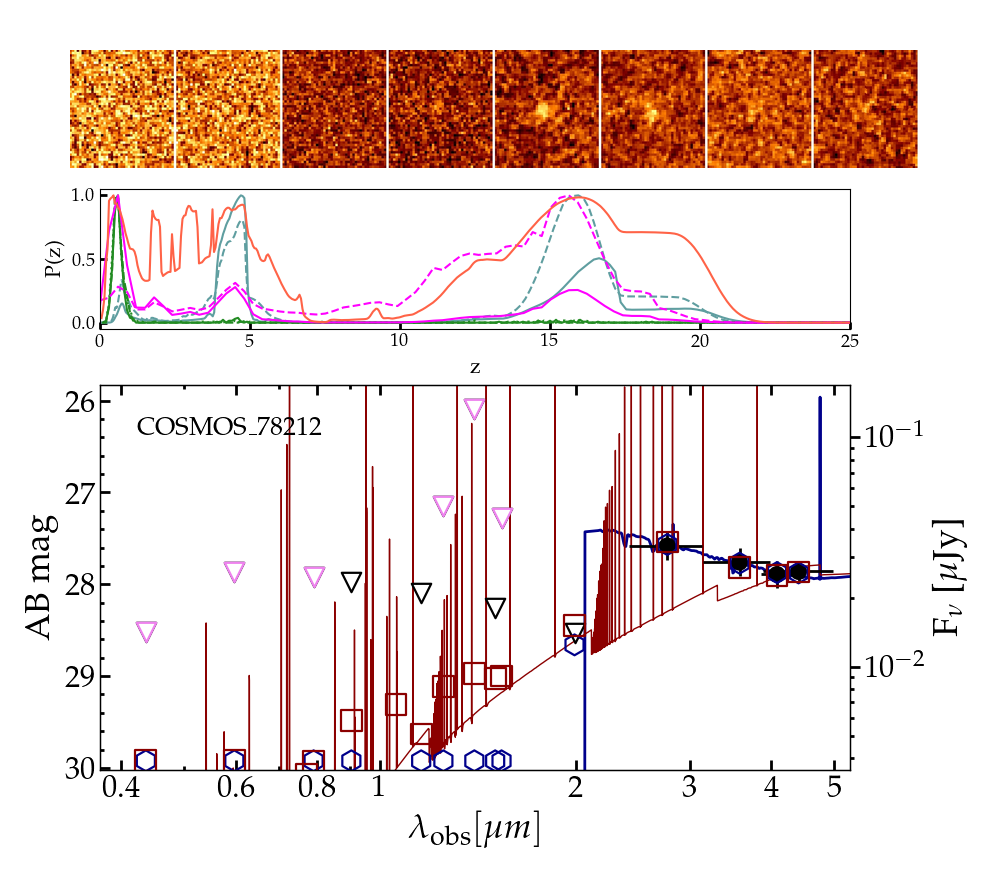}
\includegraphics[trim={0.5cm 0.5cm 0.5cm 0.1cm},clip,width=0.3\linewidth,keepaspectratio]{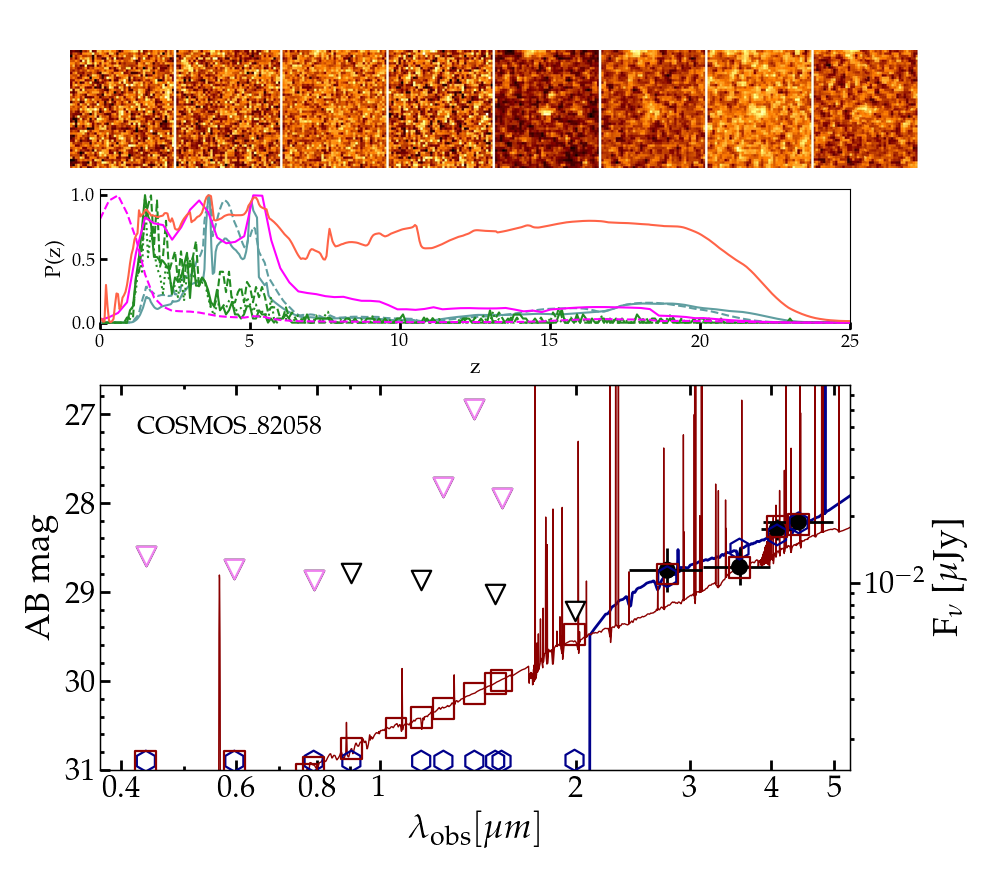}
\includegraphics[trim={0.5cm 0.5cm 0.5cm 0.1cm},clip,width=0.3\linewidth,keepaspectratio]{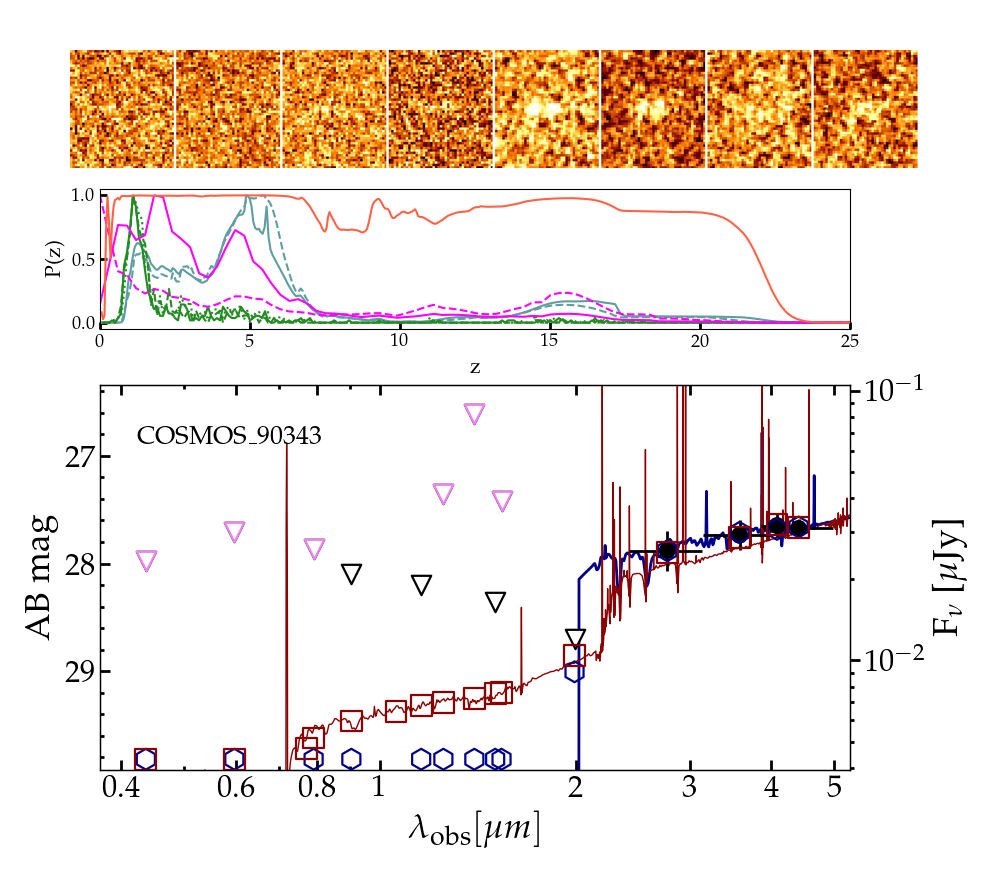}
\caption{Same as Fig.~\ref{fig_SEDs_ALL} for the extended sample of F200W-dropouts (part 1).} 
\label{fig_SEDs_InclF200W_1}
\end{figure*}

\begin{figure*}
\centering
\includegraphics[trim={0.5cm 0.5cm 0.5cm 0.1cm},clip,width=0.3\linewidth,keepaspectratio]{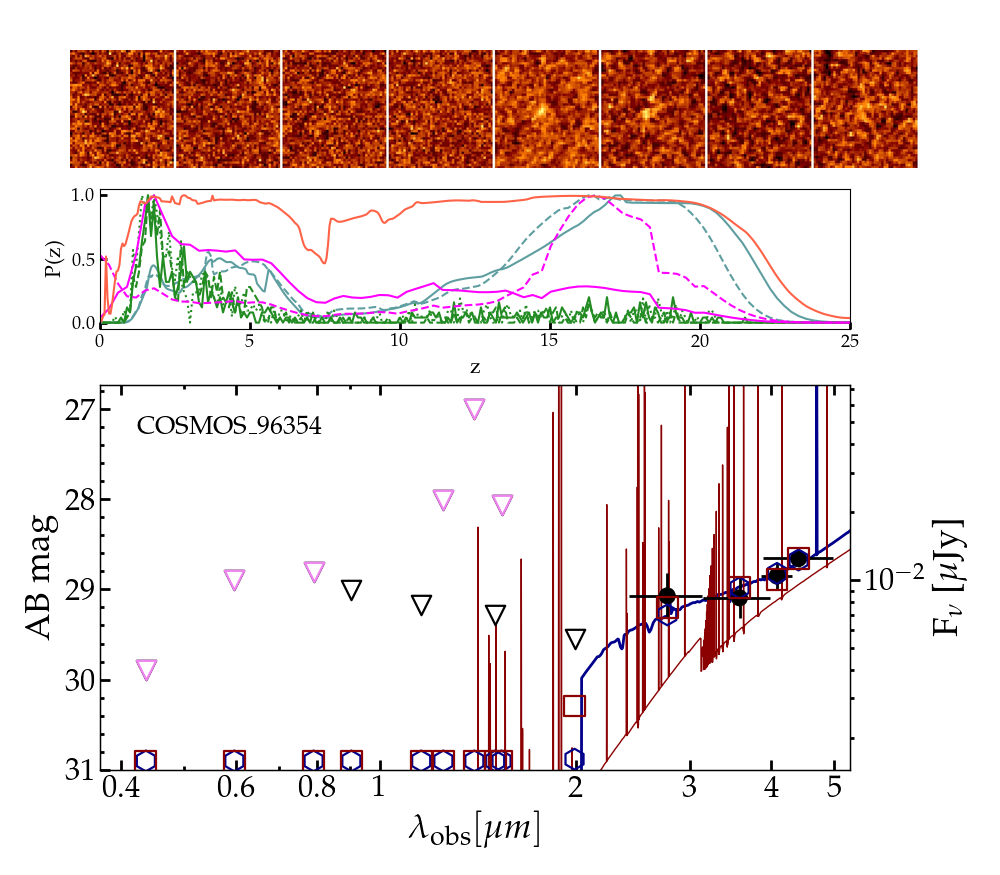}
\includegraphics[trim={0.5cm 0.5cm 0.5cm 0.1cm},clip,width=0.3\linewidth,keepaspectratio]{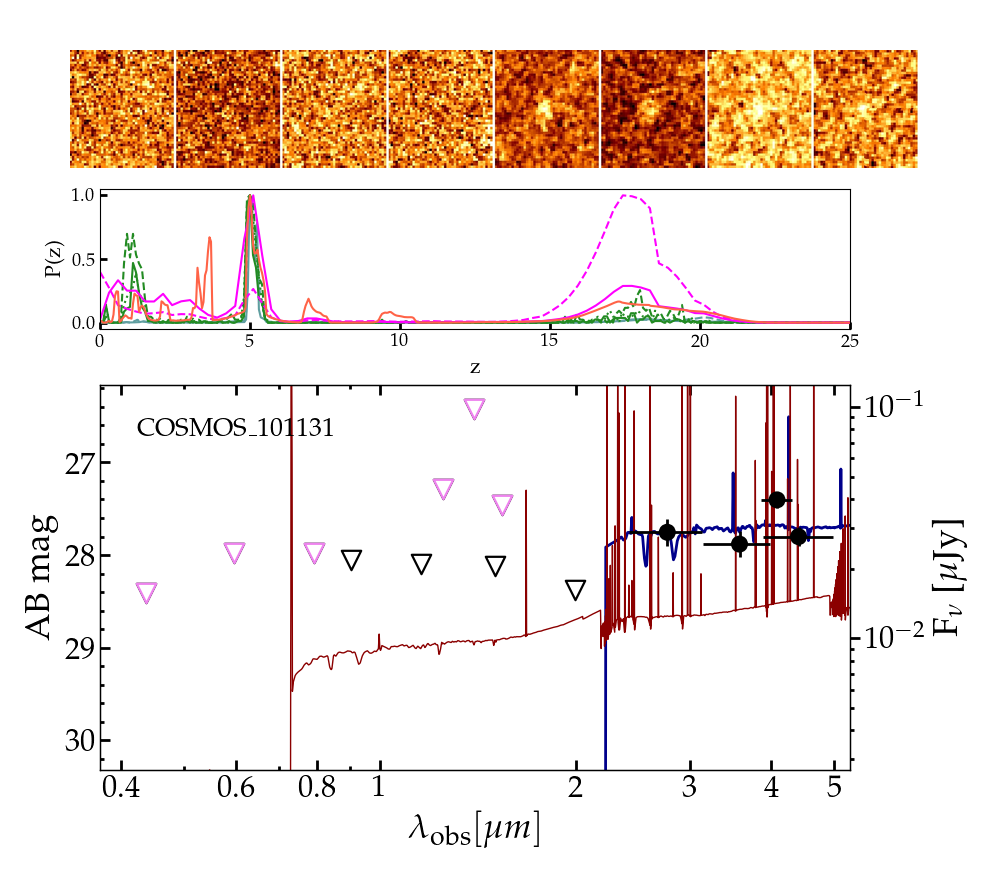}
\includegraphics[trim={0.5cm 0.5cm 0.5cm 0.1cm},clip,width=0.3\linewidth,keepaspectratio]{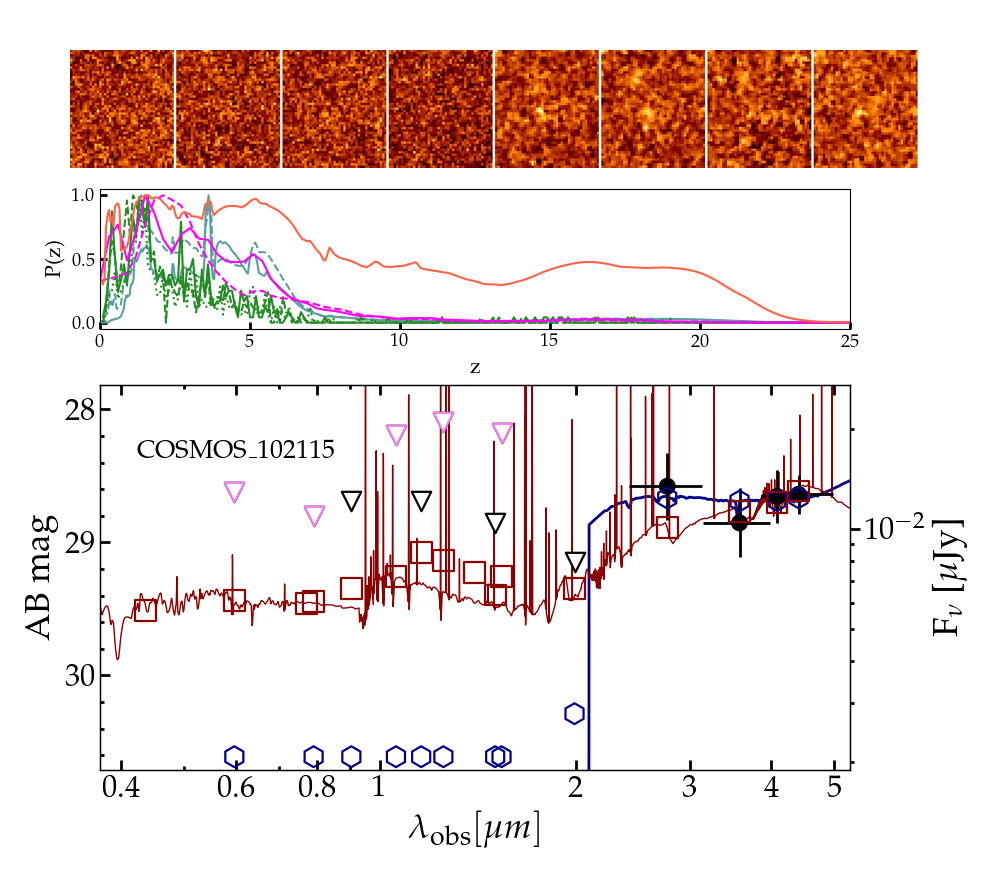}
\includegraphics[trim={0.5cm 0.5cm 0.5cm 0.1cm},clip,width=0.3\linewidth,keepaspectratio]{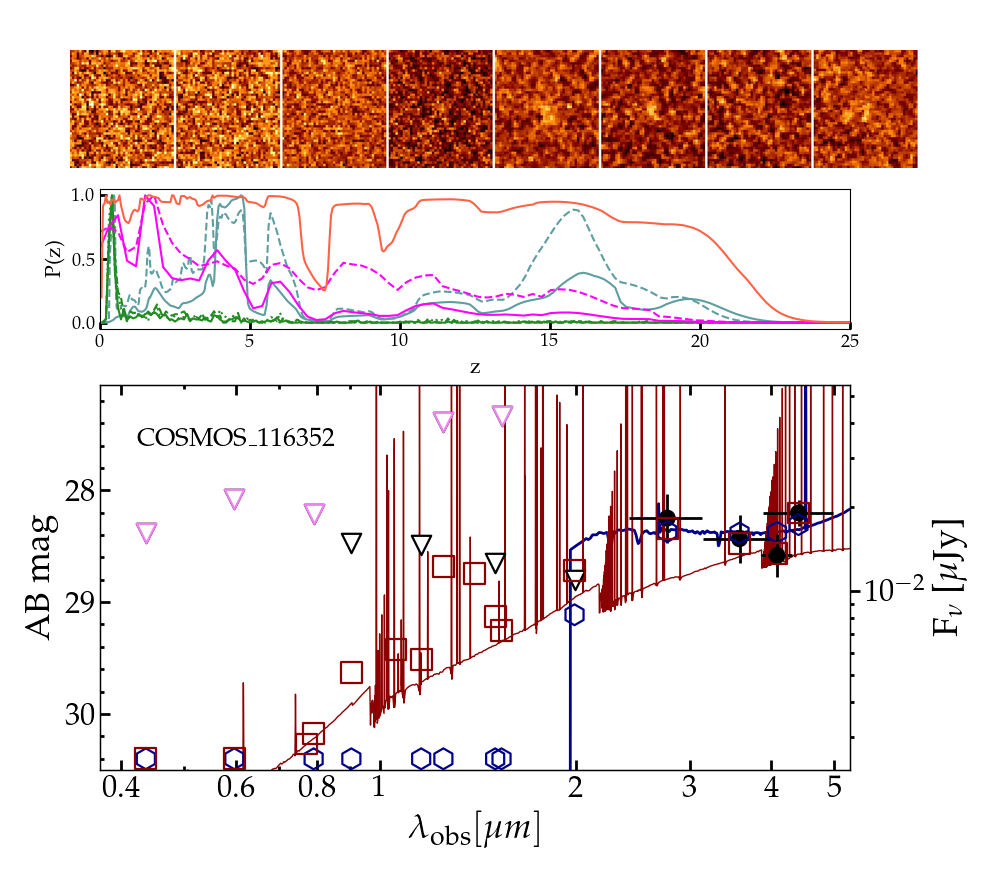}
\includegraphics[trim={0.5cm 0.5cm 0.5cm 0.1cm},clip,width=0.3\linewidth,keepaspectratio]{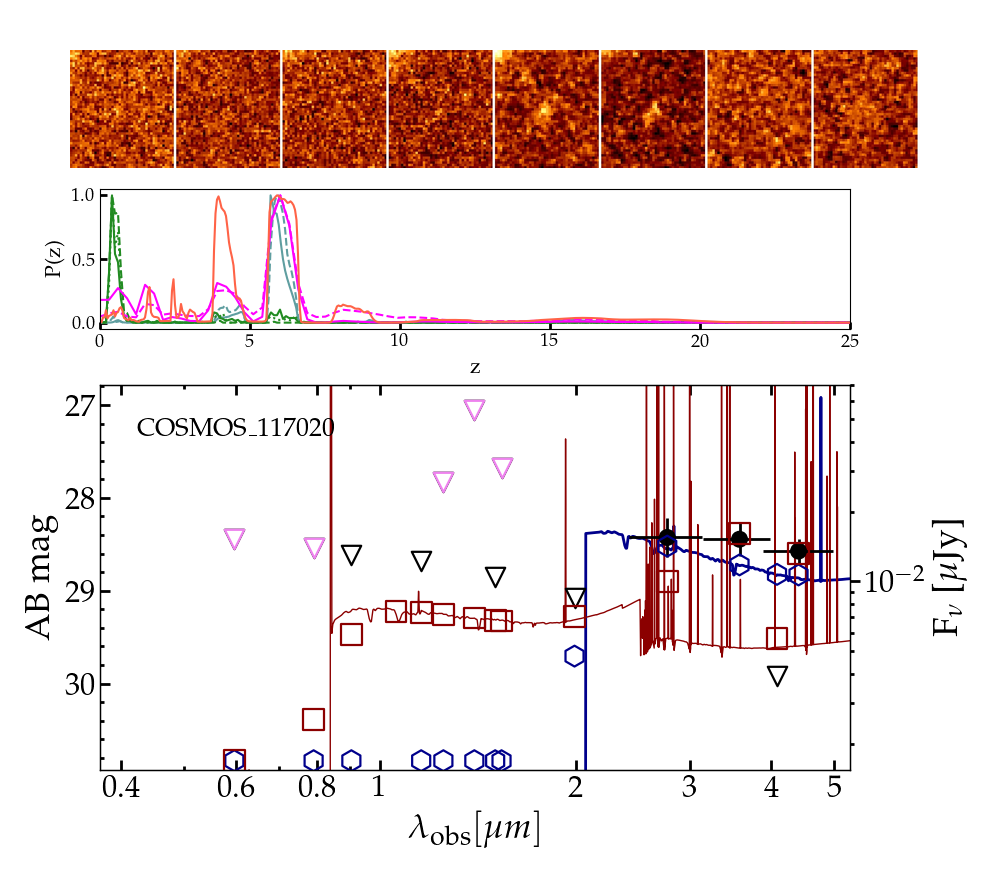}
\includegraphics[trim={0.5cm 0.5cm 0.5cm 0.1cm},clip,width=0.3\linewidth,keepaspectratio]{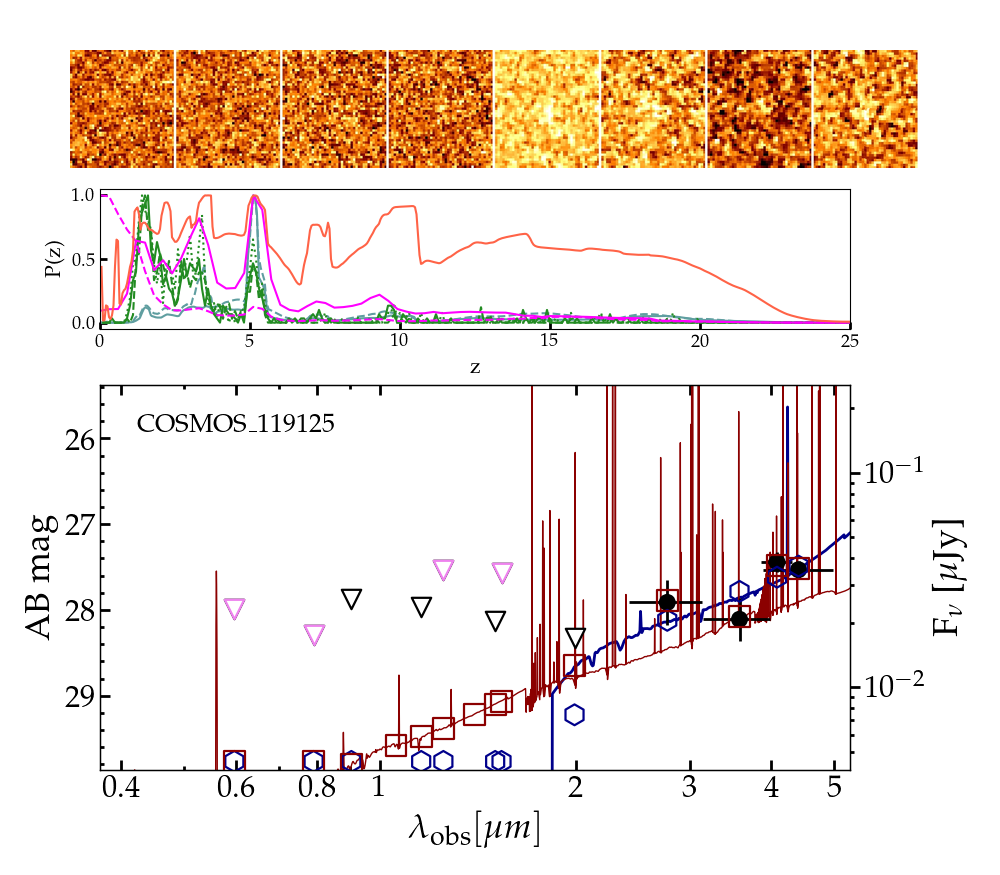}
\includegraphics[trim={0.5cm 0.5cm 0.5cm 0.1cm},clip,width=0.3\linewidth,keepaspectratio]{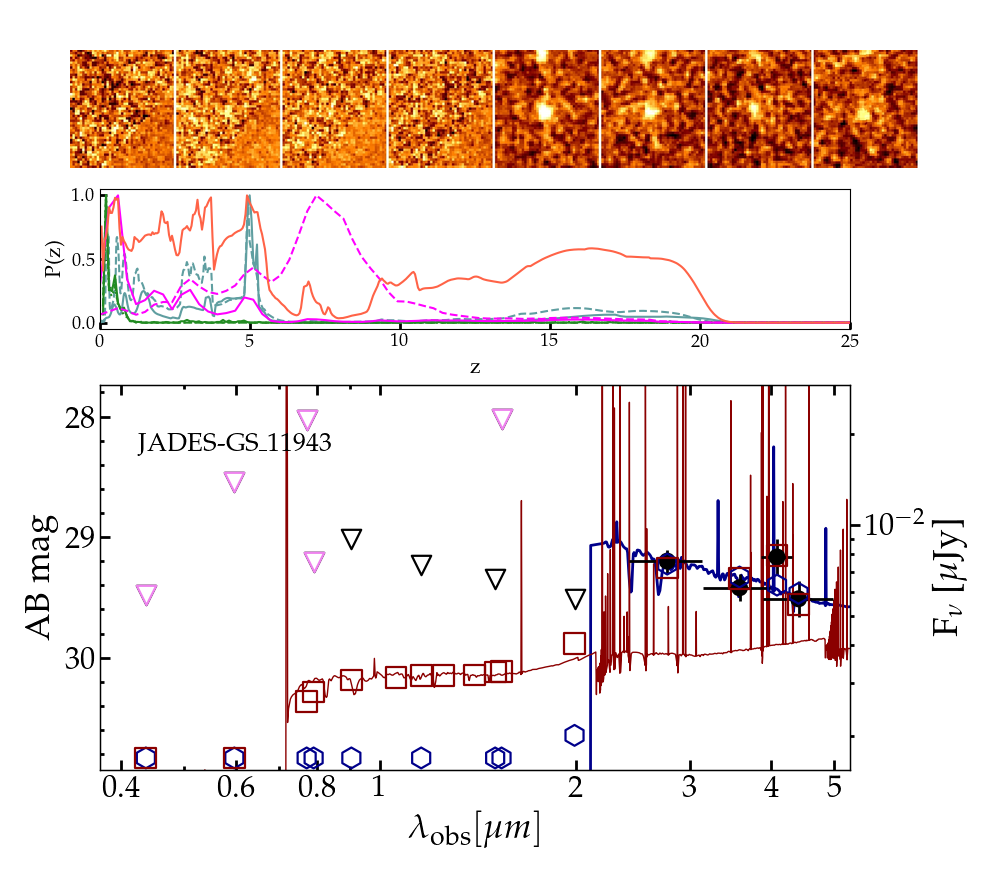}
\includegraphics[trim={0.5cm 0.5cm 0.5cm 0.1cm},clip,width=0.3\linewidth,keepaspectratio]{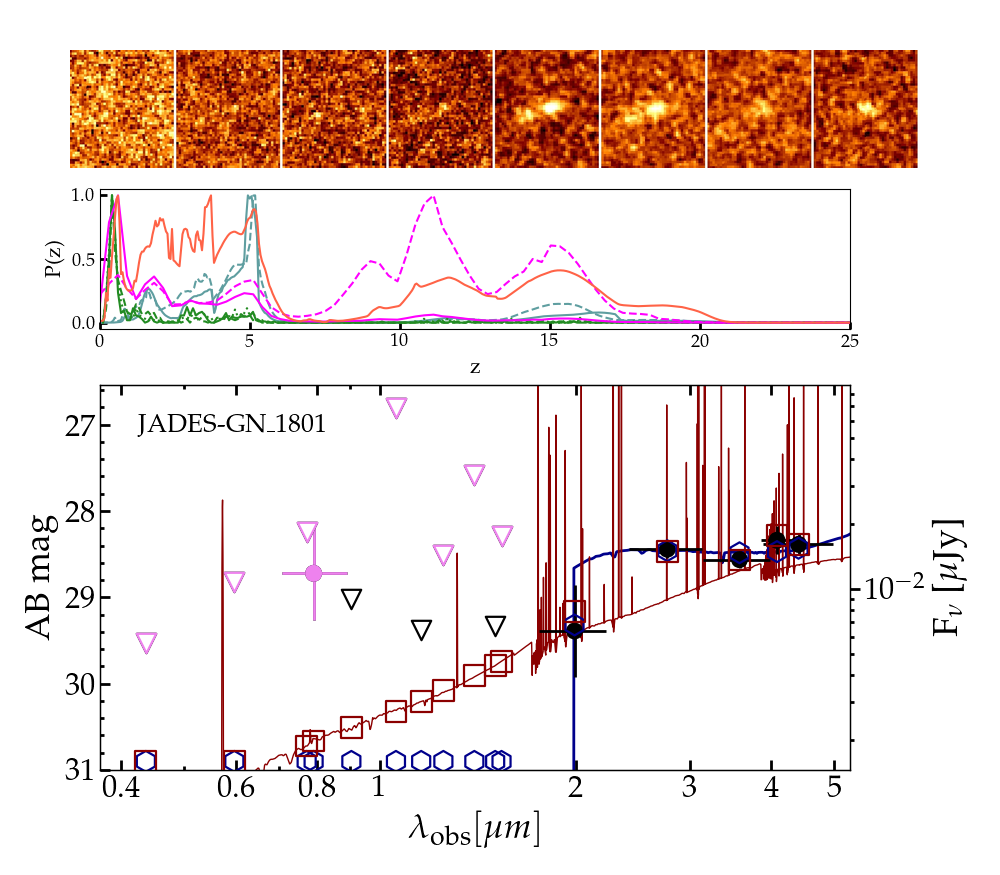}
\includegraphics[trim={0.5cm 0.5cm 0.5cm 0.1cm},clip,width=0.3\linewidth,keepaspectratio]{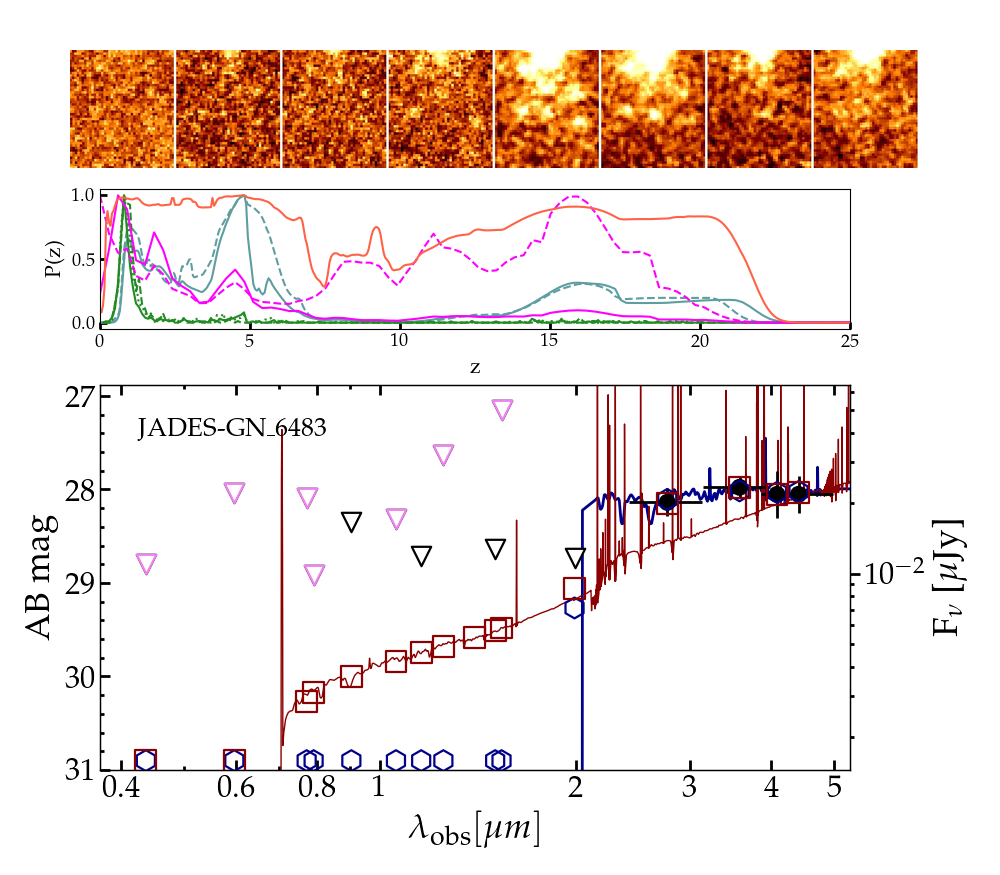}
\includegraphics[trim={0.5cm 0.5cm 0.5cm 0.1cm},clip,width=0.3\linewidth,keepaspectratio]{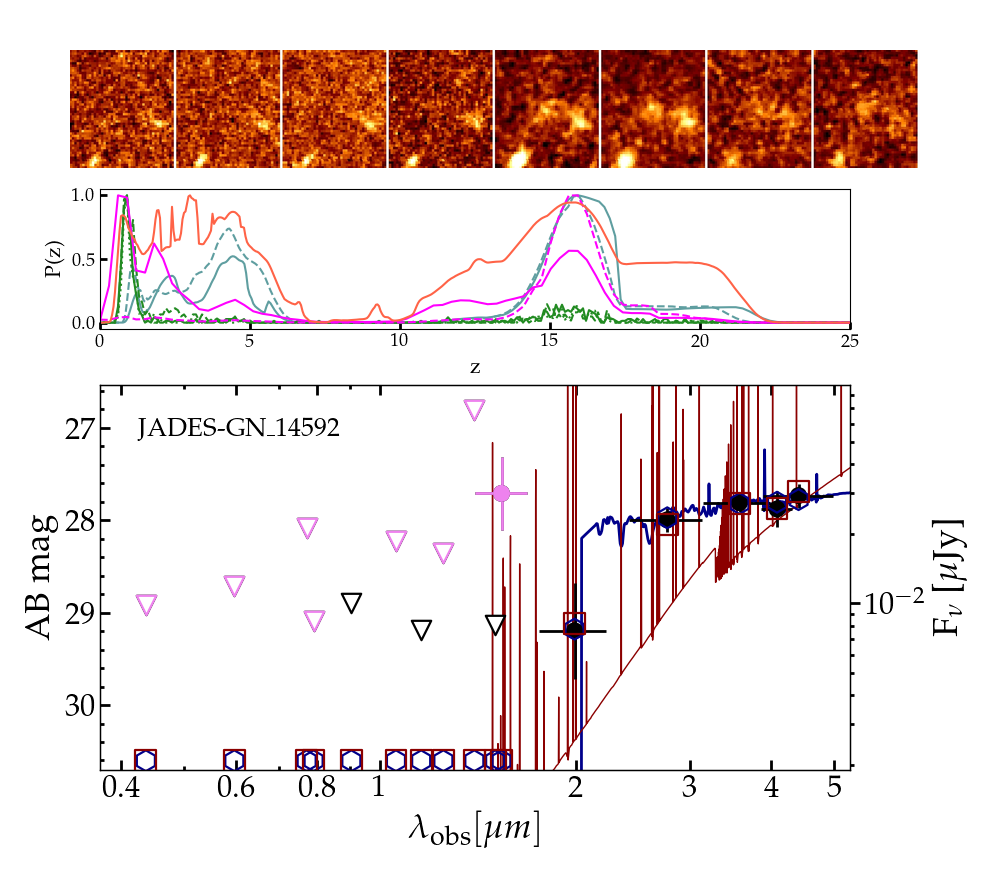}
\includegraphics[trim={0.5cm 0.5cm 0.5cm 0.1cm},clip,width=0.3\linewidth,keepaspectratio]{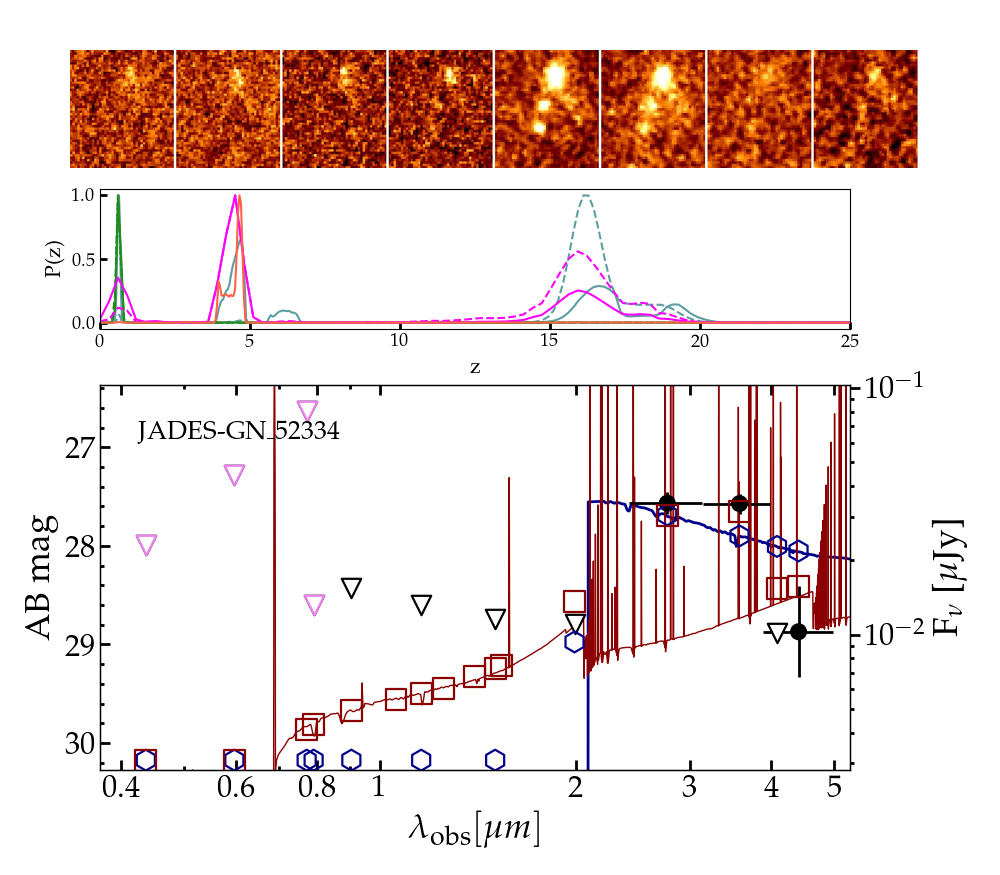}
\includegraphics[trim={0.5cm 0.5cm 0.5cm 0.1cm},clip,width=0.3\linewidth,keepaspectratio]{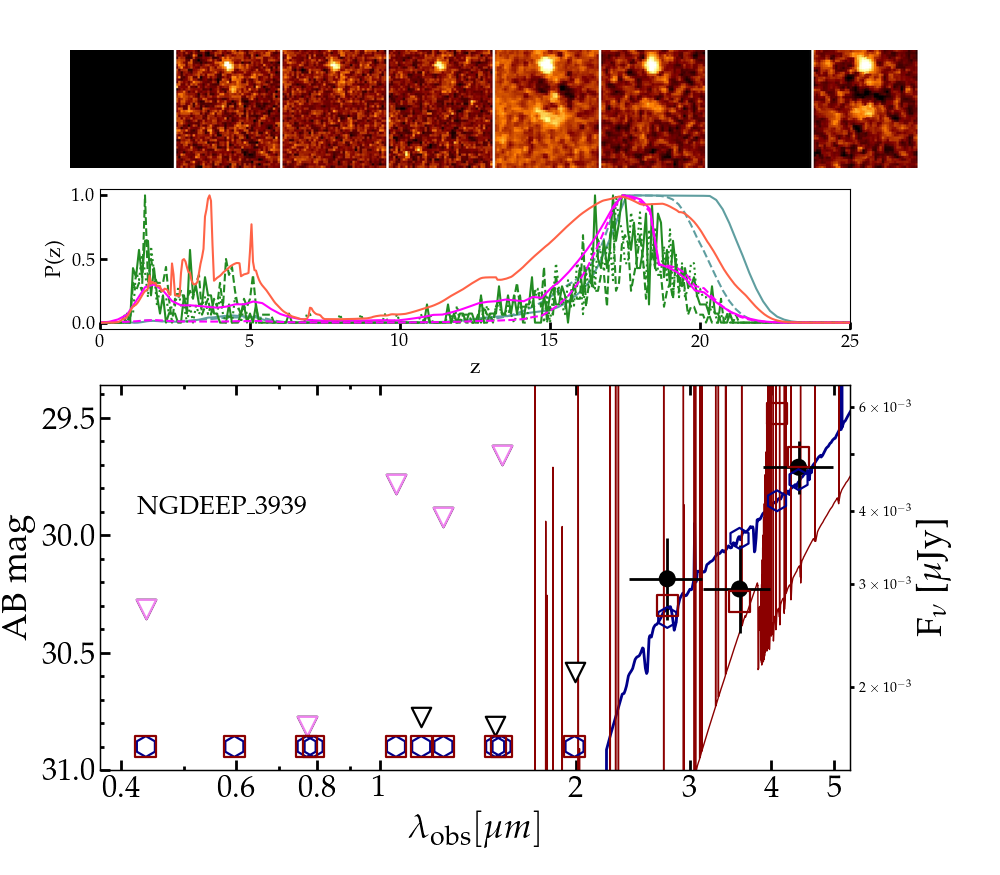}
\includegraphics[trim={0.5cm 0.5cm 0.5cm 0.1cm},clip,width=0.3\linewidth,keepaspectratio]{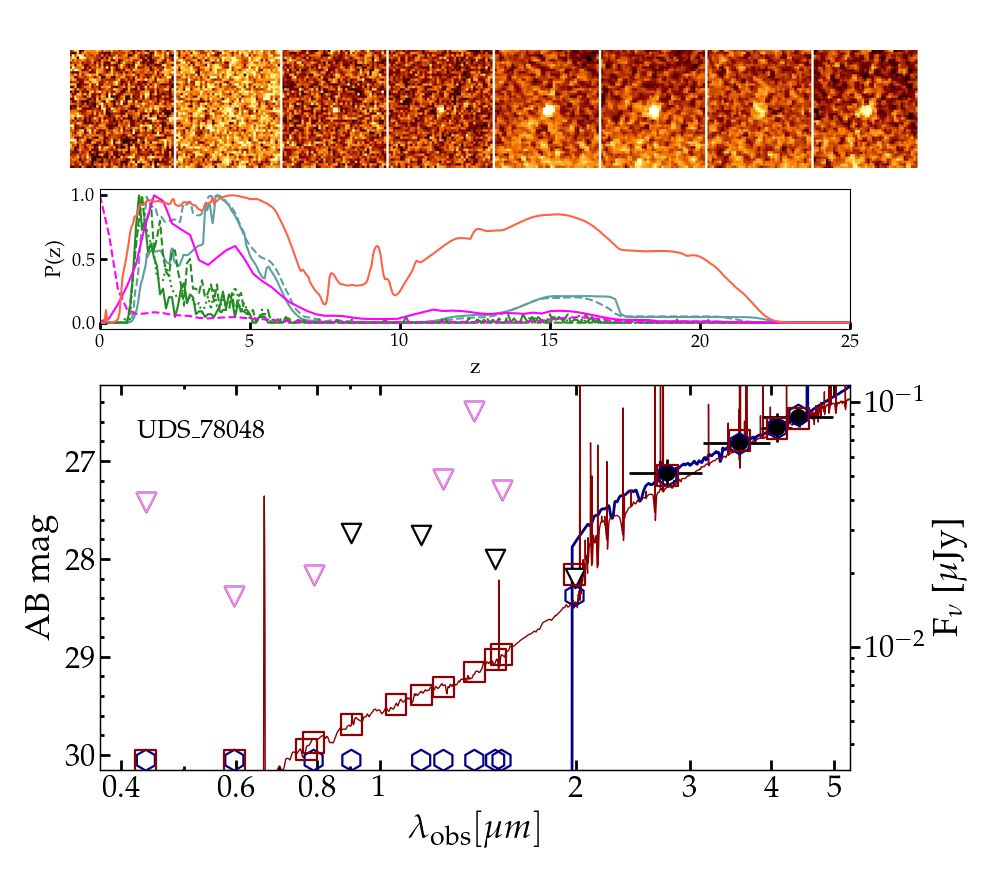}
\includegraphics[trim={0.5cm 0.5cm 0.5cm 0.1cm},clip,width=0.3\linewidth,keepaspectratio]{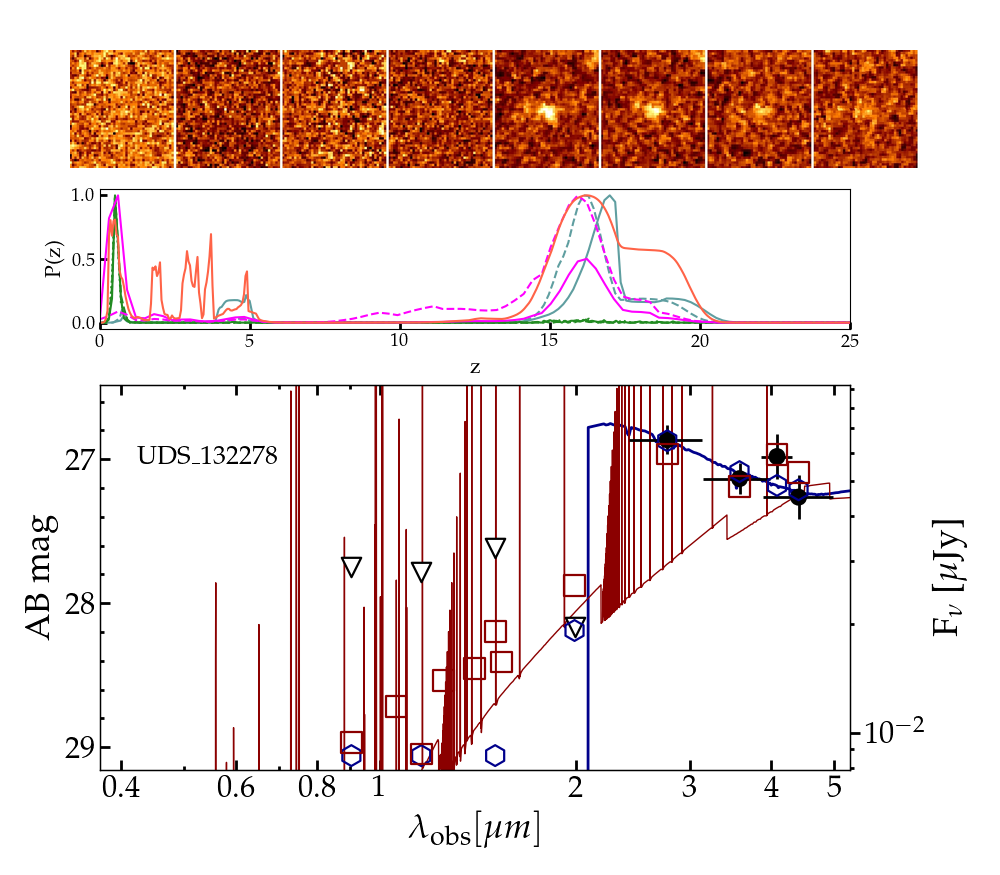}
\caption{Same as Fig.~\ref{fig_SEDs_ALL} for the extended sample of F200W-dropouts (part 2).} 
\label{fig_SEDs_InclF200W_2}
\end{figure*}

\begin{figure*}
\centering
\includegraphics[trim={0.5cm 0.5cm 0.5cm 0.1cm},clip,width=0.3\linewidth,keepaspectratio]{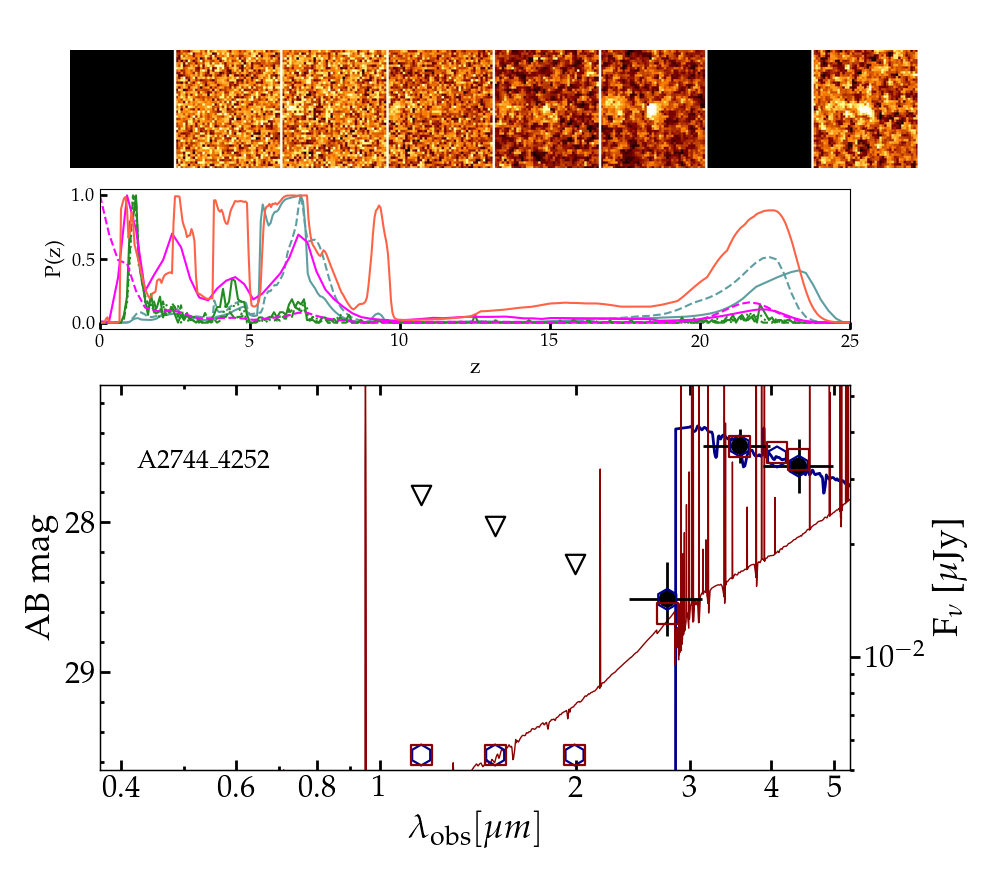}
\includegraphics[trim={0.5cm 0.5cm 0.5cm 0.1cm},clip,width=0.3\linewidth,keepaspectratio]{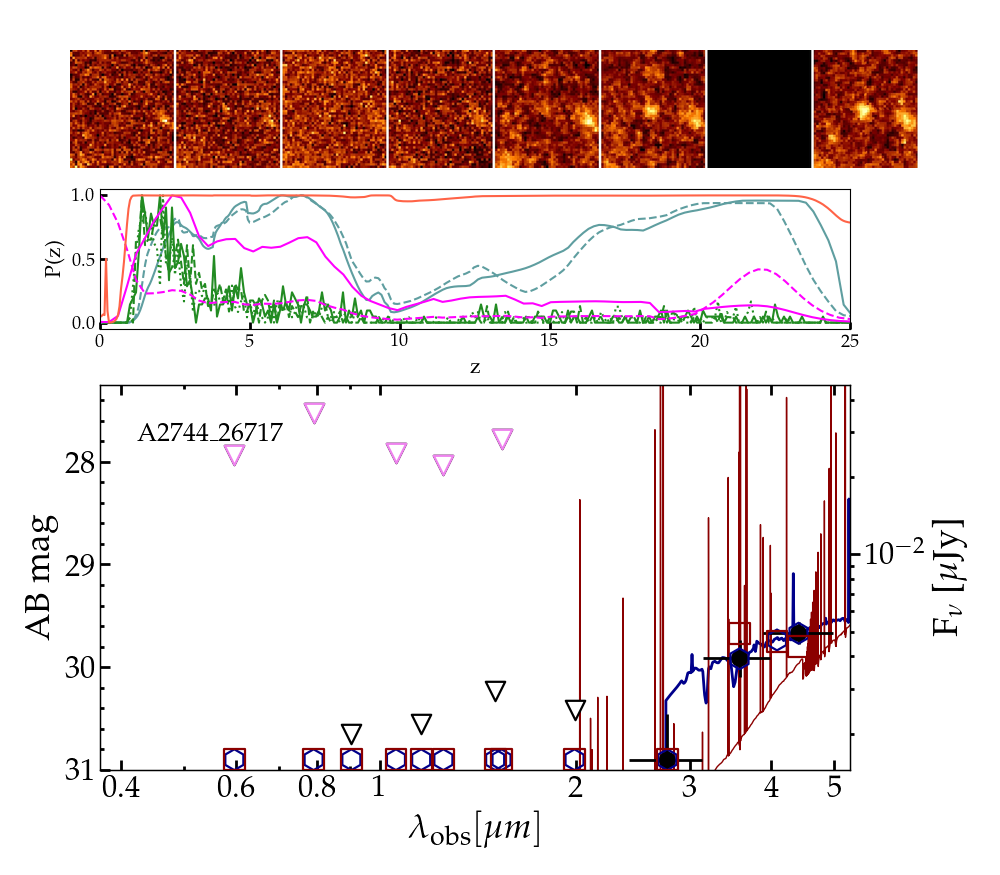}
\includegraphics[trim={0.5cm 0.5cm 0.5cm 0.1cm},clip,width=0.3\linewidth,keepaspectratio]{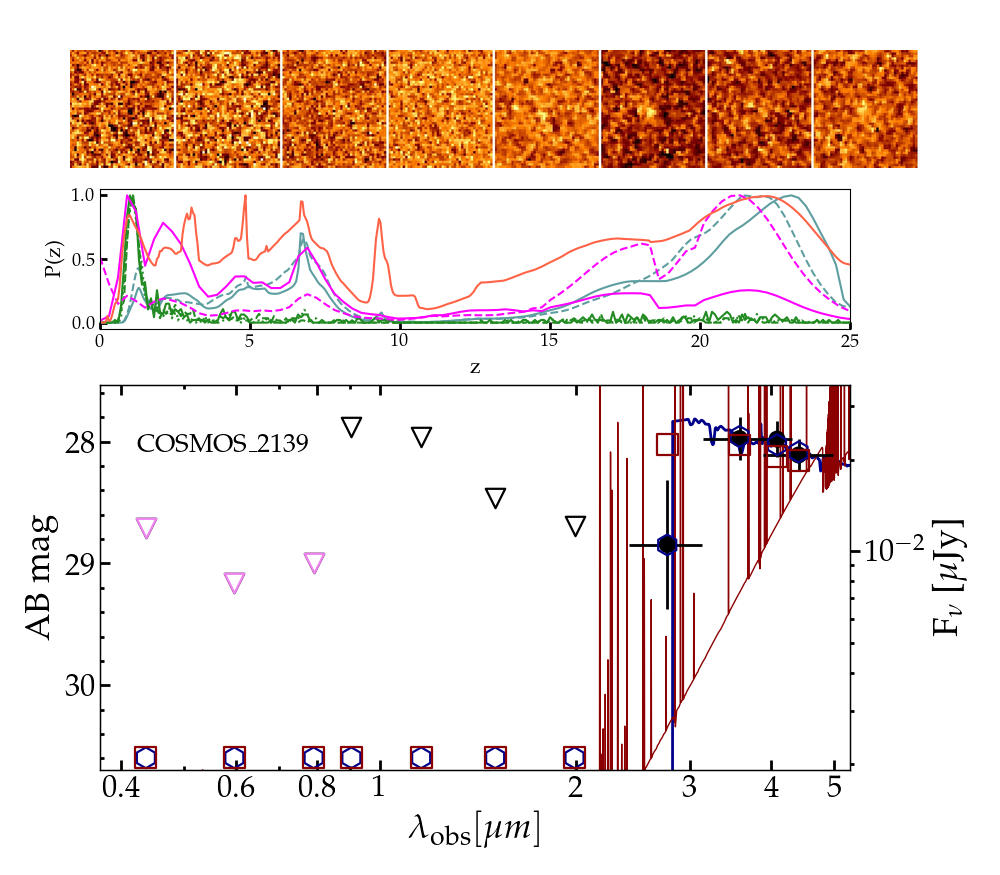}
\includegraphics[trim={0.5cm 0.5cm 0.5cm 0.1cm},clip,width=0.3\linewidth,keepaspectratio]{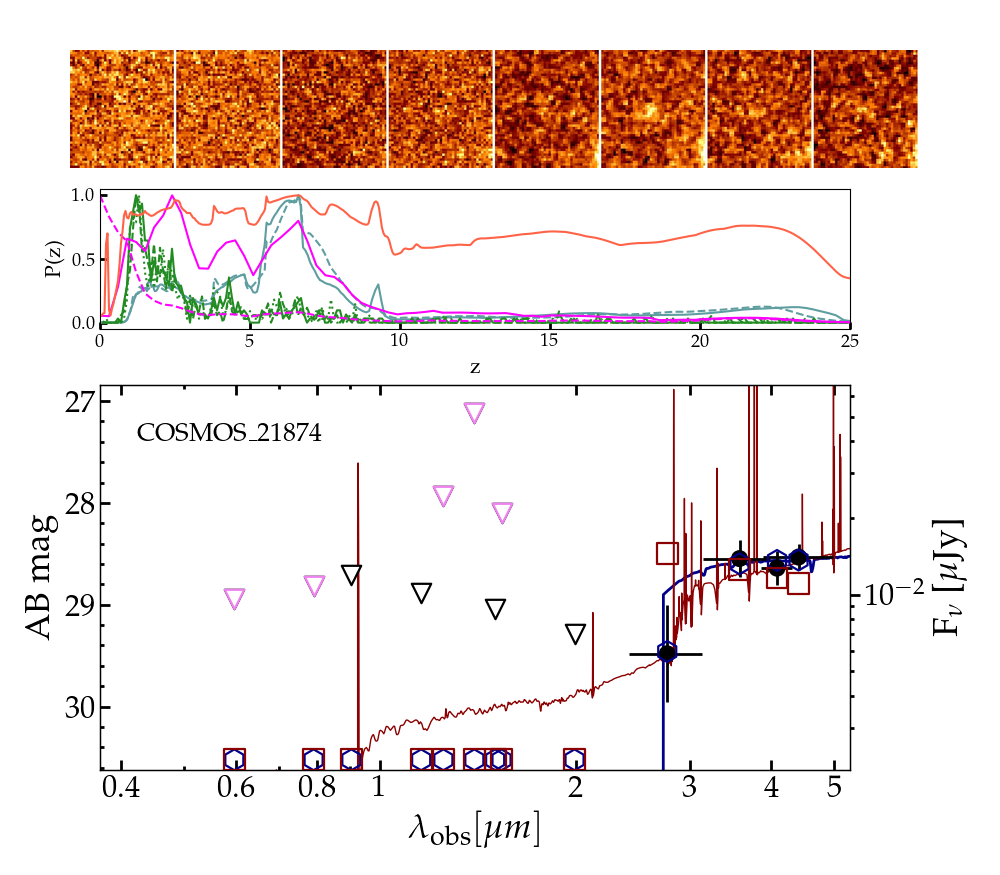}
\includegraphics[trim={0.5cm 0.5cm 0.5cm 0.1cm},clip,width=0.3\linewidth,keepaspectratio]{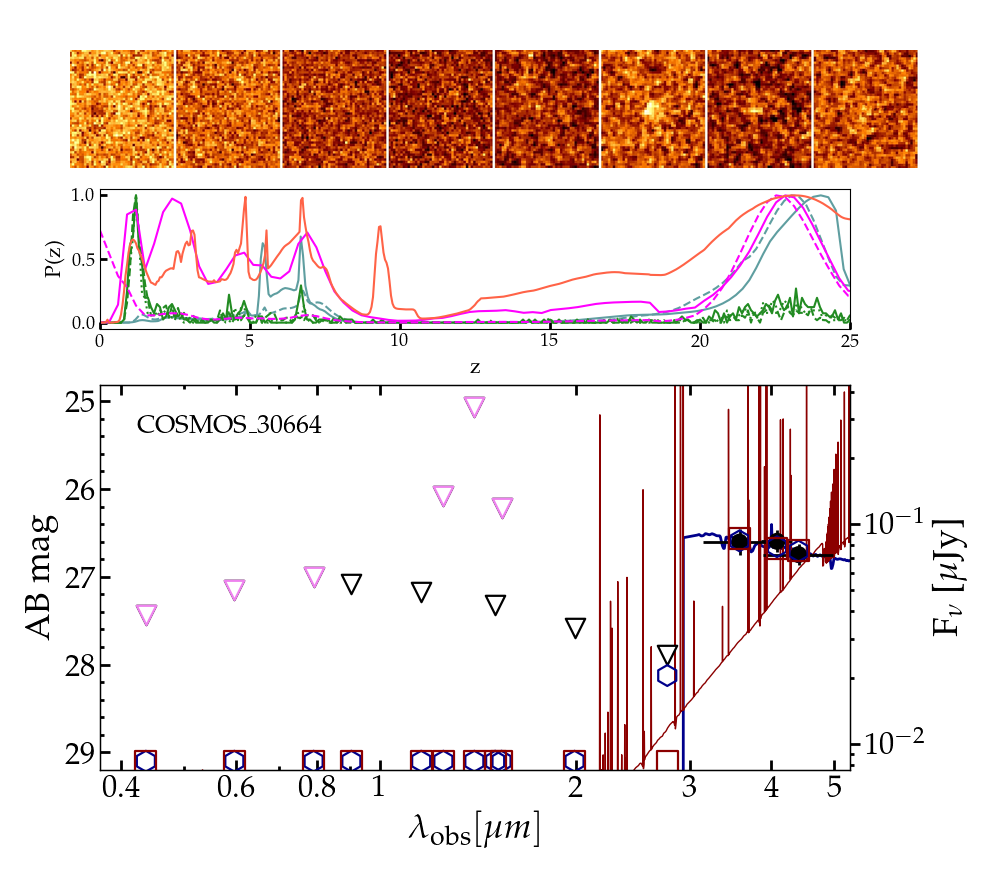}
\includegraphics[trim={0.5cm 0.5cm 0.5cm 0.1cm},clip,width=0.3\linewidth,keepaspectratio]{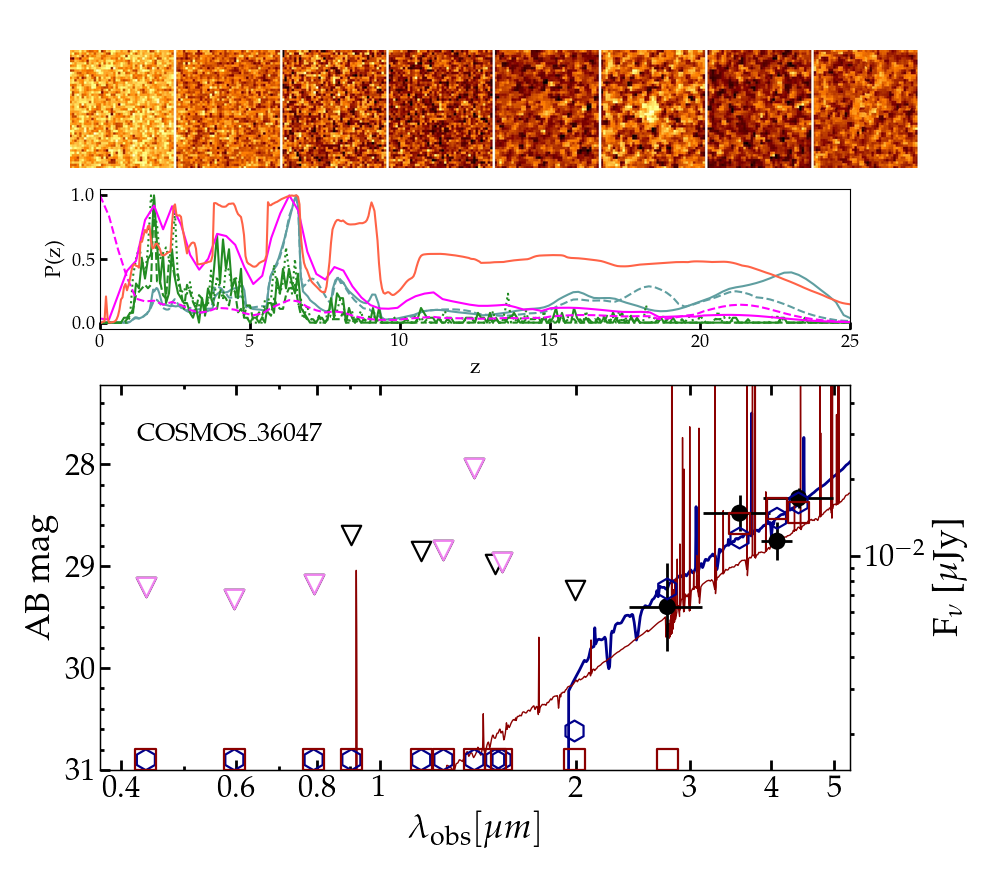}
\includegraphics[trim={0.5cm 0.5cm 0.5cm 0.1cm},clip,width=0.3\linewidth,keepaspectratio]{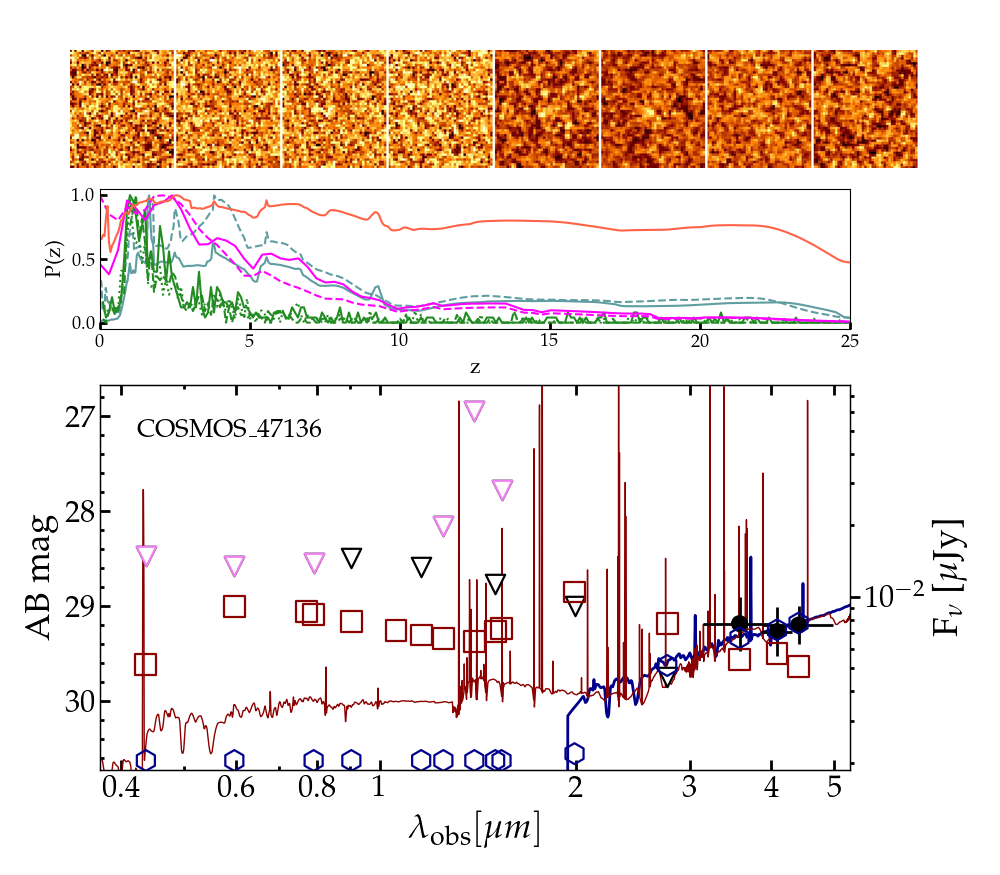}
\includegraphics[trim={0.5cm 0.5cm 0.5cm 0.1cm},clip,width=0.3\linewidth,keepaspectratio]{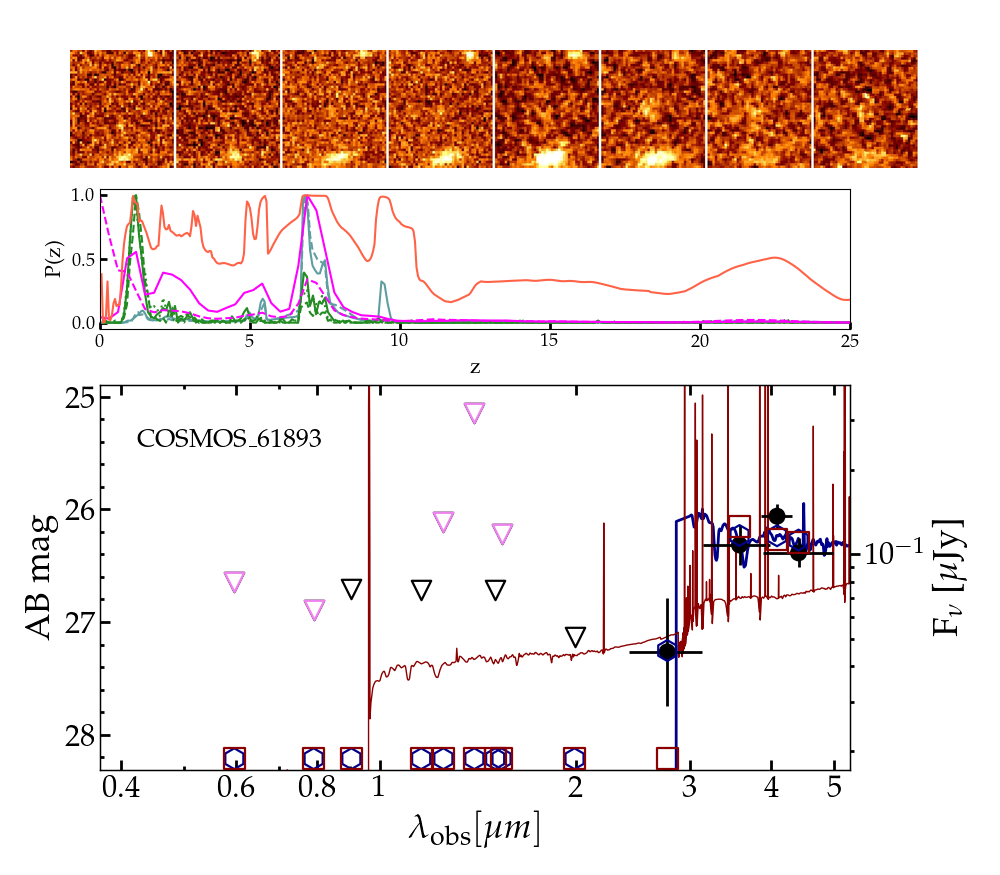}
\includegraphics[trim={0.5cm 0.5cm 0.5cm 0.1cm},clip,width=0.3\linewidth,keepaspectratio]{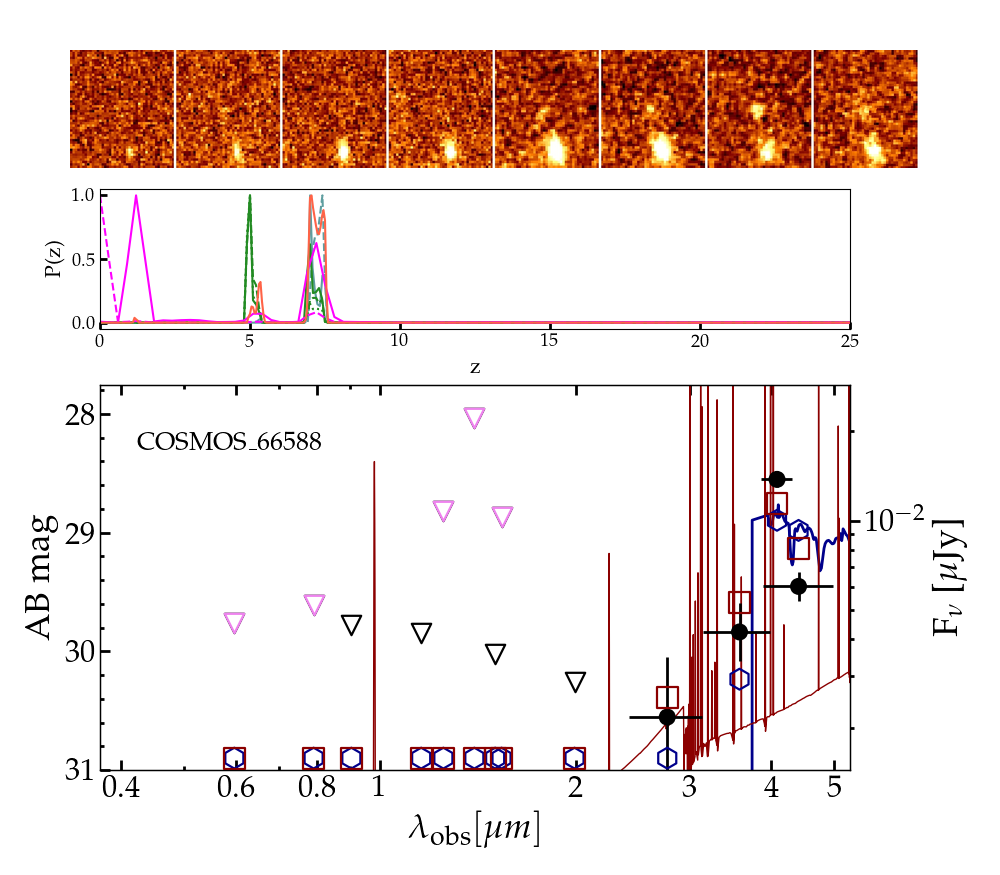}
\includegraphics[trim={0.5cm 0.5cm 0.5cm 0.1cm},clip,width=0.3\linewidth,keepaspectratio]{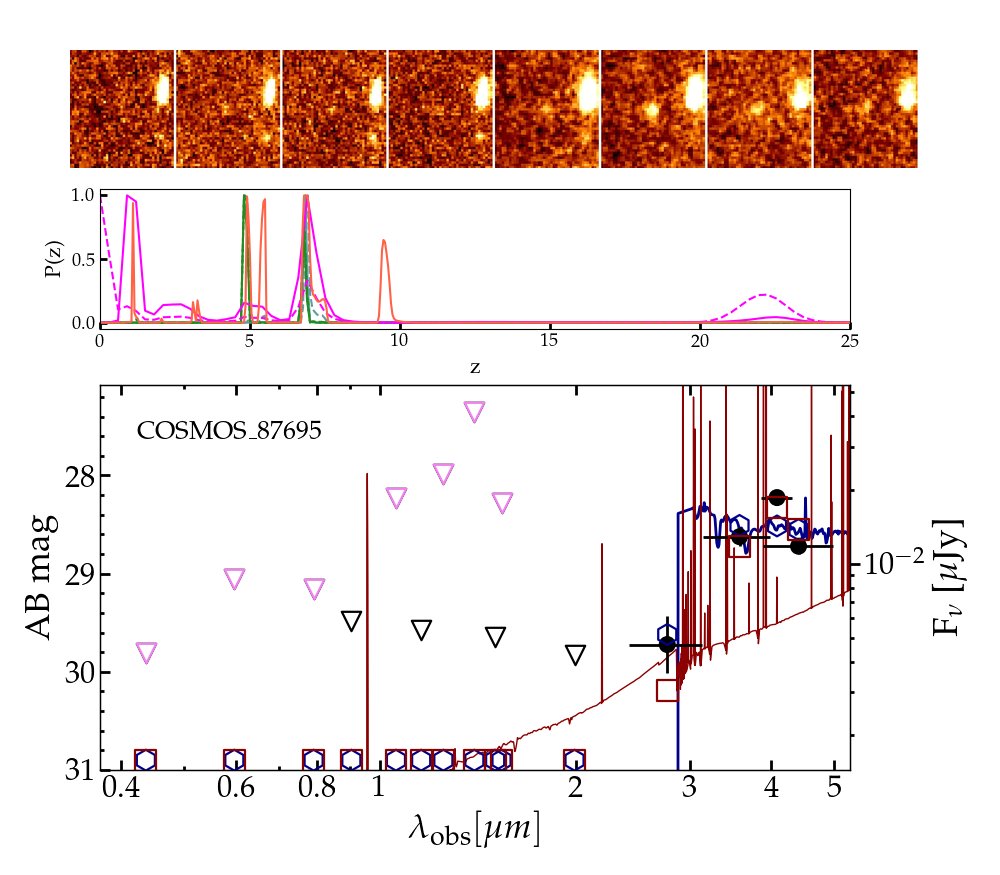}
\includegraphics[trim={0.5cm 0.5cm 0.5cm 0.1cm},clip,width=0.3\linewidth,keepaspectratio]{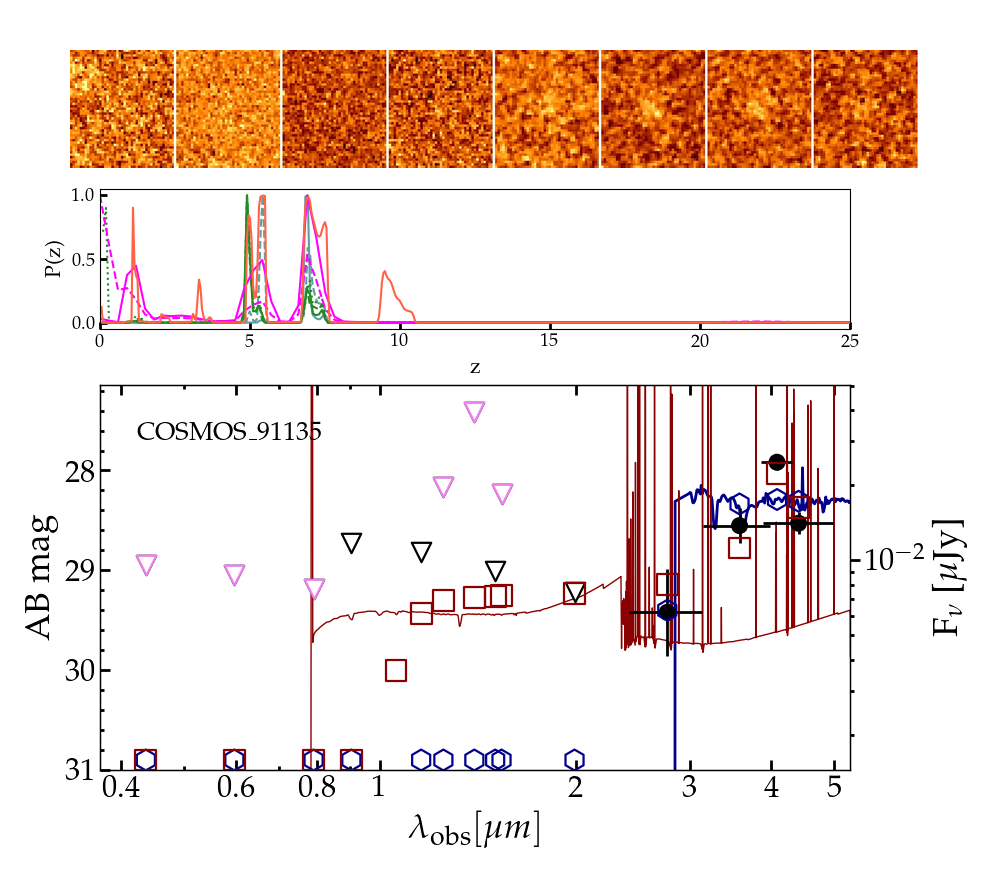}
\includegraphics[trim={0.5cm 0.5cm 0.5cm 0.1cm},clip,width=0.3\linewidth,keepaspectratio]{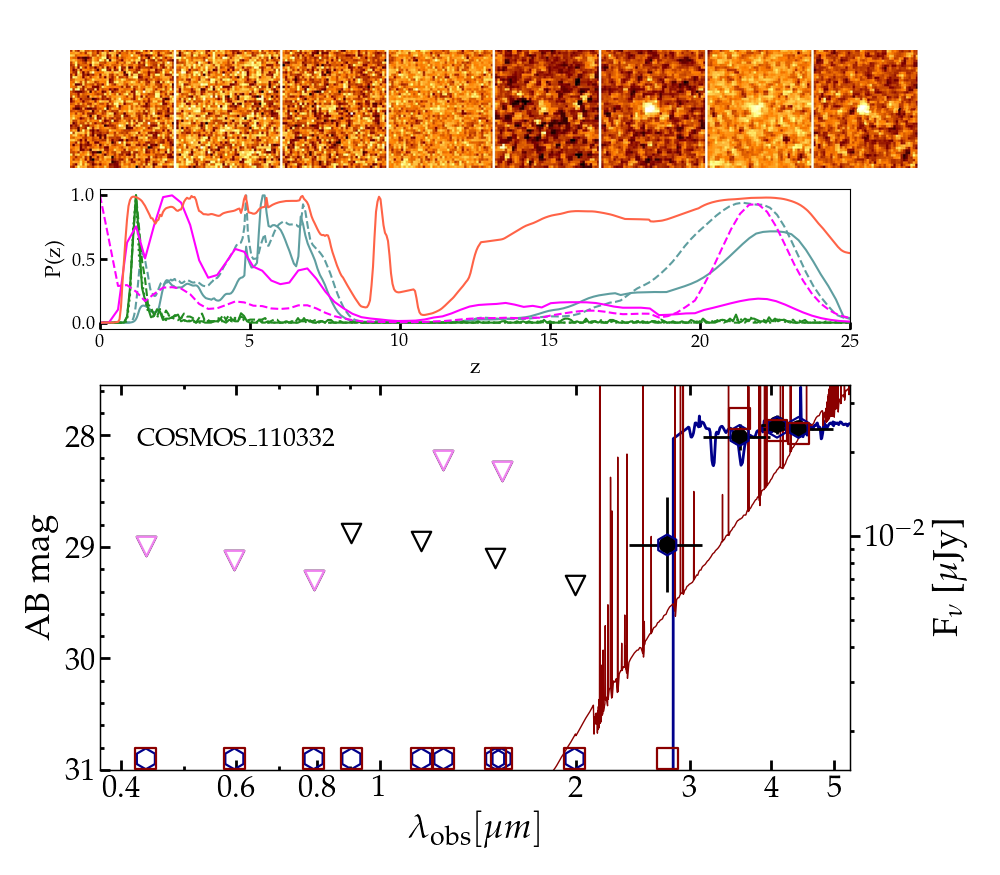}
\caption{Same as Fig.~\ref{fig_SEDs_ALL} for the extended sample of F277W-dropouts (part 1).} 
\label{fig_SEDs_InclF277W_1}
\end{figure*}

\begin{figure*}
\centering
\includegraphics[trim={0.5cm 0.5cm 0.5cm 0.1cm},clip,width=0.3\linewidth,keepaspectratio]{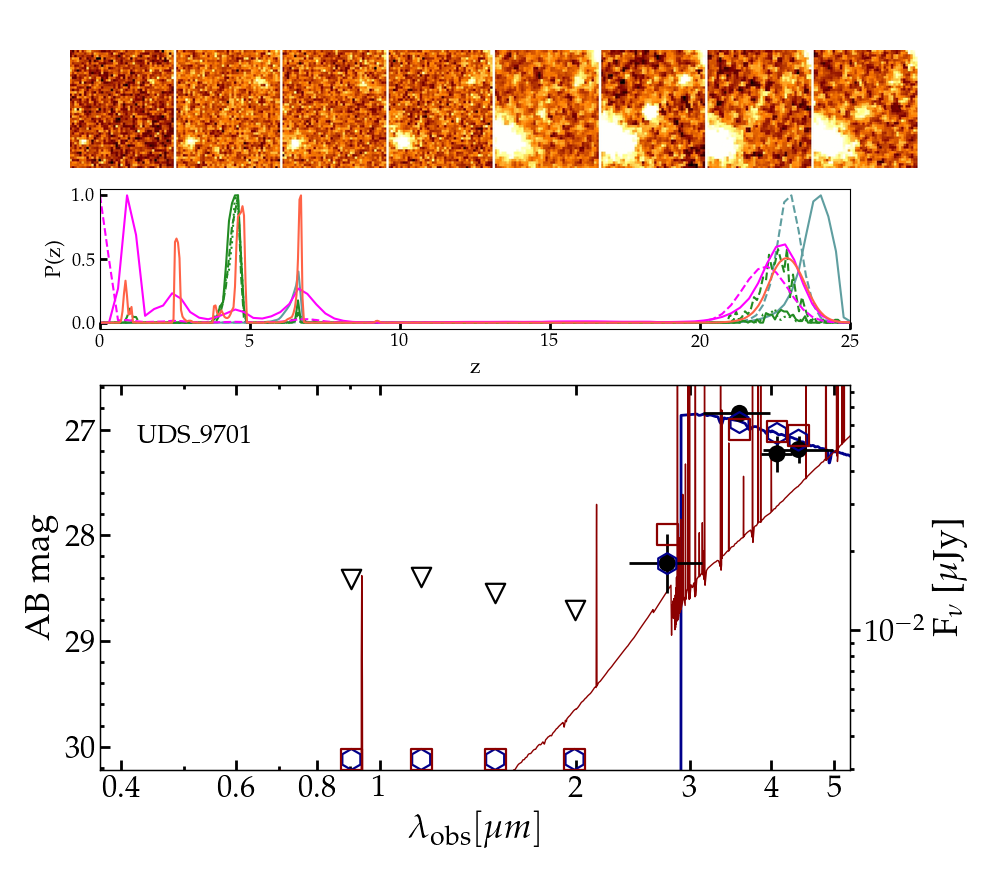}
\includegraphics[trim={0.5cm 0.5cm 0.5cm 0.1cm},clip,width=0.3\linewidth,keepaspectratio]{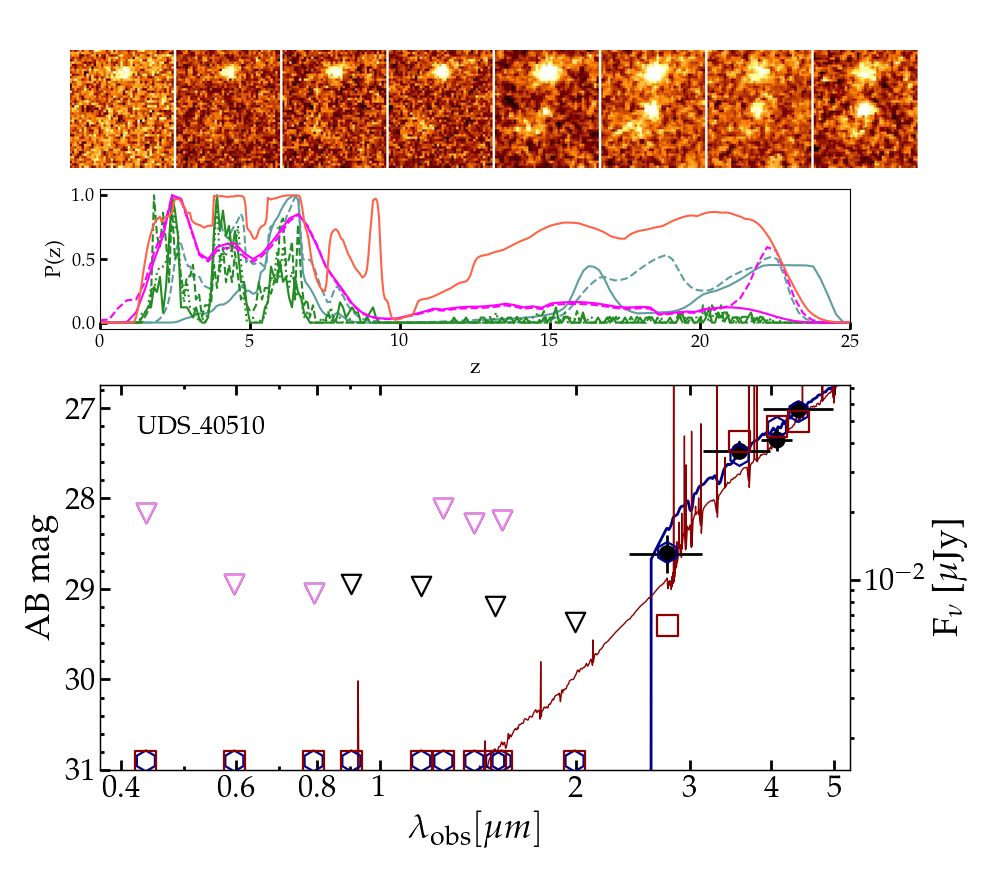}
\includegraphics[trim={0.5cm 0.5cm 0.5cm 0.1cm},clip,width=0.3\linewidth,keepaspectratio]{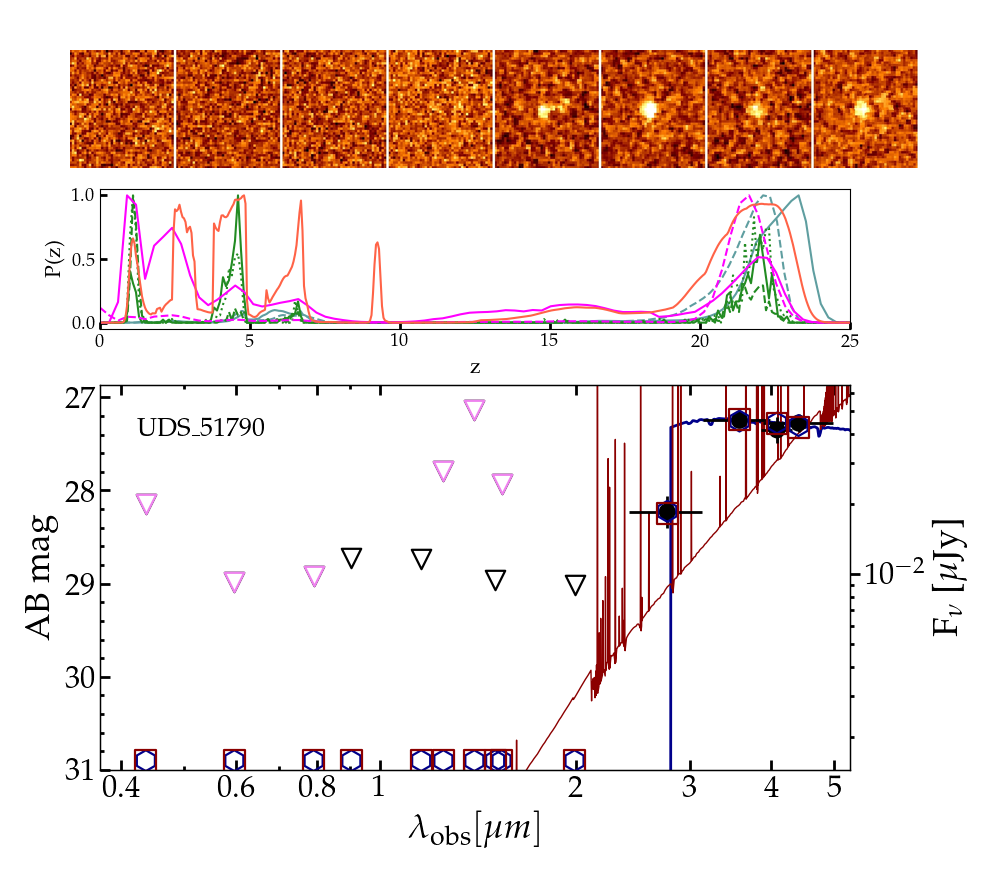}
\includegraphics[trim={0.5cm 0.5cm 0.5cm 0.1cm},clip,width=0.3\linewidth,keepaspectratio]{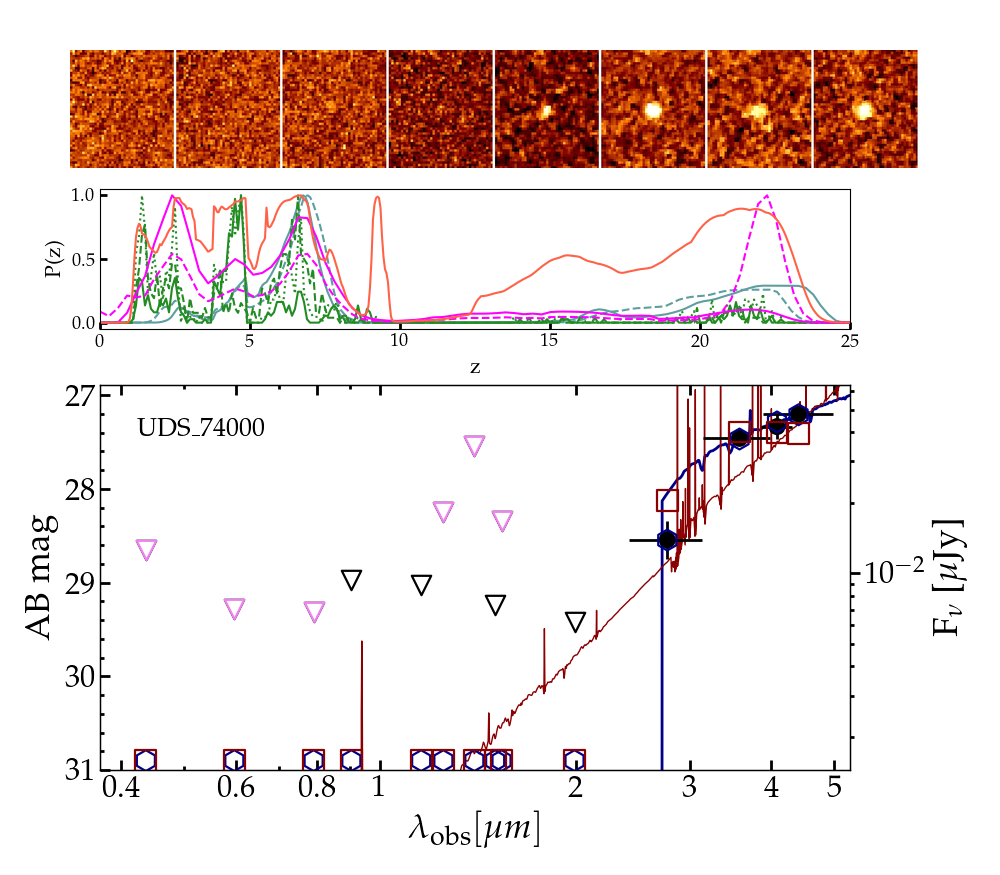}
\includegraphics[trim={0.5cm 0.5cm 0.5cm 0.1cm},clip,width=0.3\linewidth,keepaspectratio]{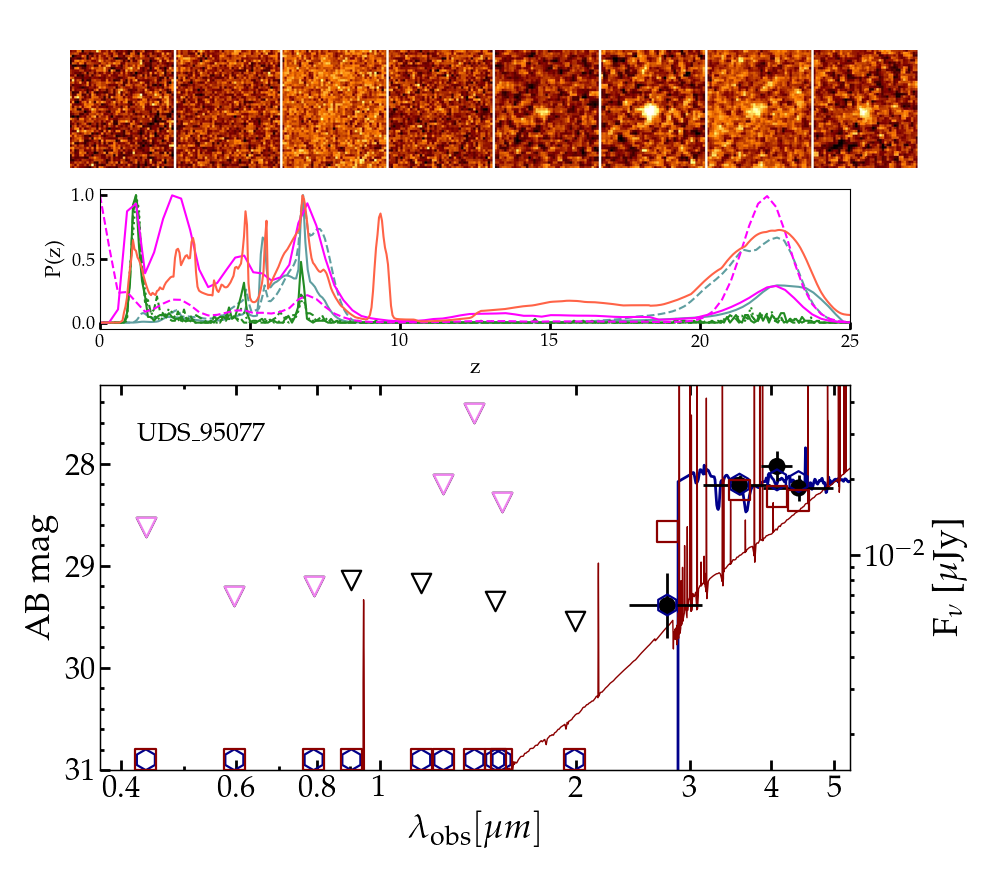}
\includegraphics[trim={0.5cm 0.5cm 0.5cm 0.1cm},clip,width=0.3\linewidth,keepaspectratio]{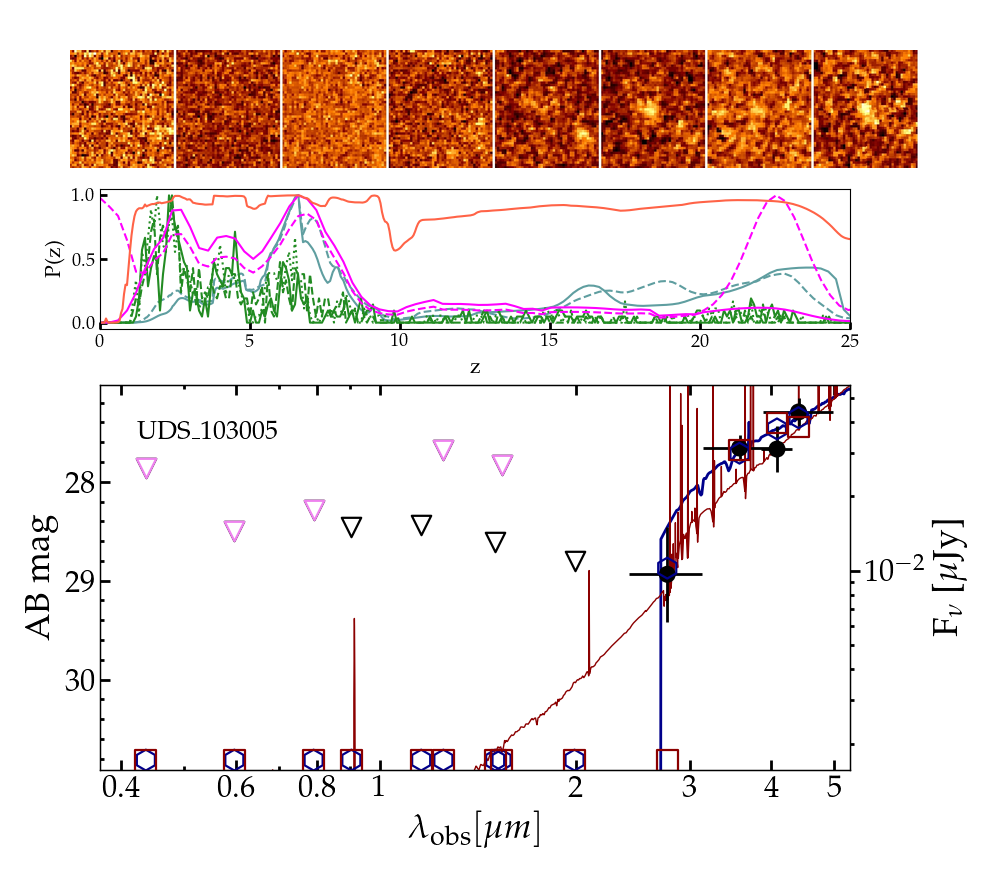}
\includegraphics[trim={0.5cm 0.5cm 0.5cm 0.1cm},clip,width=0.3\linewidth,keepaspectratio]{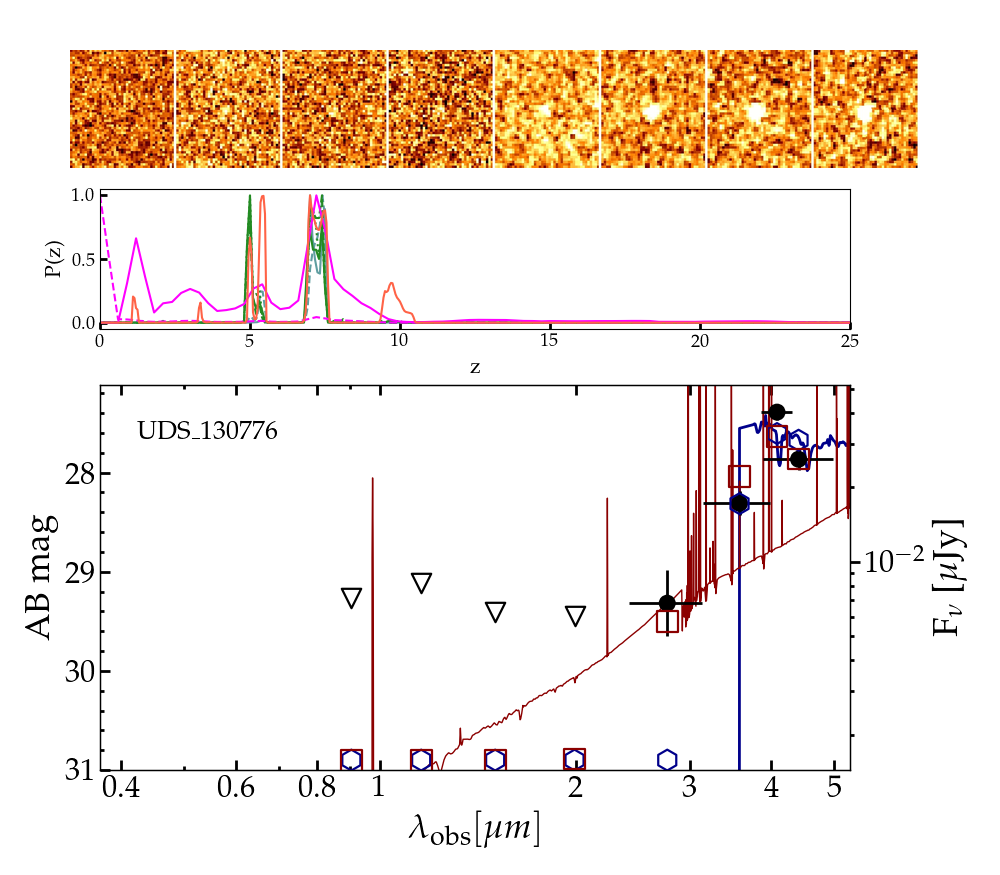}
\caption{Same as Fig.~\ref{fig_SEDs_ALL} for the extended sample of F277W-dropouts (part 2).} 
\label{fig_SEDs_InclF277W_2}
\end{figure*}

We present here the extended samples of z$\sim15-20$ (ID, coordinates and F356W magnitudes from M24 in Table~\ref{tab:ext_candidatesF200W}, SED, thumbnails and $P(z)$ in Fig.~\ref{fig_SEDs_InclF200W_1} and ~\ref{fig_SEDs_InclF200W_2}) and z$\sim20-28$ (Table~\ref{tab:ext_candidatesF277W}, Fig.~\ref{fig_SEDs_InclF277W_1} and ~\ref{fig_SEDs_InclF277W_2}) candidates. We remark that while these objects also have colours and $P(z)$ consistent with high-redshift solutions, and shall be considered of interest for potential spectroscopic follow-up, the contamination level in these samples is expected to be significant (Sect.~\ref{sec:ColorSelec}). 

\begin{table}[ht]
\caption{Extended sample of F200W dropout candidates in the ASTRODEEP-JWST fields}\label{tab:ext_candidatesF200W}
\centering
\begin{tabular}{lccc}
ID &          R.A. &        Dec &  F356W   \\

 & deg. & deg. & AB   \\
\hline
A2744\_36169 	&3.506835 	&-30.303378 	&27.81 $\pm$	0.12 	\\
COSMOS\_378 	&150.139836 	&2.162604 	&27.59 	$\pm$	0.15 	 \\
COSMOS\_10730 	&150.111759 	&2.208056 	&27.66 	$\pm$	0.22 	\\
COSMOS\_28354 	&150.076753 	&2.254899 	&27.14 	$\pm$	0.13 	 \\
COSMOS\_34866 &	150.153167 	&2.269104 	&28.66 	$\pm$	0.14	 \\
COSMOS\_45778 	&150.109207 	&2.292050 	&28.45 $\pm$		0.17 	\\
COSMOS\_47748 	&150.155793 	&2.295937 	&27.93 	$\pm$	0.21 	 \\
COSMOS\_51984 	&150.124225 	&2.304599 	&28.90 	$\pm$	0.16 	 \\
COSMOS\_66011 	&150.168328 	&2.332822 	&29.16 	$\pm$	0.20	 \\
COSMOS\_78212 	&150.155200 	&2.356336 	&27.83	$\pm$	0.15 	\\
COSMOS\_82058 	&150.088237 	&2.364684 	&28.77 	$\pm$	0.23 	 \\
COSMOS\_90343 	&150.166521 	&2.382355 	&27.66 $\pm$		0.13 	 \\
COSMOS\_96354 	&150.149768 	&2.397964 	&29.04$\pm$		0.22 	 \\
COSMOS\_101131 	&150.143521 	&2.410134 	&27.89 $\pm$		0.14 	 \\
COSMOS\_102115 	&150.192712 	&2.412584 	&28.90 $\pm$		0.26 	 \\
COSMOS\_116352 	&150.170290 	&2.456889 	&28.54 $\pm$		0.24 	\\
COSMOS\_117020 	&150.145445 	&2.459396 	&28.56 $\pm$		0.19 	 \\
COSMOS\_119125 	&150.184999 	&2.466924 	&28.04 $\pm$		0.22	 \\
UDS\_78048 	&34.397548 	&-5.178046& 	26.83 	$\pm$	0.10 	 \\
UDS\_132278 	&34.354643 	&-5.110434 	&27.15 	$\pm$	0.11 	 \\
NGDEEP\_3939 	&53.270238 	&-27.861668 	&30.15 $\pm$		0.18 \\
JADES-GS\_11943 	&53.030331 	&-27.877975 	&29.42 $\pm$		0.11	 \\
JADES-GN\_1801 	&189.238294 	&62.148230 &	28.55 $\pm$		0.10 	\\
JADES-GN\_6483 	&189.324320 	&62.165237 	&28.02	$\pm$	0.12 	 \\
JADES-GN\_14592 	&189.224961 	&62.188365 &	27.79 	$\pm$	0.09 	 \\
JADES-GN\_52334 	&188.989191 	&62.290761 	&27.54 	$\pm$	0.10 	 \\
\hline
\end{tabular}
\end{table}

\begin{table}[ht]
\caption{Extended sample of F277W dropout candidates in the ASTRODEEP-JWST fields}\label{tab:ext_candidatesF277W}
\centering
\begin{tabular}{lccc}
ID &          R.A. &        Dec &  F356W   \\
 & deg. & deg. & AB  \\
\hline
A2744\_4252 	 &3.658016 	 &-30.426589 	 &27.49$\pm$	0.12 	\\
A2744\_26717 	 &3.500812 	 &-30.354774 	 &29.89$\pm$	 	0.20 	\\
COSMOS\_2139 	 &150.132819 	 &2.175271 	 &27.97$\pm$	 	0.19 	\\
COSMOS\_21874 	 &150.077042 	 &2.238654 	 &28.56$\pm$		0.19 	\\
COSMOS\_30664 	 &150.109780 	 &2.259779 	 &26.70$\pm$		0.16	\\
COSMOS\_36047 	 &150.178087 	 &2.271629  &28.38$\pm$	 	0.19 \\
COSMOS\_47136 	 &150.133938 	 &2.294643 	 &29.32$\pm$	 	0.29 	\\
COSMOS\_61893 	 &150.065480 	 &2.324540 	 &26.36$\pm$	 	0.18 	\\
COSMOS\_66588 	 &150.122411 	 &2.333940 	 &29.75$\pm$		0.27 	\\
COSMOS\_87695 	 &150.129665 	 &2.376409 	 &28.62$\pm$	 	0.10 	\\
COSMOS\_91135 	 &150.183896 	 &2.384041 	 &28.57$\pm$	 	0.16 	\\
COSMOS\_110332 & 	150.191351 	 &2.435514 	 &27.96$\pm$	0.13 	\\
UDS\_9701 	 &34.261719 	 &-5.302606 	 &26.85$\pm$	 	0.08 	\\
UDS\_40510 	 &34.388952 	 &-5.246670 	 &27.49$\pm$	 	0.09 	\\
UDS\_51790 	 &34.247686 	 &-5.225198 	 &27.25$\pm$	 	0.08 	\\
UDS\_74000 	 &34.268062 	 &-5.184911 	 &27.48$\pm$	 	0.08 	\\
UDS\_95077 	 &34.345233 	 &-5.150057 	 &28.24$\pm$	 	0.11 	\\
UDS\_103005 	 &34.402545 	 &-5.136976 	 &27.63$\pm$	 	0.14 	\\
UDS\_130776 	 &34.434420 	 &-5.108762 	 &28.35$\pm$	 	0.14 	\\
\hline
\end{tabular}
\end{table}

\section{Spectral energy distributions of the confirmed interlopers}\label{sec:appendix-interlopers}
\begin{figure*}[!ht]
\centering
\includegraphics[trim={0.5cm 0.5cm 0.5cm 0.1cm},clip,width=0.3\linewidth,keepaspectratio]{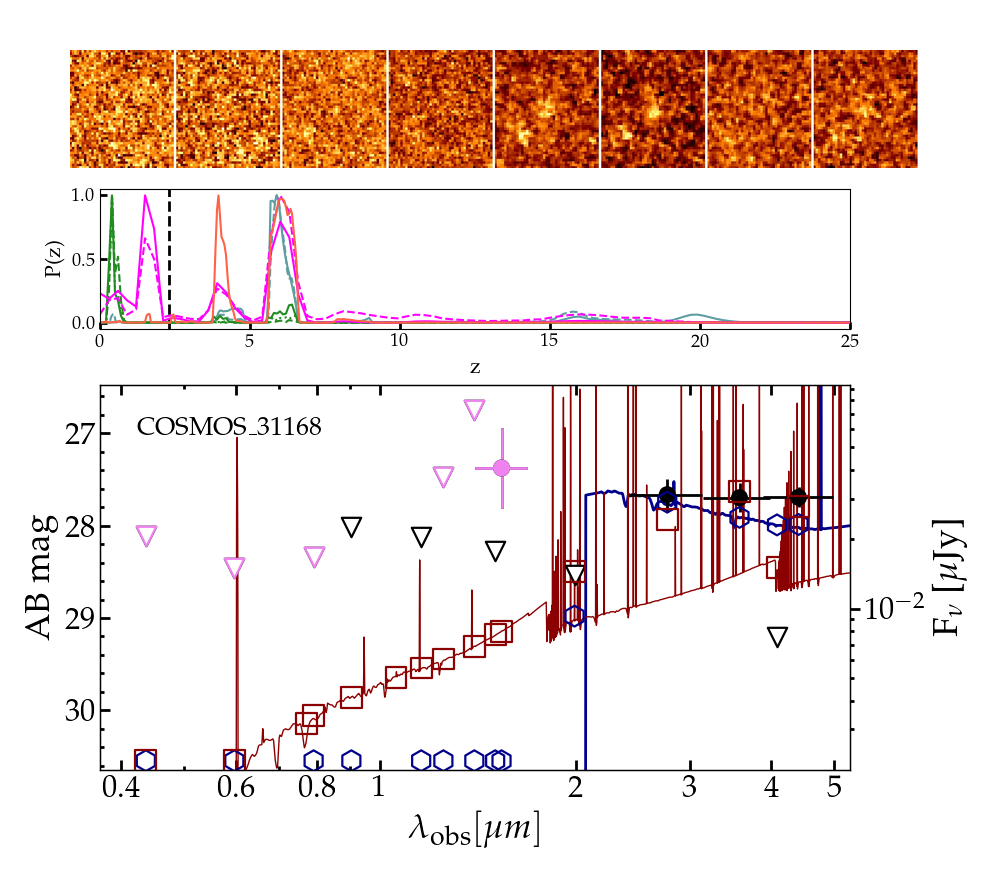}
\includegraphics[trim={0.5cm 0.5cm 0.5cm 0.1cm},clip,width=0.3\linewidth,keepaspectratio]{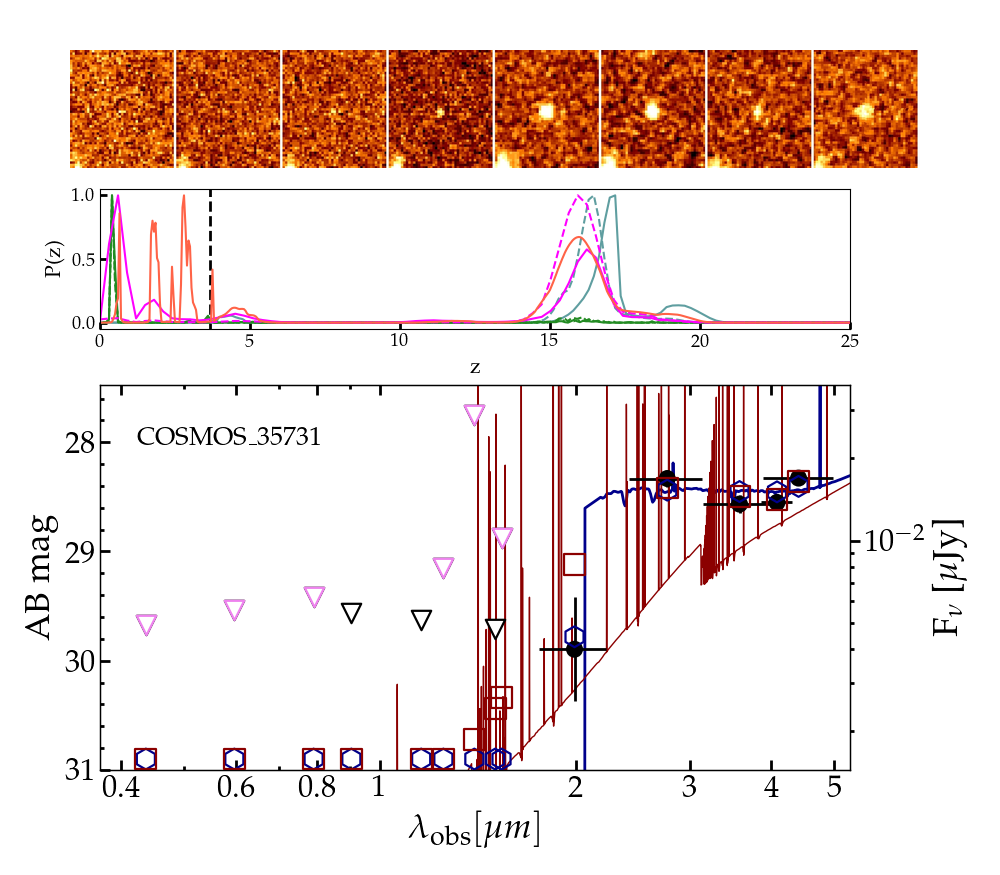}
\includegraphics[trim={0.5cm 0.5cm 0.5cm 0.1cm},clip,width=0.3\linewidth,keepaspectratio]{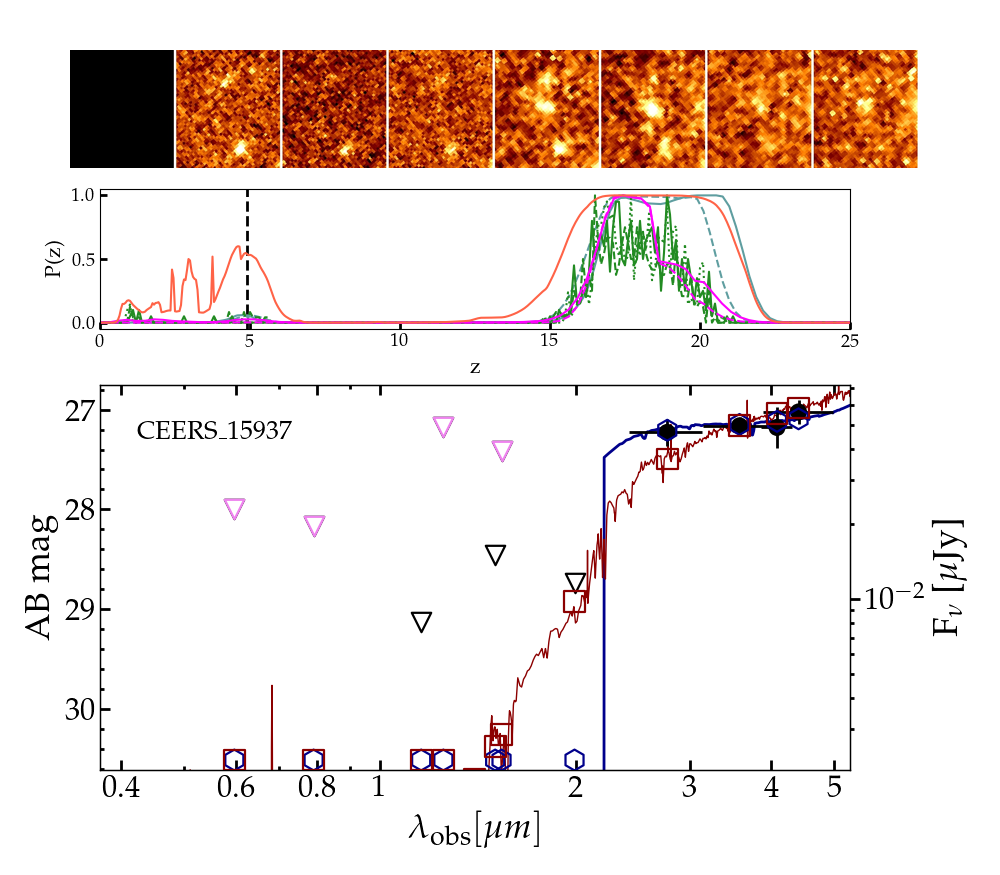}
\includegraphics[trim={0.5cm 0.5cm 0.5cm 0.1cm},clip,width=0.3\linewidth,keepaspectratio]{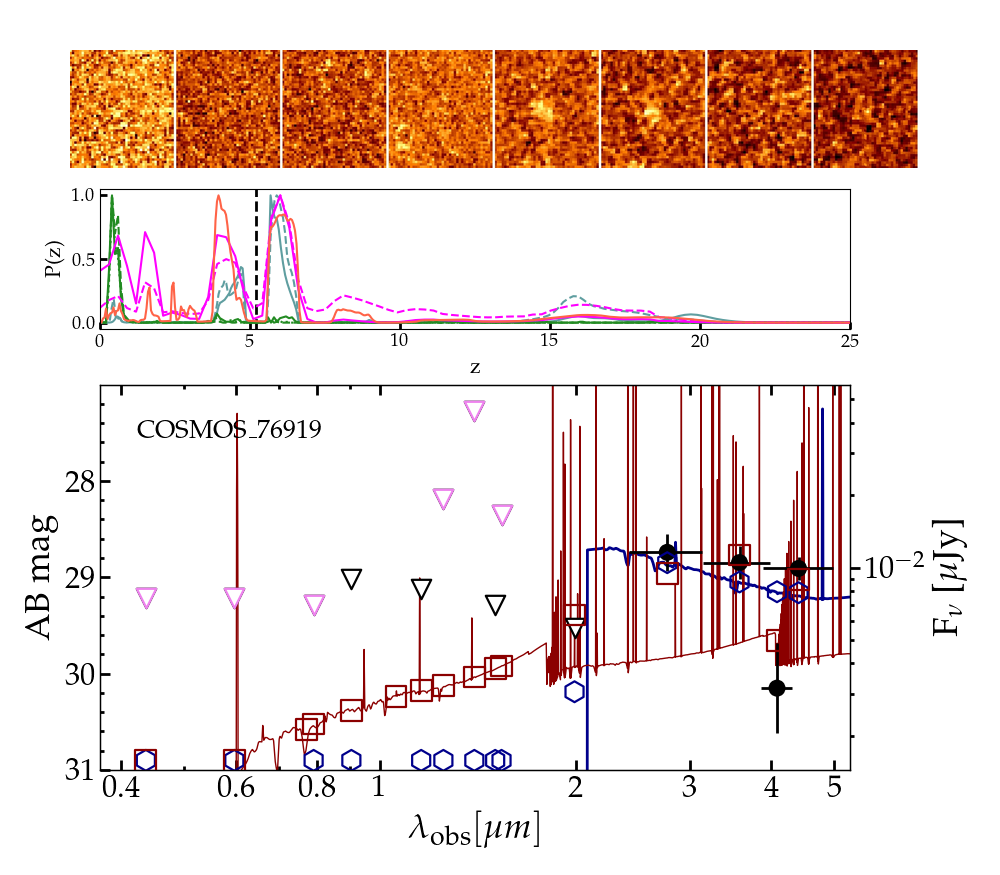}
\includegraphics[trim={0.5cm 0.5cm 0.5cm 0.1cm},clip,width=0.3\linewidth,keepaspectratio]{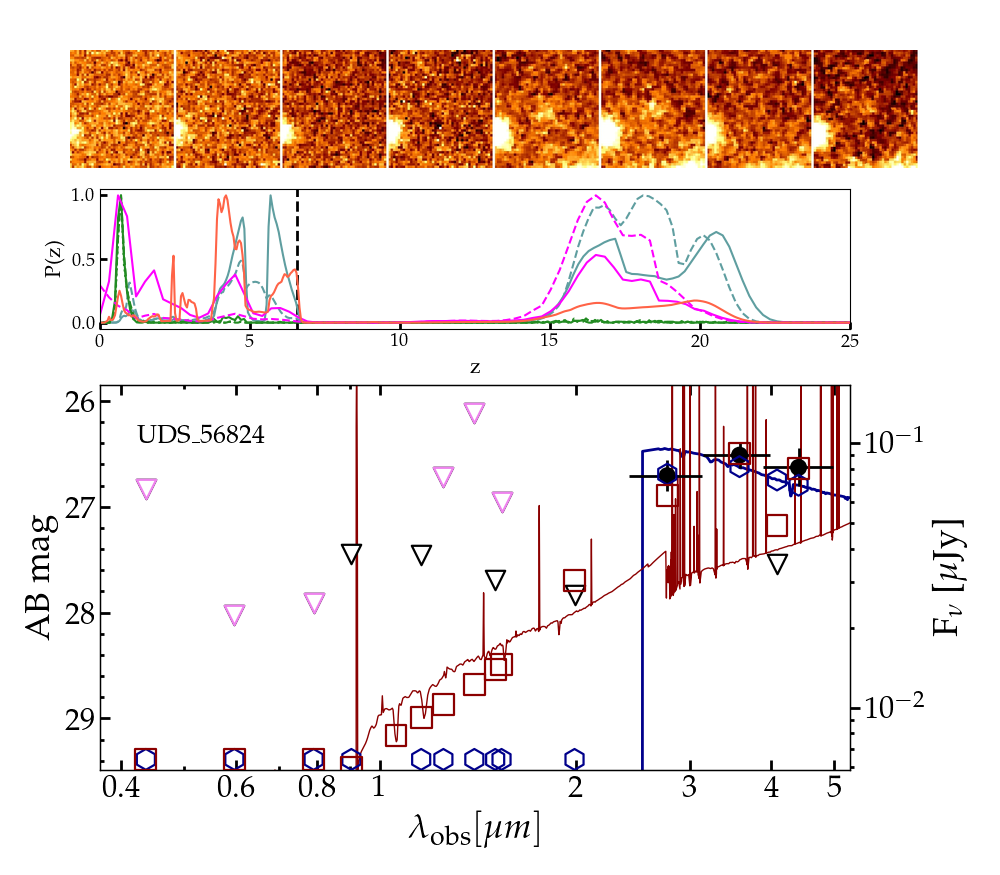}
\includegraphics[trim={0.5cm 0.5cm 0.5cm 0.1cm},clip,width=0.3\linewidth,keepaspectratio]{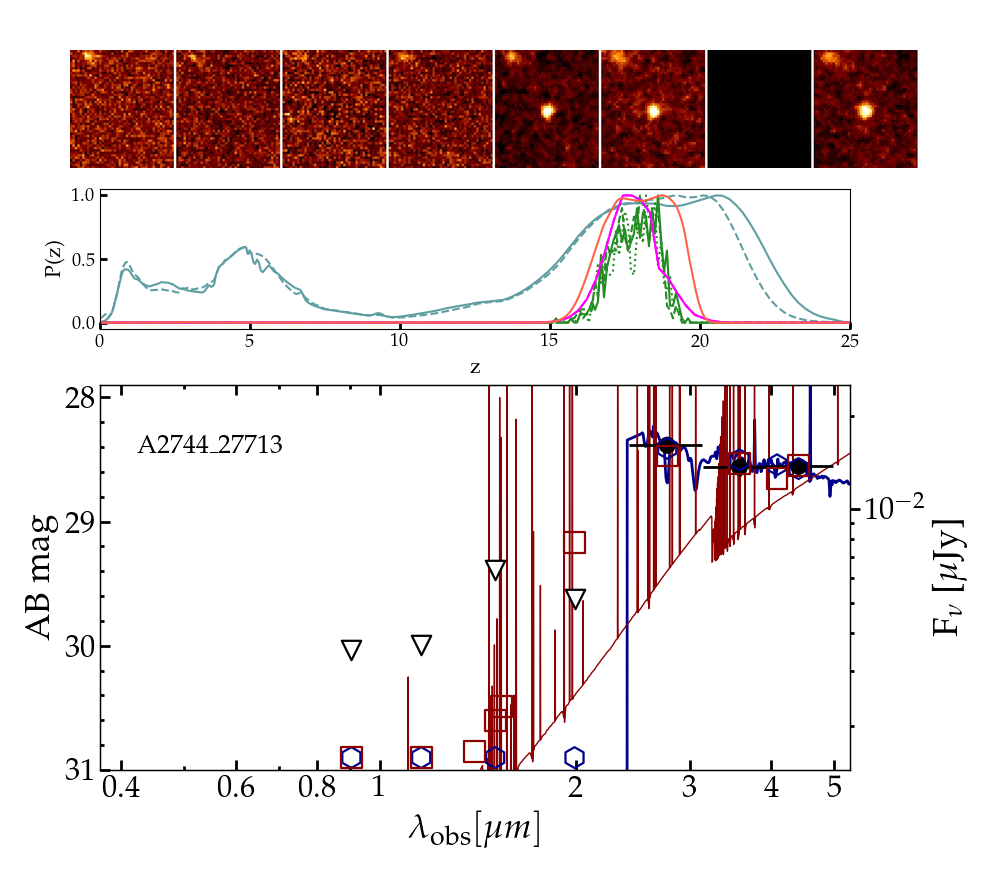}
\caption{Same as Fig.~\ref{fig_SEDs_ALL} for the confirmed interlopers of the F200W-dropout selection. The best-fit templates of objects COSMOS\_31168, COSMOS\_35731, COSMOS\_76919 and UDS\_56824 shown in red have been obtained with \textsc{zphot} after fixing the redshift at the spectroscopic values (black dashed line in the $P(z)$ panels). The $E(B-V)$ range for COSMOS\_35731, COSMOS\_76919 and UDS\_56824 has been constrained within the 1$\sigma$ range indicated by the Balmer decrement.}
\label{fig_CAPERS_SED}
\end{figure*}

Main properties (Table~\ref{tab:interlopers}) and SEDs (Fig.~\ref{fig_CAPERS_SED}) of the six known interlopers of the F200W-dropout selection. Objects COSMOS\_31168, COSMOS\_35731, COSMOS\_76919 and UDS\_56824 have been found to be at z$\sim$2-7 on the basis of NIRSpec observations carried out by the CAPERS survey (Sec.~\ref{subsec:CAPERS_interloper}).  Object A2744\_27713 is a transient which is detectable in the second-epoch GLASS-JWST observations (November 2022) when its position was covered only by NIRCam LW bands.

\begin{table*}[ht]
\caption{Confirmed interlopers of the F200W-dropout selection$^a$}\label{tab:interlopers}
\centering
\begin{tabular}{cccccccccc}
ID &          R.A. &        Dec &  z$_{spec}$ & F356W  & z$_{high}$ & z$_{low}$ & $\chi^2_{high}$ &$\chi^2_{low}$ & Type\\

 & deg. & deg. & &AB &  & & &  &\\
\hline

A2744\_27713 &  3.522866  & -30.351297  & - &  28.56 $\pm$ 0.07
 &   18.7& 2.9 &0.46 &3.28 & 1\\
CEERS\_15937 &  214.944272 &  52.835847 & 4.91 & 27.15 $\pm$ 0.09 &   17.2& 4.6 & 0.27 & 0.84 & 2\\
CEERS\_81406$^b$ & 214.914542   &  52.943021   & 4.91 &  26.52 $\pm$ 0.03 &   16.3& 4.6 &0.94 &3.90 & 3\\
COSMOS\_31168 &  150.180936 &  2.260756 & 2.29 &27.67 $\pm$   0.16 &  16.2 & 4.0& 2.85& 1.31& 3\\
COSMOS\_35731 &  150.133571 &  2.271020 & 3.66 &28.54 $\pm$   0.09 &   15.9& 2.8& 1.07& 0.88  & 3\\
COSMOS\_76919 &  150.184553 &  2.353510  & 5.18 & 28.71 $\pm$   0.18 &  16.1	& 4.0 &1.76 &0.74 & 3\\
UDS\_56824 & 34.454893 &  -5.215586  & 6.56 &26.51 $\pm$ 0.12  &   19.8& 4.2 &1.45 &1.93 & 3\\
\hline
\end{tabular}
\small \\a) ID, coordinates and F356W magnitudes from M24. The best-fit photometric redshift solutions have been obtained with \textsc{zphot} at $z>10$ ($z_{high}$) and $z<10$ ($z_{low}$). The last column report the classification discussed in Sect.~\ref{sec:SAMPLES} and  ~\ref{sec:INTERLOPERS} (1=transient; 2=quiescent; 3=dusty star-forming).\\b) Object CEERS-93316 from \citet{ArrabalHaro2023b}.
\end{table*}

\end{document}